\documentclass[aps,prx,reprint,showpacs,superscriptaddress,longbibliography]{revtex4-2}
\usepackage[english]{babel}
\usepackage{amsmath,amssymb,bbm,graphicx,color,comment}
\usepackage[bookmarks=true,colorlinks,citecolor=blue,urlcolor=blue]{hyperref}
\usepackage{dsfont}
\usepackage{soul}

\usepackage{amsthm}
\usepackage[
  ruled,            
  vlined,           
  linesnumbered,    
]{algorithm2e}

\usepackage{bbm}
\usepackage{float}

\usepackage{dcolumn}   
\usepackage{bm}        
\usepackage{filecontents}
\usepackage{lineno}

\usepackage{mathtools}
\usepackage[usenames,dvipsnames]{xcolor} 

\usepackage[export]{adjustbox}
\usepackage{adjustbox}

\usepackage{braket}
\usepackage{tikz}
\usepackage{bbold}

\usepackage{lineno}

\let\baraccent=\= \renewcommand{\=}[1]{\stackrel{#1}{=}}

\newtheorem{theorem}{Theorem}
\newtheorem{lemma}{Lemma}
\newtheorem{corollary}{Corollary}
\newtheorem{definition}{Definition}

\usepackage[normalem]{ulem}

\begin{document}

\title{Fast simulation of fermions with reconfigurable qubits}

\author{Nishad Maskara}
\affiliation{Department of Physics, Harvard University, Cambridge, MA 02138, USA}
\author{Marcin Kalinowski}
\affiliation{Department of Physics, Harvard University, Cambridge, MA 02138, USA}
\author{Daniel Gonz\'alez-Cuadra}
\affiliation{Department of Physics, Harvard University, Cambridge, MA 02138, USA}
\affiliation{Institute of Fundamental Physics IFF-CSIC, Calle Serrano 113b, 28006 Madrid, Spain}
\affiliation{Instituto de F\'isica Te\'orica UAM-CSIC, C. Nicol\'as Cabrera 13-15, Cantoblanco, 28049 Madrid, Spain}
\author{Mikhail D. Lukin}
\affiliation{Department of Physics, Harvard University, Cambridge, MA 02138, USA}

\date{\today}

\begin{abstract}
Performing large-scale, accurate quantum simulations of many-fermion systems is a central challenge in quantum science, with applications in chemistry, materials, and high-energy physics. Despite significant progress, realizing generic fermionic algorithms with qubit systems incurs significant space-time overhead, scaling as $O(N)$ for $N$ fermionic modes. 
Here we present a method for faster fermionic simulation with asymptotic space-time overhead of $O(\log(N))$ in the worst case, and $O(1)$ for circuits with additional structure, including important subroutines like the fermionic fast Fourier transform.
This exponential reduction is achieved by using reconfigurable quantum systems with non-local connectivity, mid-circuit measurement, and classical feedforward, to generate dynamical fermion-to-qubit mappings.
We apply this technique to achieve efficient compilation for key simulation tasks, including Hamiltonian simulation of the sparse Sachdev–Ye–Kitaev model and periodic materials, as well as free-fermion state-preparation.
Moreover, we show that the algorithms themselves can be adapted to use only the $O(1)$-overhead structures to further reduce resource overhead.
These techniques can lower gate counts by orders of magnitude for practical system sizes and are natively compatible with error corrected computation, making them ideal for early fault-tolerant quantum devices.
Our results tightly bound the computational gap between fermionic and qubit models and open new directions in quantum simulation algorithm design and implementation.
\end{abstract}

\maketitle

\begin{figure*}[t]
    \centering
    \includegraphics[width=0.99\linewidth]{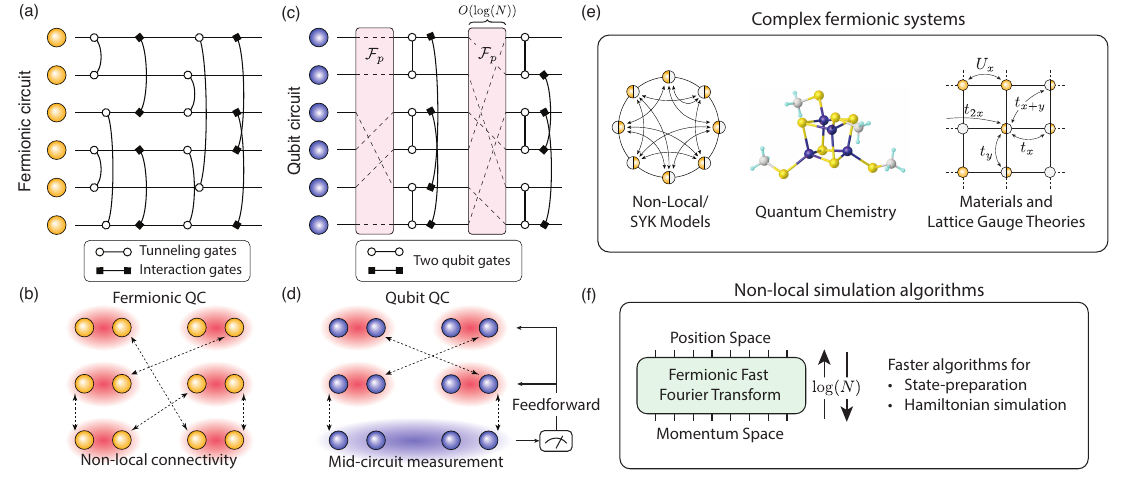}
    \caption{\textbf{Simulating arbitrary fermionic circuits with qubits.} (a) A fermionic circuit consisting of arbitrary non-local tunneling and interaction gates. (b) Such a circuit can be implemented on a quantum computer with natively fermionic degrees of freedom, such as a reconfigurable array of fermionic atoms.
    (c) Our key technical result is an algorithm to compile an arbitrary permutation of fermionic modes, with at most $O(\log(N))$ spacetime overhead. By alternating permutation and nearest-neighbor tunneling gates, this enables simulation of arbitrary circuits.  
    (d) These qubit circuits can be implemented on a quantum computer with reconfigurable connectivity. They also leverage mid-circuit measurement and fast classical feedforward, to efficiently simulate the non-local fermionic statistics. 
    (e) The ability to efficiently map arbitrary non-local fermionic circuits to qubits has signficant implications for the cost of a variety of quantum simulations of interest. We show that our approach can immediately be applied to reduce the asymptotic and practical cost of simulating complex fermionic systems, like non-local Sachdev-Ye-Kitaev (SYK) models, quantum chemistry in the plane-wave basis, and complex local lattice models like materials and lattice gauge theories.
    (f) Further, we show that the fermionic fast fourier transform (FFFT) can be implemented in qubits in depth $O(\log N)$, a factor of $N/\log N$ faster then existing implementations. We show that switching between position and momentum space, is a powerful resource for state-preparation and Hamiltonian simulation, as it can be leveraged to efficiently prepare certain long-range ordered states, and generate translation-invariant Hamiltonian terms. 
    }
    \label{fig:fig1}
\end{figure*}

Achieving accurate, large-scale simulations of interacting fermions is a central challenge in computational science with broad  impact in many  scientific domains ranging from materials science~\cite{Bauer_2020} and chemistry~\cite{mcardle_quantum_2020} to high-energy physics~\cite{di_meglio_quantum_2024} and quantum gravity~\cite{maldacena2023simplequantumdescribesblack,preskill2022physicsquantuminformation}.
Quantum computation is emerging as a promising approach for advancing such simulation capabilities~\cite{abrams_simulation_1997}, as quantum devices can natively capture challenging quantum mechanical effects including large-scale superpositions and entanglement~\cite{feynman_simulating_1982}.
However, state-of-the-art quantum computers are built from qubits, which correspond to essentially bosonic systems~\cite{altman_quantum_2021}, and techniques for simulating fermions must account for long-range correlations associated with their non-trivial statistics.
To address this challenge, a number of fermion-to-qubit mappings have been developed~\cite{jordan_uber_1928, bravyi_fermionic_2002,verstraete_mapping_2005}.
One class of mappings, introduced by Verstrate and Cirac~\cite{verstraete_mapping_2005}, can simulate circuits with fixed, local connectivity by introducing ancilla qubits and replacing each fermionic operation with a constant number of qubit operations~\cite{derby_compact_2021, chen_equivalence_2023}. These methods leverage fundamental connections to spin liquids and topological quantum matter~\cite{kitaev_anyons_2006, kells_description_2009, Evered_2025, dyrenkova_scalable_2025, Nigmatullin_compactencoding_2025}.
Another class of mappings, introduced by Bravyi and Kitaev~\cite{bravyi_fermionic_2002}, enables arbitrary connectivity, by replacing fermionic operators with $\log(N)$-sized non-local qubit operators~\cite{Jiang_optimal_ternery_2020,Havlick_operator_locality_fermions}. However, when applied to simulating parallel fermionic circuits, overlap between these operators leads to additional compilation costs and time overhead.

Many important quantum simulation algorithms centrally involve non-local fermionic circuits
~\cite{verstraete_quantum_2009, babbush_low-depth_2018,jiang_quantum_2018,kivlichan_quantum_2018,low_hamiltonian_2019}.
Non-local interactions appear in effective models relevant to chemistry and quantum gravity~\cite{mcardle_quantum_2020,maldacena2023simplequantumdescribesblack}, while non-local transformations are useful subroutines in materials and high-energy physics simulations~\cite{low_hamiltonian_2019}.
As such, significant works have focused on designing circuits that minimize the fermion-encoding overhead for specific applications~\cite{clinton_towards_2024,jiang_quantum_2018,kivlichan_quantum_2018,hashim_optimized_2021,ogorman_generalized_2019,hagge_optimal_2022,derby2021compactfermion2,paciani_quantum_2025,chiew_optimal_2025,li_accelerating_2025,chiew_discovering_2023,parella_dilme_reducing_2024,wang_ever_more_optimized_2023,steudtner_fermion_qubit_2018,Miller_grow_your_own_2023}.
Recent theoretical work have also proposed quantum computers based on reconfigurable arrays with natively fermionic atoms~\cite{gonzalez-cuadra_fermionic_2023, schuckert_fermion, ott_error-corrected_2024}, which can implement fermionic simulations without encoding overhead~\cite{Hartke_2023, Shao_2024, Bourgund_2025, Xu_2025, zache_fermion-qudit_2023, tabares_programming_2025,gkritsis_simulating_2024}.
Hence, developing fermionic simulation methods is critical for guiding the design of future quantum simulation algorithms and architectures~\cite{altman_quantum_2021}.

Here, we present a new method for fast fermion simulation which generates $O(\log(N))$ overhead per fermionic operation and maintains full parallelism (Fig.~\ref{fig:fig1}).
Our approach uses a \textit{dynamical} fermion-to-qubit mapping~\cite{kivlichan_quantum_2018}, where the encoding is modified during the computation such that the fermionic operations are local at every given step.
To rapidly switch between different encodings, we leverage non-local connectivity, ancilla qubits, mid-circuit measurements, and classical feedforward, operations which are also used extensively for manipulating quantum error correcting codes~\cite{raussendorf_measurement-based_2003,bluvstein_logical_2024}.
In addition to our general compilation result with $\log(N)$ overhead, we show that a large class of important fermionic circuits can be simulated with $O(1)$-overhead, producing effectively no overhead versus a native fermionic computer.

These  results represent a conceptual and practical advance in the simulation of fermionic systems using qubit-based quantum computers. By showing that arbitrary fermionic circuits can be implemented with negligible space-time overhead, scaling only as $O(\log(N))$, we effectively close the longstanding question of whether fermions can provide a significant computational advantage over qubits~\cite{bravyi_fermionic_2002}.
Not only is the overhead asymptotically small, but the constant prefactors for realistic fermionic circuits are modest. We illustrate this in detail for non-local Hamiltonian simulation and the fermionic fast fourier transform (FFFT).
The circuits presented here also have simple, parallel structure, making them promising for experimental realization on platforms supporting non-local or reconfigurable qubit connectivity, including neutral atom arrays~\cite{bluvstein_logical_2024,manetsch2024tweezerarray6100highly}, trapped-ions~\cite{Pino_ionCCD_2021,Moses_racetrack_2023,Iqbal_2024}, and bi-planar superconducting qubit arrays~\cite{bravyi_high_threshold_2024}.
Finally, we discuss the resource overhead of our techniques in the context of error-corrected computation. The dynamical fermionic encoding is composed of Clifford gates, making them natively compatible with fault-tolerant computation~\cite{gottesman1997stabilizercodesquantumerror, Nielsen_Chuang_2010}.

\section*{Non-local Fermionic Circuits}

\begin{figure*}
    \centering
    \includegraphics[width=\linewidth]{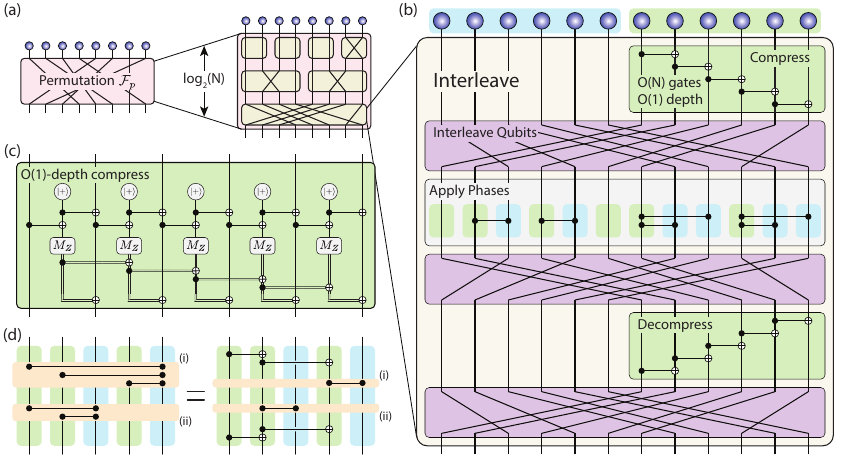}
    \caption{\textbf{Low-overhead fermionic permutations with qubits.} (a) A key ingredient in our approach to simulating fermionic circuits with qubits is implementing an encoded fermionic permutation $\mathcal{F}_p$. Here, we show how $\mathcal{F}_p$ can be implemented with $O(N)$ ancilla qubits and depth $O(\log(N))$, by applying multiple layers of \textit{interleave} operations. (b) An interleave $\mathcal{I}_p$ is defined as a fermionic permutation, where modes are partitioned into two groups $\rm A$ (blue) and $\rm B$ (green), such that the ordering within partitions is preserved. The implementation proceeds in six steps. First, a CNOT cascade is implemented on group $\rm B$. Then, qubits are permuted based on the target interleave. Next, a layer of CZ gates is applied, connecting each mode in $A$ to it's closest mode in $B$ on the left. Then the CNOT cascade is inverted, by reversing the permutation, and applying an inverse cascade. Then, the permutation is applied one more time, placing qubits in their final position.
    (c) Even though the CNOT circuit as drawn spreads information non-locally and has depth $O(N)$, standard techniques from measurement-based quantum computing can be leveraged to compile it into an $O(1)$-depth circuit using ancilla qubits, mid-circuit measurement, and non-local classical feedforward. Here, we show the compilation for the 1D CNOT cascade, although a similar decomposition with $O(1)$-depth and $O(N)$ ancilla exists for arbitrary CNOT circits with $O(N)$ gates.
    (d) The interleave effectively implements a fermionic permutation, by conjugating a single layer of CZ gates by the CNOT cascade. The effect of this conjugation is to broadcast a single CZ gate (right) into $O(N)$ gates (left), so that each mode in $\rm A$ acts pairwise with all modes in $\rm B$ to the left of the original. 
    Hence, $O(N^2)$ CZ gates are effectively generated, even though only $O(N)$ CZ and CNOT gates were applied. This ability to compress $O(N^2)$ gates into $O(N)$ gates is key to the asymptotic speedup afforded by our approach.
    }
    \label{fig:fig2}
\end{figure*}

The fermionic circuits we simulate are composed of layers of parallel two-fermion gates with arbitrary connectivity (Fig.~\ref{fig:fig1}a)~\cite{bravyi_fermionic_2002}, a model which captures parallel fermionic computation with reconfigurable connectivity (Fig.~\ref{fig:fig1}b).
When mapping to qubits, fermionic permutation operations $\mathcal{F}_p$ are inserted between each layer.
These dynamically change the encoding, so that each layer of two-fermion tunneling gates act on adjacent modes, and can be simulated with two-qubit operations (Fig.~\ref{fig:fig1}c).
Hence, the entirety of the cost associated with simulating fermion statistics is absorbed into implementing permutations.

To compile arbitrary $\mathcal{F}_p$ into qubit circuits with very low overhead, our approach leverages non-local qubit connectivity, mid-circuit measurement, fast classical computation, and non-local feedforward (Fig.~\ref{fig:fig1}d).
While existing techniques for implementing $\mathcal{F}_p$ have gate cost $O(N^2)$ and depth $O(N)$~\cite{kivlichan_quantum_2018}, we achieve an asymptotic reduction in the cost of implementing $\mathcal{F}_p$ using $O(N \log N)$ gates arranged in depth $O(\log N)$.

To highlight the utility of this technique, we show how non-local fermion operations can accelerate a variety of pratical simulation tasks (see Fig.~\ref{fig:fig1}e).
In particular, we consider simulation of Sachdev-Ye-Kitaev (SYK) models~\cite{sachdev_gapless_1993, kitaev_simple_2015, xu_sparse_2020}, which have non-local, four-body interactions, exhibit strong quantum chaos, and are believed to be holographically dual to toy models of quantum gravity.
We study sparse versions of these models, and show that using our approach, simulations can be accelerated 10-100x for relevant instances.
Other complex fermionic Hamiltonians can also be simulated using the same techniques, including sparsified models of quantum chemistry~\cite{günther2025phaseestimationpartiallyrandomized}.
Next, we show how our approach unlocks fast simulation subroutines, which were previously bottlenecked by non-local fermionic connectivities.
Specifically, we discuss in detail the fermionic fast Fourier transform (FFFT), which is an $O(\log(N))$ depth fermionic circuit that transforms from real-space to momentum-space,
and reduce the qubit overhead to $O(1)$, comparable to approaches based on native fermions~\cite{schuckert_fermion}.
Access to such an efficient FFFT has broad implications for quantum simulation algorithms, as it enables efficient preparation of arbitrary translation-invariant free fermion states, and is a core subroutine in state-of-the-art Hamiltonian simulation algorithms for chemistry, materials and high energy physics~\cite{babbush_low-depth_2018}.

\section*{Fast Fermion Permutations}

At each step of the dynamical encoding, $N$ fermionic modes are encoded into $N$ qubits via the static Jordan-Wigner (JW) encoding~\cite{jordan_uber_1928}.
The JW encoding replaces each fermionic operator with an equivalent weight $O(N)$ qubit operator, which captures the non-local correlations arising from fermionic statistics.
However, the encoding is not unique, and depends on the \textit{order} in which fermionic modes are encoded into qubits.
For each ordering, we introduce a map $m$, such that fermionic mode $i$ is mapped to qubit $m(i)$.
One of the key properties of the JW encoding, is that the cost of simulating tunneling between two fermion modes $i$ and $j$, depends on their encoded distance $|m(i) - m(j)|$~\cite{steudtner_fermion_qubit_2018}.
If the modes are adjacent, i.e. $|m(i) - m(j)| = 1$, then tunneling can be simulated with a single two-qubit gate (see methods).
Thus, by making the mapping time-dependent, we can dynamically change which tunneling operations can be performed efficiently with qubits (Fig.~\ref{fig:fig1}c).

Transformations between JW encodings are implemented using fermionic permutation operators $\mathcal{F}_p$.
In particular, for an input encoding specified by $m_{in}$, after applying $\mathcal{F}_p$, the new JW encoding is specified by $m_{out}$ such that
\begin{align}
    m_{out}(i) = p(m_{in}(i))
\end{align}
where $p$ is a function that permutes the qubit indices.
In a programmable fermionic processor, an analagous operation is performed by moving fermionic atoms~\cite{gonzalez-cuadra_fermionic_2023}, which effectively changes the connectivity of the system.
In our approach, since we are working with encoded qubits, realizing the transformation requires applying a non-trivial quantum circuit, which preserves the non-local structure of the JW mapping. 

Our main technical result is an algorithm for implementing arbitrary $\mathcal{F}_p$ with $O(N \log N)$ two-qubit gates, arranged in a circuit of depth $O(\log N)$.
The key subroutine we introduce is a fast procedure for generating \textit{interleaves} with encoded fermions, using $O(N)$ two-qubit gates arranged in $O(1)$-depth (see Fig.~\ref{fig:fig2}).
The interleave $\mathcal{I}_p$ is a special class of permutations where the modes can be partitioned into two groups, $A$ and $B$, such that the crossings only occur \textit{between} the groups.
Arbitrary permutations $\mathcal{F}_p$ can be compiled into $\log_2(N)$ interleave operation (see Fig.~\ref{fig:fig2}a), via a recursive construction based on the mergesort algorithm~\cite{knuth_art_1997}. 
To understand the intuition, imagine we have applied a permutation to the first $N/2$ and last $N/2$ modes, such that they are ordered based on their final positions. Then, a single interleave of depth-$O(1)$ is sufficient to generate the target permutation of $N$ modes. 
At the next layer of recursion, each permutation of $N/2$ modes  can be further split into arbitrary permutations of $N/4$ modes, and interleaves. During each iteration, the size of the arbitrary permutation is cut in half, so the total number of recursion layers needed is $\log_2(N)$.

Our procedure for implementing an interleave is depicted in detail in Fig.~\ref{fig:fig2}b.
We partition the modes into two contiguous groups $A$ and $B$.
In the first step of the circuit, we apply a CNOT cascade to qubits in group $B$. 
Then, we reconfigure the qubits, based on the target permutation, and apply a layer of CZ gates between each qubit in group $A$, and the qubit in $B$ closest and to its left.
Finally, we reverse the compression step by permuting qubits, applying the inverse CNOT cascade, and applying the qubit permutation one final time.
While the CNOT cascade is naively a depth $O(N)$ circuit, we can simplify it by introducing a single-ancilla per CNOT, perform a depth-2 CNOT circuit, and use measurement and classical Pauli-feedforward, to perform the same operation (see Fig.~\ref{fig:fig2}c).
We note more generally that any Clifford circuit with $O(1)$ gates per qubit can be parallelized efficiently into a constant-depth circuit with $O(1)$ ancilla's per qubit~\cite{raussendorf_measurement-based_2003,van_den_nest_graphical_2004}. Hence, the key metric we emphasize is the total number of Clifford gates per qubit.

To understand why this circuit structure realizes the target fermionic permutation, it is helpful to first understand the standard approach, a \textit{fermionic SWAP (FSWAP) network}.
These networks permute qubits, and apply CZ gates between all pairs of modes which cross. As such, they in general involve $O(N^2)$ gates~\cite{kivlichan_quantum_2018}.
In contrast, the bi-partite structure of the interleave enables our $O(1)$-overhead implementation, that effects the same transformation.
The key intution is that the CNOT cascade effectively compresses information in region $B$, such that $O(N)$ CZ gates becomes sufficient.
When the layer of CZ gates is conjugated by the CNOT cascade, it makes it so each CZ gate coupling a mode in $A$ to it's rightward-adjacent mode in $B$ is \textit{broadcasted} into $O(N)$ CZ gates coupling modes in $A$ to all rightward-modes in $B$ (see Fig.~\ref{fig:fig2}c). 
As only modes in $A$ and $B$ cross, this realizes all of the CZ's in the FSWAP network.

\begin{figure}
    \centering
    \includegraphics[width=1.0\linewidth]{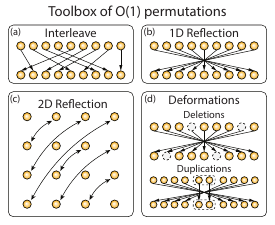}
    \caption{\textbf{Toolbox of efficient fermionic permutations}. (a) There are a variety of structured fermionic permutations that can be performed with $O(1)$-overhead, in addition to the interleave. (b) These operations include a 1D reflection, where the order of the fermionic modes is reversed. 
    (c) It also includes a more intricate 2D reflection, which switches from a row-wise ordering where the index of the mode at position $(r,c)$ is $rL + c$, to a column-wise ordering where the index is $cL+r$. None of these operations can be compiled using a constant number of the others, and hence belong to different classes. (d) Simple deformations of any fermionic permutation can also be generated with only $O(1)$ additional layer of gates (see SI). The generators of these deformations include deleting a mode and duplicating a mode. This large class of $O(1)$-overhead fermionic permutations can be used for co-designing fermion algorithms with $O(1)$-encoding overhead, like the 2D FFFT and momentum-space state preparation algorithm discussed in Fig.~\ref{fig:fig5}.}
    \label{fig:fig3}
\end{figure}

The interleave is part of a larger toolbox of $O(1)$ fermionic permutation operations we have identified (see Fig.~\ref{fig:fig3}). 
All the operations leverage the same high-level structure, where a CNOT circuit is carefully constructed to compress the FSWAP network of $O(N^2)$ CZ gates to an efficient one with $O(N)$ Clifford gates (see SI for more details). Two fundamental structures include 1D and 2D reflections, which cannot be generated using a constant number of interleaves.
In the SI, we present a systematic framework for generating efficient CNOT + CZ circuits for simulating fermionic permutations. 
This flexible toolbox provides opportunities for co-designing fermionic algorithms that effectively leverage fast, $O(1)$-depth fermionic operations.
For example, the 2D reflection is a useful subroutine for fermionic simulation in $D > 1$, and is used in our $O(1)$-overhead implementation of the 2D FFFT discussed below.

\section*{Random graphs and sparse SYK model}

One application of non-local fermionic circuits is quantum simulation of unstructured and sparse fermionic models. A particularly important class are the Sachdev-Ye-Kitaev models, which are known to exhibit quantum chaos, and conjectured to be dual to simple toy-models of gravity~\cite{sachdev_gapless_1993, kitaev_simple_2015}.
The SYK models take the general form of an interacting fermionic model
\begin{align}
    H_\text{SYK} = \frac{1}{K}\sum_{ijkl} J_{ijkl} \chi_i \chi_j \chi_k \chi_l\,,
\end{align}
where $\chi_i$ are \textit{majorana fermion} operators, which are related to standard fermionic operators by
$\chi_{2i} = (c_i + c_i^{\dagger})/\sqrt{2}$ and $\chi_{2i+1} = (c_i - c_i^{\dagger})/\sqrt{2}$.
A majorana permutation can be implemented by a simple modification of the equivalent fermionic permutation (see methods), allowing Theorem \ref{thm:circuit_reduction_mth} to be directly extended.

\begin{figure}
    \centering
    \includegraphics[width=1.0\linewidth]{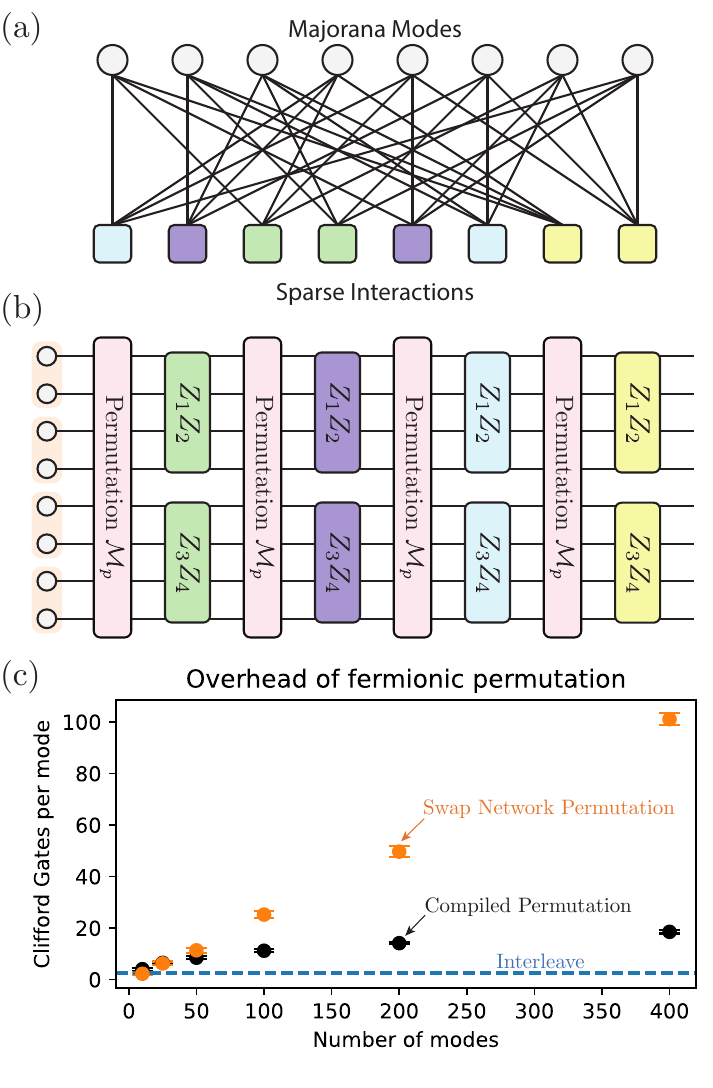}
    \caption{\textbf{Compiling non-local SYK models}. (a) We consider non-local models where each mode is involved in $d$ distinct four-majorana interactions, described by a constant-degree and $d$-colorable interaction graph. (b) A single Trotter cycle or block encoding of these Hamiltonians can be implemented by alternating Majorana permutations and two-qubit Ising interactions. (c) We use numerical optimization to compute explicit CNOT circuits for randomly sampled permutations, and present the corresponding Clifford overhead per mode after optimization (black), and for the standard FSWAP network (orange). Further, random interleaves produce constant-overhead circuits (blue dotted line), and the corresponding models still exhibit signatures of strong chaos (see methods).}
    \label{fig:fig4}
\end{figure}

To compile a circuit for Hamiltonian simulation, we consider a standard procedure where we start by dividing the interactions into non-overlapping groups $H_\text{SYK} = \sum_{\alpha} H^{\alpha}_\text{SYK}$ by solving a graph coloring problem~\cite{childs_exponential_2003,aharonov2003adiabaticquantumstategeneration} (see Fig.~\ref{fig:fig4}a), such that the $\alpha$-th step of a simulation cycle will involve $H^{\alpha}_\text{SYK}$. In a Trotter-based method, we evolve under $e^{-i H^{\alpha}_\text{SYK} dt}$ and for a QSP-based method we query a block-encoding of $H^{\alpha}_\text{SYK}$~\cite{Low_hamiltonian_simulation_2019,Babbush_linear_T}. 
For both of these approaches, we transition from step $\alpha$ to $\alpha+1$ by applying a majorana permutation, such that $H^{\alpha}_\text{SYK}$ becomes a local qubit Hamiltonian. Local four-majorana interactions take the form
\begin{align}
    \chi_{2i} \chi_{2i+1} \chi_{2i+2} \chi_{2i+3} = Z_i Z_{i+1},
\end{align}
and so by alternating majorana permutations and Ising interactions, we can implement generic quartic majorana models (see Fig.~\ref{fig:fig4}b).

\begin{figure*}
    \centering
    \includegraphics[width=\linewidth]{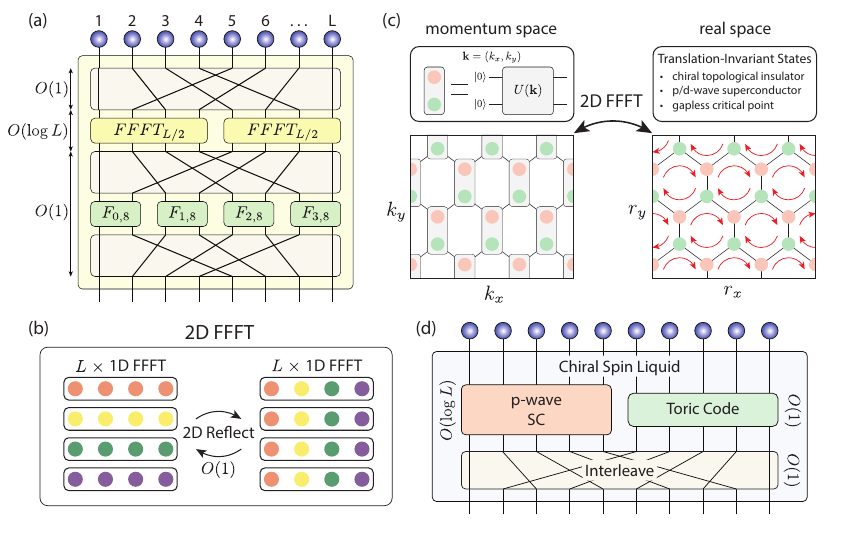}
    \caption{\textbf{FFFT and low-depth state preparation:} (a) The 1D FFFT on $L$ fermionic modes can be implemented in $\log (L)$ steps of $O(1)$ depth, each of them consisting of parallel interleave operations (see Fig.~\ref{fig:fig1}a) and one layer of entangling gates~\cite{verstraete_quantum_2009,Ferris_spectral_2014}. (b) The 2D FFFT on a $L\times L$ lattice can also be implemented efficiently. First, $L$ 1D FFFT's are applied in parallel to transform the $x$-coordinate from position to momentum space. Then, the order is modified with a $O(1)$-depth 2D reflection operation (see Fig.~\ref{fig:fig3}c), switching $x$ and $y$ coordinates.
    Then, another set of $L$ 1D FFFT's are applied to transform the $y$ coordinate. Finally, the original order is restored with another 2D reflection. (c) Translational-invariant free-fermion states can be prepared with a $O(\log L)$-depth circuit, by first preparing them in momentum space, and then transforming to real space using the FFFT supplemented with $O(1)$ permutations (see main text). In the figure, we illustrate the protocol for the case of a two-site unit cell, which can prepare e.g. a chiral topological insulator. The procedure can be equally applied to superconducting states as well as more complicated periodic structures (see SI). (d) Our techniques can be combined with fast preparation of topological states~\cite{raussendorf_measurement-based_2003}, to exactly prepare exotic phases like a non-abelian chiral spin liquid, in depth $O(\log L)$, by interleaving a p-wave superconductor with a toric code state.}
    \label{fig:fig5}
\end{figure*}

An especially interesting case for quantum simulation are \textit{sparsified} versions of the SYK model~\cite{xu_sparse_2020}, where the degree of the interaction graph $d$ is independent of the total number of modes $N$~\cite{orman_quantum_2024}. 
We consider two settings, and show how our techniques can be applied (see Fig.~\ref{fig:fig4}c).
In the first setting, we estimate the resources required to compile a randomly generated sparse-SYK instance. Numerical results suggest that despite this compilation overhead, one simulation cycle on $N=400$ modes can be performed with $\leq 20 \times d$ Clifford gates per mode per cycle.
In the second setting, we introduce a special class of SYK models, generated by inserting a random majorana \textit{interleave} between Ising interactions. 
By numerically simulating dynamics for small-size systems up to $N=14$, we observe the interleave permutations still exhibit signatures of chaos (see methods), suggesting the overall cost can be reduced to $\leq 2.5 \times d$ Clifford gates per mode per cycle. 
In contrast, if we were to implement the sparse SYK models using standard FSWAP networks, the number of Clifford gates per mode would grow linearly with system size (see Fig.~\ref{fig:fig3}c), representing multiple orders-of-magnitude higher cost for interesting instances with hundreds of modes.

\section*{Quantum simulations with the FFFT}

The ability to efficiently implement non-local fermionic circuits also unlocks new algorithms for state-preparation and Hamiltonian simulation of quantum materials.
One key subroutine we discuss is the fermionic fast fourier transform (FFFT)~\cite{verstraete_quantum_2009,Ferris_spectral_2014}, which enables fast transformation from position-space to momentum-space using a $\log(L)$-depth circuit, where $L = N^{1/D}$ for a $D$-dimensional lattice. Starting with the 1D FFFT (see Fig.~\ref{fig:fig5}a), we see that the non-local reconfiguration at each step is a special case of the interleave operation, and hence can be implemented with one application of $\mathcal{I}_p$ and $O(1)$ overhead. 
Higher-dimensional FFFT can be implemented by combining the 1D FFFT with reflection operations (see Fig.~\ref{fig:fig5}b), which are part of the expanded toolset of $O(1)$ operations (see Fig.~\ref{fig:fig3}).
This represents an exponential improvement in scaling of circuit depth with $\log(N)$ compared to prior approaches (see Table~\ref{table:2D_FFFT_asymptotics}), including state-of-the-art static encodings which efficiently generate higher-dimensional connectivity~\cite{derby_compact_2021}.

\begin{table}[h]
    \begin{tabular}{c | c c | c} 
     scalings & 1D FFFT & 2D Reflection & 2D FFFT \\ 
     \hline
     swap network~\cite{kivlichan_quantum_2018} & $O(L)$ & $O(L^2)$ & $O(L^2)$ \\ 
     compact encoding~\cite{derby_compact_2021} & $O(L)$ & N/A & $O(L)$ \\
     dynamic JW & $O(\log L)$ & O(1) & $O(\log L)$ \\
    \end{tabular}
    \caption{Asymptotic scalings of the Clifford gates per qubit required to implement a 2D FFFT, for a $N = L \times L$ square lattice. The compact encoding is a mapping with local gates in 2D, and dynamic JW represents the scheme developed in this work.}
    \label{table:2D_FFFT_asymptotics}
    \end{table}


Arbitrary, translation-invariant free-fermion states can be prepared in momentum space with constant-depth circuits, and then transformed via the FFFT to real-space, providing a remarkably efficient route to preparing various forms of long-range ordered quantum states (Fig.~\ref{fig:fig5}c). Entanglement between momentum modes related by reflection (i.e. $(k,-k)$) is also enabled using the reflections.
For example, complex states such as chiral superconductors, critical states, and fermi-surfaces, can be exactly prepared using this technique, with a total circuit depth $O(\log(L))$ (see SI).
In contrast, a device with strictly local fermionic operations in $D$-dimensions, would require at least $O(L)$ time to prepare this class of states. 
State-prep with FFFT can be further combined with techniques for preparing topologically ordered states with measurement and feedforward, to prepare exotic phases (Fig.~\ref{fig:fig5}d). For example, the non-abelian Kitaev $B$ phase~\cite{kitaev_anyons_2006}, a canonical example of a chiral spin liquid, can be prepared in $\log(L)$-depth by combining a chiral $p+ip$ superconducting state with toric-code order (see SI)~\cite{Schmoll_honeycomb_2017}.

The FFFT can also enable advanced quantum simulation algorithms. In particular, consider simulation of a generalized Hubbard Hamiltonian
\begin{align}\label{eq:generalized_hubbard}
    H &= \sum_{x,y,\sigma} t_{x-y,\sigma} c_{x,\sigma}^{\dagger} c_{y,\sigma} + \sum_{x,\sigma} U_{x,\sigma} n_{x,\sigma} \\
    &+ \sum_{x,y,\sigma,\sigma'} V_{x-y,\sigma \sigma'} n_{x,\sigma} n_{y,\sigma'} \nonumber\\
    &\equiv T + U + V \nonumber
\end{align}
consisting of three parts: a kinetic term $T$, on-site interaction term $U$, and two-fermion interaction $V$, which is also assumed to be translationally-invariant.
In Ref.~\cite{babbush_low-depth_2018}, it was shown how a \textit{generic} electronic structure Hamiltonian can be written in this form, using the plane-wave dual basis.
Further, in Ref.~\cite{low_hamiltonian_2019}, it was shown how the diagonal terms $(U+V)$ can be implemented with asymptotic circuit depth scaling as $O(\log(N))$, using a \textit{classical} fourier-transform to diagonalize the density operator $n_{x,\sigma} \rightarrow n_{k,\sigma}$.
The kinetic $T$ operator can similarly be diagonalized using the FFFT; previously this was the asymptotic bottleneck~\cite{schuckert_fermion}.
However, putting it together with our results, a single application of \eqref{eq:generalized_hubbard} (Trotter-step or block-encoding) can be implemented with asymptotic depth $O(\log(N))$, an improvement of $O(N/\log(N))$ over prior methods.

\section*{Fault-tolerant Implementation}

A major benefit of our approach to simulating fermionic circuits is that it is naturally compatible with error-correction and fault-tolerant quantum computing techniques.
In particular, the circuits used to generate $\mathcal{F}_p$ are composed entirely of CNOT and CZ gates, along with state-preparation and measurement in the Pauli basis.
All of these operations can be performed efficiently using transversal gates~\cite{gottesman1997stabilizercodesquantumerror}, or lattice surgery~\cite{fowler2019lowoverheadquantumcomputation}. Further, when transversal gates are accessible, the quantum error-correcting codes used to store the ancilla, can be prepared with $O(1)$-rounds of stabilizer measurement~\cite{cain_correlated_decoding_2024, zhou2024algorithmicfaulttolerancefast, bluvstein_logical_2024}, which leads to a significant savings in practice. In particular, as the interleave can be compiled into a depth $\leq {5}$ Clifford circuit (see Fig.~\ref{fig:fig2} and SI), it is approximately as costly as one-round of syndrome extraction, a minimal cost from the point-of-view of fault-tolerant architectures~\cite{Zhou_architecture_2025_ICSA}.

As we discussed above, the FFFT will be a particularly powerful primitive in fermionic simulation. 
In between layers of interleaves, the FFFT also involves single-qubit rotations, and $\sqrt{\text{SWAP}}$ gates, which are simple to implement in physical-qubit architectures, but more costly in the fault-tolerant setting. These non-Clifford operations, need to be compiled into a error-correctable gateset, usually by adding either the T or CCZ gate. 
Interestingly, the non-Clifford gates in the FFFT have a large amount of structure as well.
In the SI, we further show that by synthesizing these non-Clifford operations in parallel, the fault-tolerant compilation overhead can be kept minimal, with one CCZ gate required on average per non-Clifford operation.
Hence, we conjecture that even on a fault-tolerant computer, if one were using a FSWAP network to simulate the FFFT, the $O(N^2)$ Clifford gate cost per layer would rapidly become the dominant source of overhead.
In contrast, our technique equalizes the Clifford and non-Clifford cost, as both contribute $O(N)$ gates per layer.

\section*{Discussion and Outlook}

Access to fast, non-local fermionic permutations can unlock a variety of new possibilities in implementing and designing quantum simulation algorithms.
Even though nature is fundamentally local, the utility of the FFFT for state-preparation illustrates that quantum algorithms need not emulate nature to achieve best results~\cite{Poulin_spectral_2018}.
Interesting areas for future explorations include improving ansatz for state preparation e.g. by preparing fermionic tensor networks states~\cite{corboz_fermionic_2009, dai_fermionic_2025}, improving techniques for measuring fermionic observables by compressing matchgate shadow circuits~\cite{wan_matchgate_2023}, and characterizing the full family of states which are preparable in $O(\log(L))$ depth using dynamical encodings.
It would also be interesting to understand the connections with measurement-based techniques for preparing complex topological orders~\cite{Tantivasadakarn_2024,Tantivasadakarn_shortest_2023,tantivasadakarn_hierarchy_2023,verresen2022efficientlypreparingschrodingerscat,bravyi2022adaptiveconstantdepthcircuitsmanipulating,Lu_meas_shortcut_2022,Iqbal_2024,Ren_adaptive_2025}.
Early applications of this technique could include simulating time-evolution of 2D Hubbard-type models from various low-energy trial wavefunctions with different patterns of long-range order, such as d-wave superconductivity and spin-liquid order~\cite{Christos_dwave_2023,Arovas_hubbard_2022}.
Adding perturbations like doping, next-nearest-neighbor tunneling, and spin-spin interactions, could enable quantum simulators to probe even more classically challenging regimes with minimal additional cost~\cite{Qin_hubbard_2022}.

An important practical consideration is the connectivity of the underlying quantum computing architecture. 
While various computational devices support non-local connectivity, often the structure of the non-local interactions is restricted.
For example neutral-atom arrays support parallel non-local connectivity, which is generated by effectively interleaving two arrays stored in independent sets of tweezers. 
As such, compiling arbitrary permutations into parallel reconfigurations already comes with $O(\log N)$ overhead~\cite{Tan_DFPNA_2024}, so fermionic simulation incurs \textit{no} additional asymptotic overhead.
Similar device-specific considerations hold for other architectures as well~\cite{Moses_racetrack_2023,bravyi_high_threshold_2024}. 
Hence, for practical fermionic simulations, it will be important to co-design many aspects to benefit most from efficient $O(1)$-overhead fermionic permutations, spanning the hardware-native qubit permutations, fault-tolerant logic gates, and fermionic simulation algorithm.


\section*{Methods}

\subsection*{Dynamical Fermion-to-qubit Mapping}

Our key result shows that any parallel fermionic circuit can be mapped to an equivalent qubit circuit with at most logarithmic overhead.
Our starting point is a simple description of the fermionic circuit $C_F$.

\begin{definition}[Fermionic computation]
    A fermionic computation $C_F$ on $N$-modes starts with the initial vacuum state $\vert 0 \rangle^{n}$.
    Then, alternating layers of two-mode tunneling gates $h_{ij}$ and two-mode interactions $v_{ij}$ are applied. During each layer, each mode is involved in at most one tunneling and one interaction respectively. Finally, each mode is measured in the occupation number basis.
    The depth $T$ of $C_F$ is given by the number of alternating layers of gates. 
    Generalized tunneling gates take the form $h_{ij} = \exp\left[-i (\alpha c_i^{\dagger} c_j + \beta c_i^{\dagger} c_j^{\dagger} + h.c.)\right]$ and interaction gates take the form $v_{ij} = \exp\left[-i (\gamma n_i n_j + \delta_i n_i + \delta_j n_j)\right]$, where $\{c_i,c_j\}=\{c_i^{\dagger},c_j^{\dagger}\}=0,\{c_i^{\dagger},c_j\}=\delta_{ij}$ are canonical fermionic operators and $n_i = c_i^{\dagger}c_i$ is the density operator. 
\end{definition}
Even though this definition only involves density-density interactions, this fermionic gateset is universal~\cite{bravyi_fermionic_2002}.
For example, notice that the operator $n_i n_j = c_i^{\dagger} c_i c_j^{\dagger} c_j$ can be straightforwardly transformed into an arbitrary parity-preserving four-fermion term by conjugation with two-fermion terms $h_{ij}$.

Our goal is to simulate $C_F$ using a qubit-based device. Therefore, we define an analogous model for qubit-based computation $C_Q$.
In particular, on the qubit-computation side, we also need the system to support mid-circuit measurement, fast non-local (i.e. $O(N)$ depth) classical computation, and conditional Pauli gates~\cite{bluvstein2025architecturalmechanismsuniversalfaulttolerant,Iqbal_2024}.

\begin{definition}[Qubit computation]
    A qubit computation $C_Q$ with $N$-modes starts with $N + N_A$ qubits in the initial product state $\vert 0 \rangle^N$ respectively, where $N_A$ is the number of ancilla.
    During each gate layer, each mode is involved in at most one two-qubit gate $u_{ij}$.
    During each measurement layer, any subset of qubits can be measured in the computational (i.e. $Z$) basis. 
    Following a measurement layer, measured qubits are reset to $\vert 0 \rangle$, and Pauli gates 
    $\{\sigma^x, \sigma^y, \sigma^z \} \equiv \{ X, Y, Z \}$ can be conditionally applied based on the measurement outcomes. We further assume that any classical computation involving $O(N)$ classical logic operations can be performed fast, in the time between the measurement and the conditional Pauli gate, and do not contribute to the cost. 
\end{definition}

The key challenge to simulating a fermionic computation $C_F$ with a qubit computation $C_Q$ is efficiently capturing the non-trivial fermionic statistics. Natively, the qubit operators obey a different algebra $[\sigma^{\alpha}_i, \sigma^{\beta}_j] = \epsilon_{\alpha \beta \gamma} \delta_{ij} \sigma_i^{\gamma}$, so qubit operators on different sites \textit{commute} while fermionic operators \textit{anti-commute}. To encode fermions into qubits, we use the Jordan-Wigner (JW) encoding~\cite{jordan_uber_1928}.
However, the definition of the JW encoding is not unique, as it depends on the choice of \textit{ordering} of the underlying modes. 
There are $N!$ distinct possible JW encodings to choose from, and we can identify a particular encoding by introducing a function $m(i)$, which defines an ordering of the indexes $i=1,...,N$.
For a given $m$, we define the JW encoding map, which maps fermionic operators to qubit operators, as $J_{m}$,
\begin{align}
    c_i &\xrightarrow{J_m} \left(\frac{X_i - i Y_i}{2}\right) \prod_{i' \in L(i)} Z_{i'} = \sigma_i^- \prod_{i' \in L(i)} Z_{i'} \\
    c_i^{\dagger} &\xrightarrow{J_m} \left(\frac{X_i + i Y_i}{2}\right) \prod_{i' \in L(i)} Z_{i'} = \sigma_i^+ \prod_{i' \in L(i)} Z_{i'}\,.
\end{align}
Here, $L(i) = \{j | m(j) < m(i),j=1,...,N \}$ is the set of all indexes positioned to the left of $i$.
It can be checked that $J_m[c_i]$ and $J_m[c_i^{\dagger}]$ obey fermionic anti-commutation relations.
Importantly, the qubit representation is non-local, and in this way mediate the long-range pairwise fermionic statistics.

\newcommand{\setdiff}{\;\Delta\;}

While any JW can be used to encode fermions, the cost of implementing a given two-qubit tunneling gate $h_{ij}$ depends sensitively on the choice of ordering.
Observe that a two-fermion operator $c_i^{\dagger} c_j$ is equivalent to the qubit operator
\begin{align}
    c_i^{\dagger} c_j \xrightarrow{J_m} (-1)^{m(i) > m(j)} \sigma_i^{+} \sigma_j^{-} \prod_{k \in L(i) \setdiff L(j)} Z_k
\end{align}
where $A \setdiff B = (A \cup B) - (A \cap B)$ is the set-difference operator, selecting all elements in $A$ or $B$ but not both. 
The operator weight of $J_m[c_i^{\dagger} c_j]$ is related to the distance between the two sites in the encoding, i.e. $|J_m[c_i^{\dagger} c_j]| = |m(j) - m(i)| + 1$.
Since, the cost of implementing $e^{-i h_{ij}}$ scales naively \textit{linearly} with the operator weight, this could be as large as $O(N)$ in the worst-case. However, if $|m(j)-m(i)|=1$, then $c_i^{\dagger} c_j$ maps to a two-qubit operator, and $e^{-i h_{ij}}$ can be implemented with a two-qubit gate.

In contrast, the cost of the density-density operator $v_{ij}$ does not depend on the particular encoding $m$.
This is because the density operators are local in the qubits
\begin{align}
    n_i = c_i^{\dagger} c_i &\xrightarrow{J_m} \frac{1-Z_i}{2}
\end{align}
and so the $e^{-i v_{ij}}$ can be straightforwardly replaced with a simple two-qubit gate.

Therefore, we propose a compilation strategy for mapping $C_F$ to $C_Q$, where the particular Jordan-Wigner encoding we are using is \textit{dynamically changed} during execution of the circuit.
To facilitate this, we introduce a fermionic permutation operation $\mathcal{F}_p$, to transform from one encoding to the other. 
\begin{definition}[Fermionic Permutation]\label{def:fermionic_permutation}
    A fermionic permutation operator $\mathcal{F}_p$ transforms a JW encoding, defined by the ordering $m_0$ to another JW encoding $m_1$. The permutation $p$ is chosen such that $p(m_0(i)) = m_1(i)$. Equivalently, $p(i) = m_1(m_0^{-1}(i))$. Let $J_{m_0}[c_i]$ be the initial JW encoding of a fermionic operator $c_i$, and $J_{m_1}[c_i]$ be the final JW encoding. Then, $\mathcal{F}_p$ should satisfy
    \begin{align}
        J_{m_1}[c_i] = \mathcal{F}_p (J_{m_0}[c_i])\mathcal{F}_p^{\dagger}.
    \end{align}    
\end{definition}

Assuming access to $\mathcal{F}_p$, we present a simple compilation algorithm for mapping a fermionic circuit $C_F$ to an equivalent qubit circuit $C_Q$.

\begin{algorithm}[H]\label{alg:fermion_to_qubit}
\caption{Fermion to qubit circuit mapping}
\KwIn{Fermionic circuit $C_F$ of depth $T$ acting on $n$ modes}
\KwOut{Qubit circuit $C_Q$}
\DontPrintSemicolon        
Initialize $C_Q$\;
\For{$\ell \gets 1$ \KwTo $T$}{
    $S_{\mathrm{tun}} \leftarrow$ tunneling gates in layer $\ell$\;
    $pairs \leftarrow [(i,j) \;\mathbf{for}\; h_{ij} \in S_{\mathrm{tun}}]$ \tcp*{List of pairs of interacting indexes}
    $o \leftarrow \mathrm{flatten}(pair\_list)$\tcp*{List of indexes where interacting modes are adjacent}
    $m_\ell \leftarrow o^{-1}$ \tcp*{Ordering where interacting modes are mapped to adjacent positions}
    \If{$\ell > 1$}{
        $p(i) \gets m_\ell\bigl(m_{\ell-1}^{-1}(i)\bigr)$\;
        append $\mathcal{F}_p$ to $C_Q$\tcp*{Change encoding between layers}
    }
    \ForEach{$h_{ij} \in S_{\mathrm{tun}}$}{
        append $J_{m_\ell}\!\left[e^{-i h_{ij}}\right]$ to $C_Q$ \tcp*{Add tunneling gates}
    }
    $S_{\mathrm{int}} \leftarrow$ interaction gates in layer $\ell$\;
    \ForEach{$v_{ij} \in S_{\mathrm{int}}$}{
        append $J_{m_\ell}\!\left[e^{-i v_{ij}}\right]$ to $C_Q$ \tcp*{Add interaction gates}
    }
}
\Return $C_Q$\;
\end{algorithm}

\hspace{0.2in}

Intuitively, this algorithm generates an ordering at each gate layer that ensures the tunneling operations at each layer are between adjacent modes, and hence the qubit-gates remain 2-local. 
The depth of the qubit circuit produced by this technique depends primarily on the cost of $\mathcal{F}_p$. 
As such, the key technical result of our work, is an efficient compilation of $\mathcal{F}_p$ into arbitrary qubit operations, as described in Figure~\ref{fig:fig2}. 
We present a more formal proof in Lemma~\ref{lem:fermionic_permutation}.

Taken together, these ingredients naturally lead to our main theorem, a result bounding the overhead of fermionic computation.

\begin{theorem}{\label{thm:circuit_reduction_mth}}
    Given an arbitrary fermionic circuit $C_F$ consisting of $N$ modes and $T$ layers of pairwise gates,  there is an analogous circuit $C_Q$ consisting of $O(N)$ qubits that can simulate $C_F$ in depth $O(T \log(N))$.
\end{theorem}

\begin{proof}
    The proof follows from Lemma~\ref{lem:fermionic_permutation}, and Algorithm~\ref{alg:fermion_to_qubit}.
    First note that in the circuit $C_Q$ returned by Algorithm~\ref{alg:fermion_to_qubit}, each layer of tunneling gates (lines 10-11) and each layer of interaction gates (lines 13-14) can be implemented with a single layer of two-qubit gates each. For the tunneling gates, this is because $m_l$ was defined to ensure that interacting modes are positioned next to each other, i.e. $|m_l(i) - m_l(j)| = 1$ for all interacting $i,j$.
    Next, by the definition of $C_F$, each mode is involved in only one tunneling and one interaction in each layer. Hence, these gates all commute and can be applied simultaneously. Together, these ultimately contribute $2T$ layers of gates.
    The rest of the cost comes from $\mathcal{F}_p$, which is upper-bounded by Lemma~\ref{lem:fermionic_permutation}. There are only $T-1$ applications of $\mathcal{F}_p$, each with $O(N)$ ancilla qubits and depth $O(\log(N))$, and so the total circuit depth in $C_Q$ scales as $O(T \log(N))$.
\end{proof}

\subsection*{Compiling arbitrary fermionic permutations}

Asymptotically, the cost of Algorithm~\ref{alg:fermion_to_qubit} will be limited by the cost of fermionic permutation $\mathcal{F}_p$.
While the FSWAP network approach requires $O(N^2)$ gates arranged in a depth $O(N)$ circuit, our approach to fast fermionic simulation only requires $O(N \log N)$ gates arranged in a depth $O(\log N)$ circuit.

In the SI, we rigorously prove that the interleave circuit of Figure.~\ref{fig:fig2}b satisfies Def.~\ref{def:fermionic_permutation}, resulting in the following Lemma.

\begin{lemma}\label{lem:interleave}
Given a fermionic permutation $\mathcal{F}_p$ where $p$ is an interleave  permutation, there is a circuit consisting of $O(N)$ gates that can be implemented with $O(N)$ ancillas and depth $O(1)$, using measurement and instantaneous $O(N)$-depth classical processing. Formally, $p$ is an interleave, if its modes can be partitioned into two groups $A$ and $B$, such that for all $a, a' \in A$ satisfying $a < a'$, $p(a) < p(a')$, and similarly for all $b,b' \in B$.
\end{lemma}

Using this result, we prove that an \textit{arbitrary} permutation can be generated with low-overhead, by compiling it into $\log(N)$ interleave operations, as depicted in Figure.~\ref{fig:fig2}a.

\begin{lemma}\label{lem:fermionic_permutation}
Given an arbitrary fermionic permutation $\mathcal{F}_p$ of $N$ modes, there is a circuit consisting of $O(N \log (N))$ two-qubit Clifford gates, that can be implemented with $O(N)$ ancilla's, depth $O(\log (N))$, and fast $O(N)$-depth classical processing, that exactly realizes $\mathcal{F}_p$.
\end{lemma}

\begin{proof}
To compile $\mathcal{F}_p$, we start by showing that any permutation $p$ can be decomposed into a sequence of $\log(N)$ interleave permutations $p_i$.
From Lemma~\ref{lem:interleave}, each interleave permutation can be implemented with $O(N)$ gates arranged into a depth $O(1)$ gate with measurement and $O(N)$-depth instantaneous classical processing. 
Our proof relies on constructing a sequence of $k = \lceil \log_2(N) \rceil$ permutations $p_i$, which generate the target permutation $p$ when composed,
\begin{align}
    p = p_0 \circ p_1 \circ ... \circ p_{k-1}.
\end{align}
For notational simplicity, we assume $N = 2^k$ is a power-of-two.

To start, we decompose $p$ into two parts,
\begin{align}
    p = p_0 \circ \overline{p_1}
\end{align}
where $\overline{p_1}$ is defined to be a permutation which separately sorts modes $0,...,N/2-1$ and $N/2,...,N-1$.
In particular, the following simple algorithm lets us determine $\overline{p_1}$ from $p$.
\begin{enumerate}
    \item Generate the array $O_A = [p(0),p(1),...,p(N/2-1)]$ and $O_B = [p(N/2),p(N/2+1),...,p(N-1)]$.
    \item Determine the orderings, $o_A = \mathrm{argsort}(O_A)^{-1}$ and $o_B = \mathrm{argsort}(O_B)^{-1}$.
    \item Define the permutation
    \begin{align}
        \overline{p_{1}}(i)&=\begin{cases}
            o_{A}(i) & i<N/2\\
            o_{B}(i)+N/2 & i\geq N/2
        \end{cases}.
    \end{align}
\end{enumerate}
Importantly, this choice of decomposition guarantees that $p_0$ is of interleave type.
By definition, we know the composition of the two functions satisfies
\begin{align}
    p(i) = p_0(\overline{p_1}(i)).
\end{align}
Further, for each pair of indexes $i, j < N/2$ satisfying $\overline{p_1}(i) < \overline{p_1}(j)$, we are guaranteed by construction that $p(i) < p(j)$.
To see this, assume for contradiction that $p(i) > p(j)$. Then, $o_A(i) > o_A(j)$, and  $\overline{p_1}(i) > \overline{p_1}(j)$.

Similarly, for each pair of indexes $i,j \geq N/2$ satisfying $\overline{p_1}(i) < \overline{p_1}(j)$, we are guaranteed by construction that $p(i) < p(j)$.
Thus, $p_0$ is of interleave type, since the order of the first $N/2$ modes and last $N/2$ modes is preserved.

Since $\overline{p_1}$ permutes only $N/2$ modes at a time, we can apply this procedure recursively. At the next step, we further decompose $\overline{p_1}$  as
\begin{align}
    \overline{p_1} = p_1 \circ \overline{p_2}
\end{align}
where $\overline{p_2}$ permutes groups of $N/4$ modes at a time.
Thus, $p_1$ becomes a permutation of interleave type on $N/2$ modes.
After $k$ recursion layers, $N/2^k = 1$ is a single mode, and $\overline{p_k} = \mathds{1}$ is the trivial permutation.
Therefore, $k = \log_2(N)$ interleave layers is enough compile an arbitrary permutation.
\end{proof}

\subsection*{Majorana Permutation}

\begin{figure*}
    \centering
    \includegraphics[width=1.0\linewidth]{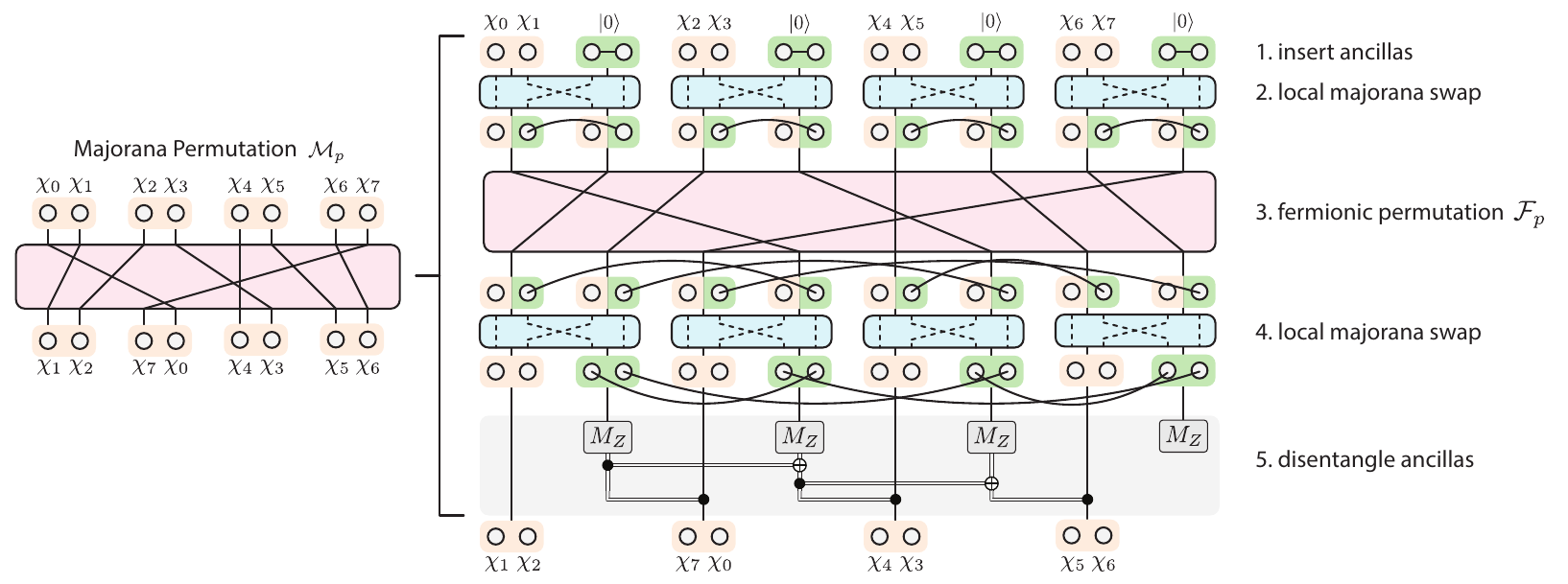}
    \caption{\textbf{Majorana permutation protocol.} Consider a system of $N$ qubits simulating fermions via the Jordan-Wigner encoding. There are $2N$ majorana operators that form the fermionic operations. Here, our goal is to realize a majorana permutation $\mathcal{M}_p$ that non-locally shuffles the majorana operators. An efficient protocol to implement $\mathcal{M}_p$ is depicted on the right. (1) In the first step, we insert $N$ ancilla qubits, prepared in $\vert 0 \rangle$. Each ancilla has associated two majorana modes, which are paired to each other. (2) In the second step, we apply local two-qubit gates between each qubit $i$ and its adjacent ancilla $i+1$. In particular, we evolve under $e^{-i \frac{\pi}{4} X_{i} X_{i+1}}$, which locally permutes the four majorana modes, distributing the encoded $2N$ majorana modes (colored orange) into $2N$ independent qubits. (3) Then, an analogous fermionic permutation $\mathcal{F}_p$ is used to transform the $2N$ qubits. (4) The local majorana swap is then inverted, to separate the encoded modes from the ancilla. Note the permutation generates a long-range entangled stabilizer state within the ancilla register, as evidenced by structure of majorana pairings. Hence, the local fermion number is no longer well-defined, and due to the jordan-wigner encoding, the ancilla is entangled with the system. (5) To decouple the ancilla, we can simply measure in the $Z$ basis to determine whether the fermion mode is occupied, and apply a simple feed-forward circuit to effectively perform an inverse interleave and remove the ancillas from the system.}
    \label{fig:majorana_permutation}
\end{figure*}

Next, we extend the fermionic permutation technique to generate arbitrary \textit{majorana permutations}. This is a useful subroutine for various quantum simulation tasks, including simulation of SYK models with arbitrary four-fermion interactions.

In the Jordan-Wigner encoding with $m(i)=i$, majorana fermions take the form
\begin{align}
    \chi_{2i} = X_i \prod_{i'<i} Z_{i'}, 
    \chi_{2i+1} = Y_i \prod_{i' < i} Z_{i'}.
\end{align}
A majorana permutation $\mathcal{M}_p$ is defined by a permutation of $2N$ modes, and should map $\chi_{2i}$ to $\chi_{p(2i)}$.
The problem of implementing $\mathcal{M}_p$ of $2N$ majorana modes can be reduced to realizing $\mathcal{F}_p$ of $2N$ fermion modes, using the circuit gadget in Figure.~\ref{fig:majorana_permutation}.
The procedure works by introducing a set of $N$ ancilla fermions modes initialized in a trivial state, and then applying a layer of two-qubit gates $U_{lms}$ that effectively \textit{swaps} all odd-index majorana modes with the even-index majorana of the ancilla modes.
\begin{align}
    U_{lms} &= \prod_{i=0}^{N-1} \exp\left( -i \frac{\pi}{2} \chi_{4i+1} \chi_{4i+2} \right) \\
    &= \prod_{i=0}^{N-1} \exp\left( -i \frac{\pi}{2} X_{2i} X_{2i+1} \right)
\end{align}
This ensures that each fermionic mode supports exactly one of the relevant majorana modes.
Then, $\mathcal{F}_p$ is applied, and the majorana swap gate is applied again, to separate the target modes from the ancilla system.
Finally, the ancilla system is measured, and a feedforward correction applied, to account for pairs of fermions created in the ancilla system.
A more detailed definition and proof of this procedure is presented in the SI.

\subsection*{SYK model from random efficient permutations}

\begin{figure*}
    \centering
    \includegraphics[width=1.0\linewidth]{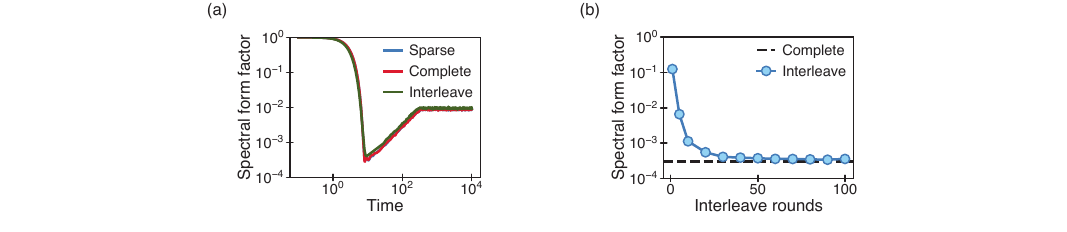}
    \caption{\textbf{Interleave-SYK Hamiltonians and their properties.} (a) Spectral form factor (SFF) of SYK Hamiltonians ($N$=14) for the sparse, complete, and interleave models. (b) Minimum values (dips) of I-SYK form factors as a function of interleave permutation rounds. The asymptotic value is that of a fully-connected SYK model.
    }
    \label{fig:si_syk}
\end{figure*}

The SYK model is a paradigmatic example of a disordered chaotic model with interesting connections to quantum gravity.
The Hamiltonian is given by,
\begin{align}
    H_{\rm SYK} = \sum_{i,j,k,l}J_{ijkl}\,\chi_i \chi_j\chi_k\chi_l,\label{seq:syk}
\end{align}
where $\chi_i$ are Majorana modes, and the coupling strengths $J$ are drawn from a normal distribution $\mathcal{N}(0,\sigma^2)$ with $\sigma^2\,{=}6J^2/N^3$. 

Recently, it has been shown~\cite{orman_quantum_2024} that the spectral properties of the SYK Hamiltonian ensembles are maintained even in the sparse limit, where the number of terms in Eq.~\eqref{seq:syk} is $O(N)$, instead of $O(N^4)$. 
By applying majorana permutations, trotterized versions~\cite{Childs_trotter_2021} or block-encodings~\cite{Low_hamiltonian_simulation_2019} of sparse SYK models with $O(N)$ terms can be achieved in $\log(N)$-depth, using a modification of the compilation described in Alg.~\ref{alg:fermion_to_qubit}.

However, we can simplify the mapping further. In Fig.~\ref{fig:si_syk}, we provide numerical evidence that these properties can be maintained for a Hamiltonian ensemble generated by the repeated application of random interleave procedures. This suggests that the SYK model, and its connections to quantum gravity, can be efficiently studied on qubit systems, with only $O(1)$-overhead coming from simulating fermions with qubits. More detailed numerics or analytics are required to confirm that this strategy generalizes to larger systems.

\subsection*{Acknowledgments}
We would like to thank 
Dolev Bluvstein, Pablo Bonilla, Madelyn Cain, Charles Cao, Simon Evered, Johannes Feldmeier, Alexandra Geim, Jakob Günther, Jeongwan Haah, Alexander Jahn, Lev Kendrick, Jin Ming Koh, Christian Kokail, Sophie Li, Shayan Majiday, Tom Manovitz, Robert Ott, Hannes Pichler, Alexander Schukert, Monika Schleier-Smith, Stefan Ostermann, Nazlı Uğur Köylüoğlu, Susanne Yelin, Shengtao Wang, Qian Xu, Torsten Zache, Peter Zoller, Zhelun Zhang, and Hengyun Zhou
for interesting and insightful discussions.
We acknowledge financial support from US Department of Energy (DOE Quantum Systems Accelerator Center, grant number DE-AC02-05CH11231, and the QUACQ program, grant number DE-SC0025572), IARPA and the Army Research Office, under the Entangled Logical Qubits program (Cooperative Agreement Number W911NF-23-2-0219), the DARPA MeasQuIT program (grant number HR0011-24-9-0359), the Center for Ultracold Atoms (a NSF Physics Frontiers Center, PHY-1734011), the National Science Foundation (grant numbers PHY-2012023 and  CCF-2313084),  the Wellcome Leap Quantum for Bio program, and QuEra Computing. 
D.G.-C. acknowledges support from the European Union's Horizon Europe program under the Marie Sk{\l}odowska Curie Action PROGRAM (Grant No. 101150724), and through the Ramón y Cajal Program (RYC2023-044201-I), financed by MICIU/AEI/10.13039/501100011033 and by the FSE+.

\bibliographystyle{nature_arxiv}
\bibliography{FermionCircuitstoQubitCircuits}

\begin{thebibliography}{100}
\expandafter\ifx\csname url\endcsname\relax
  \def\url#1{\texttt{#1}}\fi
\expandafter\ifx\csname urlprefix\endcsname\relax\def\urlprefix{URL }\fi
\providecommand{\bibinfo}[2]{#2}
\providecommand{\eprint}[2][]{\url{#2}}

\bibitem{Bauer_2020}
\bibinfo{author}{Bauer, B.}, \bibinfo{author}{Bravyi, S.}, \bibinfo{author}{Motta, M.} \& \bibinfo{author}{Chan, G. K.-L.}
\newblock \bibinfo{title}{Quantum algorithms for quantum chemistry and quantum materials science}.
\newblock \href {https://doi.org/10.1021/acs.chemrev.9b00829}{\emph{\bibinfo{journal}{Chem. Rev.}} \textbf{\bibinfo{volume}{120}}, \bibinfo{pages}{12685--12717}} (\bibinfo{year}{2020}).

\bibitem{mcardle_quantum_2020}
\bibinfo{author}{McArdle, S.}, \bibinfo{author}{Endo, S.}, \bibinfo{author}{Aspuru-Guzik, A.}, \bibinfo{author}{Benjamin, S.~C.} \& \bibinfo{author}{Yuan, X.}
\newblock \bibinfo{title}{Quantum computational chemistry}.
\newblock \href {https://doi.org/10.1103/RevModPhys.92.015003}{\emph{\bibinfo{journal}{Rev. Mod. Phys.}} \textbf{\bibinfo{volume}{92}}, \bibinfo{pages}{015003}} (\bibinfo{year}{2020}).

\bibitem{di_meglio_quantum_2024}
\bibinfo{author}{Di~Meglio, A.} \emph{et~al.}
\newblock \bibinfo{title}{Quantum {Computing} for {High}-{Energy} {Physics}: {State} of the {Art} and {Challenges}}.
\newblock \href {https://doi.org/10.1103/PRXQuantum.5.037001}{\emph{\bibinfo{journal}{PRX Quantum}} \textbf{\bibinfo{volume}{5}}, \bibinfo{pages}{037001}} (\bibinfo{year}{2024}).

\bibitem{maldacena2023simplequantumdescribesblack}
\bibinfo{author}{Maldacena, J.}
\newblock \bibinfo{title}{A simple quantum system that describes a black hole}.
\newblock Preprint at \href {https://arxiv.org/abs/2303.11534} {https://arxiv.org/abs/2303.11534} (\bibinfo{year}{2023}).

\bibitem{preskill2022physicsquantuminformation}
\bibinfo{author}{Preskill, J.}
\newblock \bibinfo{title}{The physics of quantum information}.
\newblock Preprint at \href {https://arxiv.org/abs/2208.08064} {https://arxiv.org/abs/2208.08064} (\bibinfo{year}{2022}).

\bibitem{abrams_simulation_1997}
\bibinfo{author}{Abrams, D.~S.} \& \bibinfo{author}{Lloyd, S.}
\newblock \bibinfo{title}{Simulation of {Many}-{Body} {Fermi} {Systems} on a {Universal} {Quantum} {Computer}}.
\newblock \href {https://doi.org/10.1103/PhysRevLett.79.2586}{\emph{\bibinfo{journal}{Phys. Rev. Lett.}} \textbf{\bibinfo{volume}{79}}, \bibinfo{pages}{2586--2589}} (\bibinfo{year}{1997}).

\bibitem{feynman_simulating_1982}
\bibinfo{author}{Feynman, R.~P.}
\newblock \bibinfo{title}{Simulating physics with computers}.
\newblock \href {https://doi.org/10.1007/BF02650179}{\emph{\bibinfo{journal}{Int. J. Theor. Phys.}} \textbf{\bibinfo{volume}{21}}, \bibinfo{pages}{467--488}} (\bibinfo{year}{1982}).

\bibitem{altman_quantum_2021}
\bibinfo{author}{Altman, E.} \emph{et~al.}
\newblock \bibinfo{title}{Quantum {Simulators}: {Architectures} and {Opportunities}}.
\newblock \href {https://doi.org/10.1103/PRXQuantum.2.017003}{\emph{\bibinfo{journal}{PRX Quantum}} \textbf{\bibinfo{volume}{2}}, \bibinfo{pages}{017003}} (\bibinfo{year}{2021}).

\bibitem{jordan_uber_1928}
\bibinfo{author}{Jordan, P.} \& \bibinfo{author}{Wigner, E.}
\newblock \bibinfo{title}{Über das {Paulische} Äquivalenzverbot}.
\newblock \href {https://doi.org/10.1007/BF01331938}{\emph{\bibinfo{journal}{Z. Phys.}} \textbf{\bibinfo{volume}{47}}, \bibinfo{pages}{631--651}} (\bibinfo{year}{1928}).

\bibitem{bravyi_fermionic_2002}
\bibinfo{author}{Bravyi, S.~B.} \& \bibinfo{author}{Kitaev, A.~Y.}
\newblock \bibinfo{title}{Fermionic {Quantum} {Computation}}.
\newblock \href {https://doi.org/https://doi.org/10.1006/aphy.2002.6254}{\emph{\bibinfo{journal}{Ann. Phys. (N. Y.)}} \textbf{\bibinfo{volume}{298}}, \bibinfo{pages}{210--226}} (\bibinfo{year}{2002}).

\bibitem{verstraete_mapping_2005}
\bibinfo{author}{Verstraete, F.} \& \bibinfo{author}{Cirac, J.~I.}
\newblock \bibinfo{title}{Mapping local {Hamiltonians} of fermions to local {Hamiltonians} of spins}.
\newblock \href {https://doi.org/10.1088/1742-5468/2005/09/P09012}{\emph{\bibinfo{journal}{J. Stat. Mech.: Theory Exp.}} \textbf{\bibinfo{volume}{2005}}, \bibinfo{pages}{P09012}} (\bibinfo{year}{2005}).

\bibitem{derby_compact_2021}
\bibinfo{author}{Derby, C.} \& \bibinfo{author}{Klassen, J.}
\newblock \bibinfo{title}{A {Compact} {Fermion} to {Qubit} {Mapping}}.
\newblock \href {https://doi.org/10.1103/PhysRevB.104.035118}{\emph{\bibinfo{journal}{Phys. Rev. B}} \textbf{\bibinfo{volume}{104}}, \bibinfo{pages}{035118}} (\bibinfo{year}{2021}).

\bibitem{chen_equivalence_2023}
\bibinfo{author}{Chen, Y.-A.} \& \bibinfo{author}{Xu, Y.}
\newblock \bibinfo{title}{Equivalence between {Fermion}-to-{Qubit} {Mappings} in two {Spatial} {Dimensions}}.
\newblock \href {https://doi.org/10.1103/PRXQuantum.4.010326}{\emph{\bibinfo{journal}{PRX Quantum}} \textbf{\bibinfo{volume}{4}}, \bibinfo{pages}{010326}} (\bibinfo{year}{2023}).

\bibitem{kitaev_anyons_2006}
\bibinfo{author}{Kitaev, A.}
\newblock \bibinfo{title}{Anyons in an exactly solved model and beyond}.
\newblock \href {https://doi.org/https://doi.org/10.1016/j.aop.2005.10.005}{\emph{\bibinfo{journal}{Ann. Phys. (N. Y.)}} \textbf{\bibinfo{volume}{321}}, \bibinfo{pages}{2--111}} (\bibinfo{year}{2006}).

\bibitem{kells_description_2009}
\bibinfo{author}{Kells, G.}, \bibinfo{author}{Slingerland, J.~K.} \& \bibinfo{author}{Vala, J.}
\newblock \bibinfo{title}{Description of {Kitaev}’s honeycomb model with toric-code stabilizers}.
\newblock \href {https://doi.org/10.1103/PhysRevB.80.125415}{\emph{\bibinfo{journal}{Phys. Rev. B}} \textbf{\bibinfo{volume}{80}}, \bibinfo{pages}{125415}} (\bibinfo{year}{2009}).

\bibitem{Evered_2025}
\bibinfo{author}{Evered, S.~J.} \emph{et~al.}
\newblock \bibinfo{title}{Probing topological matter and fermion dynamics on a neutral-atom quantum computer}.
\newblock Preprint at \href {https://arxiv.org/abs/2501.18554} {https://arxiv.org/abs/2501.18554} (\bibinfo{year}{2025}).

\bibitem{dyrenkova_scalable_2025}
\bibinfo{author}{Dyrenkova, E.}, \bibinfo{author}{Laflamme, R.} \& \bibinfo{author}{Vasmer, M.}
\newblock \bibinfo{title}{Scalable {Simulation} of {Fermionic} {Encoding} {Performance} on {Noisy} {Quantum} {Computers}}.
\newblock Preprint at \href {https://arxiv.org/abs/2506.06425} {https://arxiv.org/abs/2506.06425} (\bibinfo{year}{2025}).

\bibitem{Nigmatullin_compactencoding_2025}
\bibinfo{author}{Nigmatullin, R.} \emph{et~al.}
\newblock \bibinfo{title}{Experimental demonstration of breakeven for a compact fermionic encoding}.
\newblock \href {https://doi.org/10.1038/s41567-025-02931-8}{\emph{\bibinfo{journal}{Nat. Phys.}} \textbf{\bibinfo{volume}{21}}, \bibinfo{pages}{1319–1325}} (\bibinfo{year}{2025}).

\bibitem{Jiang_optimal_ternery_2020}
\bibinfo{author}{Jiang, Z.}, \bibinfo{author}{Kalev, A.}, \bibinfo{author}{Mruczkiewicz, W.} \& \bibinfo{author}{Neven, H.}
\newblock \bibinfo{title}{Optimal fermion-to-qubit mapping via ternary trees with applications to reduced quantum states learning}.
\newblock \href {https://doi.org/10.22331/q-2020-06-04-276}{\emph{\bibinfo{journal}{Quantum}} \textbf{\bibinfo{volume}{4}}, \bibinfo{pages}{276}} (\bibinfo{year}{2020}).

\bibitem{Havlick_operator_locality_fermions}
\bibinfo{author}{Havlíček, V.}, \bibinfo{author}{Troyer, M.} \& \bibinfo{author}{Whitfield, J.~D.}
\newblock \bibinfo{title}{Operator locality in the quantum simulation of fermionic models}.
\newblock \href {https://doi.org/10.1103/physreva.95.032332}{\emph{\bibinfo{journal}{Phys. Rev. A}} \textbf{\bibinfo{volume}{95}}} (\bibinfo{year}{2017}).

\bibitem{verstraete_quantum_2009}
\bibinfo{author}{Verstraete, F.}, \bibinfo{author}{Cirac, J.~I.} \& \bibinfo{author}{Latorre, J.~I.}
\newblock \bibinfo{title}{Quantum circuits for strongly correlated quantum systems}.
\newblock \href {https://doi.org/10.1103/PhysRevA.79.032316}{\emph{\bibinfo{journal}{Phys. Rev. A}} \textbf{\bibinfo{volume}{79}}, \bibinfo{pages}{032316}} (\bibinfo{year}{2009}).

\bibitem{babbush_low-depth_2018}
\bibinfo{author}{Babbush, R.} \emph{et~al.}
\newblock \bibinfo{title}{Low-{Depth} {Quantum} {Simulation} of {Materials}}.
\newblock \href {https://doi.org/10.1103/PhysRevX.8.011044}{\emph{\bibinfo{journal}{Phys. Rev. X}} \textbf{\bibinfo{volume}{8}}, \bibinfo{pages}{011044}} (\bibinfo{year}{2018}).

\bibitem{jiang_quantum_2018}
\bibinfo{author}{Jiang, Z.}, \bibinfo{author}{Sung, K.~J.}, \bibinfo{author}{Kechedzhi, K.}, \bibinfo{author}{Smelyanskiy, V.~N.} \& \bibinfo{author}{Boixo, S.}
\newblock \bibinfo{title}{Quantum {Algorithms} to {Simulate} {Many}-{Body} {Physics} of {Correlated} {Fermions}}.
\newblock \href {https://doi.org/10.1103/PhysRevApplied.9.044036}{\emph{\bibinfo{journal}{Phys. Rev. Appl.}} \textbf{\bibinfo{volume}{9}}, \bibinfo{pages}{044036}} (\bibinfo{year}{2018}).

\bibitem{kivlichan_quantum_2018}
\bibinfo{author}{Kivlichan, I.~D.} \emph{et~al.}
\newblock \bibinfo{title}{Quantum {Simulation} of {Electronic} {Structure} with {Linear} {Depth} and {Connectivity}}.
\newblock \href {https://doi.org/10.1103/PhysRevLett.120.110501}{\emph{\bibinfo{journal}{Phys. Rev. Lett.}} \textbf{\bibinfo{volume}{120}}, \bibinfo{pages}{110501}} (\bibinfo{year}{2018}).

\bibitem{low_hamiltonian_2019}
\bibinfo{author}{Low, G.~H.} \& \bibinfo{author}{Wiebe, N.}
\newblock \bibinfo{title}{Hamiltonian simulation in the interaction picture}.
\newblock Preprint at \href {https://arxiv.org/abs/1805.00675} {https://arxiv.org/abs/1805.00675} (\bibinfo{year}{2019}).

\bibitem{clinton_towards_2024}
\bibinfo{author}{Clinton, L.} \emph{et~al.}
\newblock \bibinfo{title}{Towards near-term quantum simulation of materials}.
\newblock \href {https://doi.org/10.1038/s41467-023-43479-6}{\emph{\bibinfo{journal}{Nat. Commun.}} \textbf{\bibinfo{volume}{15}}, \bibinfo{pages}{211}} (\bibinfo{year}{2024}).

\bibitem{hashim_optimized_2021}
\bibinfo{author}{Hashim, A.} \emph{et~al.}
\newblock \bibinfo{title}{Optimized fermionic {SWAP} networks with equivalent circuit averaging for {QAOA}}.
\newblock Preprint at \href {https://arxiv.org/abs/2111.04572} {https://arxiv.org/abs/2111.04572} (\bibinfo{year}{2021}).

\bibitem{ogorman_generalized_2019}
\bibinfo{author}{O'Gorman, B.}, \bibinfo{author}{Huggins, W.~J.}, \bibinfo{author}{Rieffel, E.~G.} \& \bibinfo{author}{Whaley, K.~B.}
\newblock \bibinfo{title}{Generalized swap networks for near-term quantum computing}.
\newblock Preprint at \href {https://arxiv.org/abs/1905.05118} {https://arxiv.org/abs/1905.05118} (\bibinfo{year}{2019}).

\bibitem{hagge_optimal_2022}
\bibinfo{author}{Hagge, T.}
\newblock \bibinfo{title}{Optimal fermionic swap networks for {Hubbard} models}.
\newblock Preprint at \href {https://arxiv.org/abs/2001.08324} {https://arxiv.org/abs/2001.08324} (\bibinfo{year}{2022}).

\bibitem{derby2021compactfermion2}
\bibinfo{author}{Derby, C.} \& \bibinfo{author}{Klassen, J.}
\newblock \bibinfo{title}{A compact fermion to qubit mapping part 2: Alternative lattice geometries}.
\newblock Preprint at \href {https://arxiv.org/abs/2101.10735} {https://arxiv.org/abs/2101.10735} (\bibinfo{year}{2021}).

\bibitem{paciani_quantum_2025}
\bibinfo{author}{Paciani, G.~D.}, \bibinfo{author}{Homeier, L.}, \bibinfo{author}{Halimeh, J.~C.}, \bibinfo{author}{Aidelsburger, M.} \& \bibinfo{author}{Grusdt, F.}
\newblock \bibinfo{title}{Quantum simulation of fermionic non-{Abelian} lattice gauge theories in $(2+1)${D} with built-in gauge protection}.
\newblock Preprint at \href {https://arxiv.org/abs/2506.14747} {https://arxiv.org/abs/2506.14747} (\bibinfo{year}{2025}).

\bibitem{chiew_optimal_2025}
\bibinfo{author}{Chiew, M.}, \bibinfo{author}{Ibrahim, C.}, \bibinfo{author}{Safro, I.} \& \bibinfo{author}{Strelchuk, S.}
\newblock \bibinfo{title}{Optimal fermion-qubit mappings via quadratic assignment}.
\newblock Preprint at \href {https://arxiv.org/abs/2504.21636} {https://arxiv.org/abs/2504.21636} (\bibinfo{year}{2025}).

\bibitem{li_accelerating_2025}
\bibinfo{author}{Li, Q.-S.} \emph{et~al.}
\newblock \bibinfo{title}{Accelerating {Fermionic} {System} {Simulation} on {Quantum} {Computers}}.
\newblock \href {https://doi.org/10.1103/PhysRevA.111.052606}{\emph{\bibinfo{journal}{Phys. Rev. A}} \textbf{\bibinfo{volume}{111}}, \bibinfo{pages}{052606}} (\bibinfo{year}{2025}).

\bibitem{chiew_discovering_2023}
\bibinfo{author}{Chiew, M.} \& \bibinfo{author}{Strelchuk, S.}
\newblock \bibinfo{title}{Discovering optimal fermion-qubit mappings through algorithmic enumeration}.
\newblock \href {https://doi.org/10.22331/q-2023-10-18-1145}{\emph{\bibinfo{journal}{Quantum}} \textbf{\bibinfo{volume}{7}}, \bibinfo{pages}{1145}} (\bibinfo{year}{2023}).

\bibitem{parella_dilme_reducing_2024}
\bibinfo{author}{Parella-Dilmé, T.} \emph{et~al.}
\newblock \bibinfo{title}{Reducing {Entanglement} with {Physically} {Inspired} {Fermion}-{To}-{Qubit} {Mappings}}.
\newblock \href {https://doi.org/10.1103/prxquantum.5.030333}{\emph{\bibinfo{journal}{PRX Quantum}} \textbf{\bibinfo{volume}{5}}, \bibinfo{pages}{030333}} (\bibinfo{year}{2024}).

\bibitem{wang_ever_more_optimized_2023}
\bibinfo{author}{Wang, Q.}, \bibinfo{author}{Cian, Z.-P.}, \bibinfo{author}{Li, M.}, \bibinfo{author}{Markov, I.~L.} \& \bibinfo{author}{Nam, Y.}
\newblock \bibinfo{title}{Ever more optimized simulations of fermionic systems on a quantum computer}.
\newblock Preprint at \href {https://arxiv.org/abs/2303.03460} {https://arxiv.org/abs/2303.03460} (\bibinfo{year}{2023}).

\bibitem{steudtner_fermion_qubit_2018}
\bibinfo{author}{Steudtner, M.} \& \bibinfo{author}{Wehner, S.}
\newblock \bibinfo{title}{Fermion-to-qubit mappings with varying resource requirements for quantum simulation}.
\newblock \href {https://doi.org/10.1088/1367-2630/aac54f}{\emph{\bibinfo{journal}{New J. Phys.}} \textbf{\bibinfo{volume}{20}}, \bibinfo{pages}{063010}} (\bibinfo{year}{2018}).

\bibitem{Miller_grow_your_own_2023}
\bibinfo{author}{Miller, A.}, \bibinfo{author}{Zimborás, Z.}, \bibinfo{author}{Knecht, S.}, \bibinfo{author}{Maniscalco, S.} \& \bibinfo{author}{García-Pérez, G.}
\newblock \bibinfo{title}{Bonsai algorithm: Grow your own fermion-to-qubit mappings}.
\newblock \href {https://doi.org/10.1103/prxquantum.4.030314}{\emph{\bibinfo{journal}{PRX Quantum}} \textbf{\bibinfo{volume}{4}}} (\bibinfo{year}{2023}).

\bibitem{gonzalez-cuadra_fermionic_2023}
\bibinfo{author}{González-Cuadra, D.} \emph{et~al.}
\newblock \bibinfo{title}{Fermionic quantum processing with programmable neutral atom arrays}.
\newblock \href {https://doi.org/10.1073/pnas.2304294120}{\emph{\bibinfo{journal}{Proc. Natl. Acad. Sci. U.S.A.}} \textbf{\bibinfo{volume}{120}}, \bibinfo{pages}{e2304294120}} (\bibinfo{year}{2023}).

\bibitem{schuckert_fermion}
\bibinfo{author}{Schuckert, A.}, \bibinfo{author}{Crane, E.}, \bibinfo{author}{Gorshkov, A.~V.}, \bibinfo{author}{Hafezi, M.} \& \bibinfo{author}{Gullans, M.~J.}
\newblock \bibinfo{title}{Fault-tolerant fermionic quantum computing}.
\newblock Preprint at \href {https://arxiv.org/abs/2411.08955} {https://arxiv.org/abs/2411.08955} (\bibinfo{year}{2025}).

\bibitem{ott_error-corrected_2024}
\bibinfo{author}{Ott, R.} \emph{et~al.}
\newblock \bibinfo{title}{Error-corrected fermionic quantum processors with neutral atoms}.
\newblock \href {https://doi.org/10.1103/zkpl-hh28}{\emph{\bibinfo{journal}{Phys. Rev. Lett.}} \textbf{\bibinfo{volume}{135}}, \bibinfo{pages}{090601}} (\bibinfo{year}{2025}).

\bibitem{Hartke_2023}
\bibinfo{author}{Hartke, T.}, \bibinfo{author}{Oreg, B.}, \bibinfo{author}{Turnbaugh, C.}, \bibinfo{author}{Jia, N.} \& \bibinfo{author}{Zwierlein, M.}
\newblock \bibinfo{title}{Direct observation of nonlocal fermion pairing in an attractive fermi-hubbard gas}.
\newblock \href {https://doi.org/10.1126/science.ade4245}{\emph{\bibinfo{journal}{Science}} \textbf{\bibinfo{volume}{381}}, \bibinfo{pages}{82--86}} (\bibinfo{year}{2023}).

\bibitem{Shao_2024}
\bibinfo{author}{Shao, H.-J.} \emph{et~al.}
\newblock \bibinfo{title}{Antiferromagnetic phase transition in a {3D} fermionic {Hubbard} model}.
\newblock \href {https://doi.org/10.1038/s41586-024-07689-2}{\emph{\bibinfo{journal}{Nature}} \textbf{\bibinfo{volume}{632}}, \bibinfo{pages}{267--272}} (\bibinfo{year}{2024}).

\bibitem{Bourgund_2025}
\bibinfo{author}{Bourgund, D.} \emph{et~al.}
\newblock \bibinfo{title}{Formation of individual stripes in a mixed-dimensional cold-atom fermi--hubbard system}.
\newblock \href {https://doi.org/10.1038/s41586-024-08270-7}{\emph{\bibinfo{journal}{Nature}} \textbf{\bibinfo{volume}{637}}, \bibinfo{pages}{57--62}} (\bibinfo{year}{2025}).

\bibitem{Xu_2025}
\bibinfo{author}{Xu, M.} \emph{et~al.}
\newblock \bibinfo{title}{A neutral-atom hubbard quantum simulator in the cryogenic regime}.
\newblock \href {https://doi.org/10.1038/s41586-025-09112-w}{\emph{\bibinfo{journal}{Nature}} \textbf{\bibinfo{volume}{642}}, \bibinfo{pages}{909--915}} (\bibinfo{year}{2025}).

\bibitem{zache_fermion-qudit_2023}
\bibinfo{author}{Zache, T.~V.}, \bibinfo{author}{González-Cuadra, D.} \& \bibinfo{author}{Zoller, P.}
\newblock \bibinfo{title}{Fermion-qudit quantum processors for simulating lattice gauge theories with matter}.
\newblock \href {https://doi.org/10.22331/q-2023-10-16-1140}{\emph{\bibinfo{journal}{Quantum}} \textbf{\bibinfo{volume}{7}}, \bibinfo{pages}{1140}} (\bibinfo{year}{2023}).

\bibitem{tabares_programming_2025}
\bibinfo{author}{Tabares, C.}, \bibinfo{author}{Kokail, C.}, \bibinfo{author}{Zoller, P.}, \bibinfo{author}{González-Cuadra, D.} \& \bibinfo{author}{González-Tudela, A.}
\newblock \bibinfo{title}{Programming optical-lattice {Fermi}-{Hubbard} quantum simulators}.
\newblock Preprint at \href {https://arxiv.org/abs/2502.05067} {https://arxiv.org/abs/2502.05067} (\bibinfo{year}{2025}).

\bibitem{gkritsis_simulating_2024}
\bibinfo{author}{Gkritsis, F.} \emph{et~al.}
\newblock \bibinfo{title}{Simulating chemistry with fermionic optical superlattices}.
\newblock \href {https://doi.org/10.1103/PRXQuantum.6.010318}{\emph{\bibinfo{journal}{PRX Quantum}} \textbf{\bibinfo{volume}{6}}, \bibinfo{pages}{010318}} (\bibinfo{year}{2025}).

\bibitem{raussendorf_measurement-based_2003}
\bibinfo{author}{Raussendorf, R.}, \bibinfo{author}{Browne, D.~E.} \& \bibinfo{author}{Briegel, H.~J.}
\newblock \bibinfo{title}{Measurement-based quantum computation on cluster states}.
\newblock \href {https://doi.org/10.1103/PhysRevA.68.022312}{\emph{\bibinfo{journal}{Phys. Rev. A}} \textbf{\bibinfo{volume}{68}}, \bibinfo{pages}{022312}} (\bibinfo{year}{2003}).

\bibitem{bluvstein_logical_2024}
\bibinfo{author}{Bluvstein, D.} \emph{et~al.}
\newblock \bibinfo{title}{Logical quantum processor based on reconfigurable atom arrays}.
\newblock \href {https://doi.org/10.1038/s41586-023-06927-3}{\emph{\bibinfo{journal}{Nature}} \textbf{\bibinfo{volume}{626}}, \bibinfo{pages}{58--65}} (\bibinfo{year}{2024}).

\bibitem{manetsch2024tweezerarray6100highly}
\bibinfo{author}{Manetsch, H.~J.} \emph{et~al.}
\newblock \bibinfo{title}{A tweezer array with 6100 highly coherent atomic qubits}.
\newblock Preprint at \href {https://arxiv.org/abs/2403.12021} {https://arxiv.org/abs/2403.12021} (\bibinfo{year}{2024}).

\bibitem{Pino_ionCCD_2021}
\bibinfo{author}{Pino, J.~M.} \emph{et~al.}
\newblock \bibinfo{title}{Demonstration of the trapped-ion quantum ccd computer architecture}.
\newblock \href {https://doi.org/10.1038/s41586-021-03318-4}{\emph{\bibinfo{journal}{Nature}} \textbf{\bibinfo{volume}{592}}, \bibinfo{pages}{209–213}} (\bibinfo{year}{2021}).

\bibitem{Moses_racetrack_2023}
\bibinfo{author}{Moses, S.} \emph{et~al.}
\newblock \bibinfo{title}{A race-track trapped-ion quantum processor}.
\newblock \href {https://doi.org/10.1103/physrevx.13.041052}{\emph{\bibinfo{journal}{Phys. Rev. X}} \textbf{\bibinfo{volume}{13}}, \bibinfo{pages}{041052}} (\bibinfo{year}{2023}).

\bibitem{Iqbal_2024}
\bibinfo{author}{Iqbal, M.} \emph{et~al.}
\newblock \bibinfo{title}{Topological order from measurements and feed-forward on a trapped ion quantum computer}.
\newblock \href {https://doi.org/10.1038/s42005-024-01698-3}{\emph{\bibinfo{journal}{Commun. Phys.}} \textbf{\bibinfo{volume}{7}}, \bibinfo{pages}{205}} (\bibinfo{year}{2024}).

\bibitem{bravyi_high_threshold_2024}
\bibinfo{author}{Bravyi, S.} \emph{et~al.}
\newblock \bibinfo{title}{High-threshold and low-overhead fault-tolerant quantum memory}.
\newblock \href {https://doi.org/10.1038/s41586-024-07107-7}{\emph{\bibinfo{journal}{Nature}} \textbf{\bibinfo{volume}{627}}, \bibinfo{pages}{778--782}} (\bibinfo{year}{2024}).

\bibitem{gottesman1997stabilizercodesquantumerror}
\bibinfo{author}{Gottesman, D.}
\newblock \bibinfo{title}{Stabilizer codes and quantum error correction}.
\newblock Preprint at \href {https://arxiv.org/abs/quant-ph/9705052} {https://arxiv.org/abs/quant-ph/9705052} (\bibinfo{year}{1997}).

\bibitem{Nielsen_Chuang_2010}
\bibinfo{author}{Nielsen, M.~A.} \& \bibinfo{author}{Chuang, I.~L.}
\newblock \emph{\bibinfo{title}{Quantum Computation and Quantum Information: 10th Anniversary Edition}} (\bibinfo{publisher}{Cambridge University Press}, \bibinfo{year}{2010}).

\bibitem{sachdev_gapless_1993}
\bibinfo{author}{Sachdev, S.} \& \bibinfo{author}{Ye, J.}
\newblock \bibinfo{title}{Gapless spin-fluid ground state in a random quantum {Heisenberg} magnet}.
\newblock \href {https://doi.org/10.1103/PhysRevLett.70.3339}{\emph{\bibinfo{journal}{Phys. Rev. Lett.}} \textbf{\bibinfo{volume}{70}}, \bibinfo{pages}{3339--3342}} (\bibinfo{year}{1993}).

\bibitem{kitaev_simple_2015}
\bibinfo{author}{Kitaev, A.}
\newblock \bibinfo{title}{A simple model of quantum holography} (\bibinfo{year}{2015}).

\bibitem{xu_sparse_2020}
\bibinfo{author}{Xu, S.}, \bibinfo{author}{Susskind, L.}, \bibinfo{author}{Su, Y.} \& \bibinfo{author}{Swingle, B.}
\newblock \bibinfo{title}{A sparse model of quantum holography}.
\newblock Preprint at \href {https://arxiv.org/abs/2008.02303} {https://arxiv.org/abs/2008.02303} (\bibinfo{year}{2020}).

\bibitem{günther2025phaseestimationpartiallyrandomized}
\bibinfo{author}{Günther, J.} \emph{et~al.}
\newblock \bibinfo{title}{Phase estimation with partially randomized time evolution}.
\newblock Preprint at \href {https://arxiv.org/abs/2503.05647} {https://arxiv.org/abs/2503.05647} (\bibinfo{year}{2025}).

\bibitem{knuth_art_1997}
\bibinfo{author}{Knuth, D.~E.}
\newblock \emph{\bibinfo{title}{The art of computer programming}} (\bibinfo{address}{Reading, Mass}, \bibinfo{year}{1997}), \bibinfo{edition}{3rd ed} edn.

\bibitem{van_den_nest_graphical_2004}
\bibinfo{author}{Van Den~Nest, M.}, \bibinfo{author}{Dehaene, J.} \& \bibinfo{author}{De~Moor, B.}
\newblock \bibinfo{title}{Graphical description of the action of local {Clifford} transformations on graph states}.
\newblock \href {https://doi.org/10.1103/PhysRevA.69.022316}{\emph{\bibinfo{journal}{Phys. Rev. A}} \textbf{\bibinfo{volume}{69}}, \bibinfo{pages}{022316}} (\bibinfo{year}{2004}).

\bibitem{childs_exponential_2003}
\bibinfo{author}{Childs, A.~M.} \emph{et~al.}
\newblock \bibinfo{title}{Exponential algorithmic speedup by quantum walk}.
\newblock In \emph{\bibinfo{booktitle}{Proceedings of the thirty-fifth annual {ACM} symposium on {Theory} of computing}}, \bibinfo{pages}{59--68} (\bibinfo{year}{2003}).
\newblock \urlprefix\url{http://arxiv.org/abs/quant-ph/0209131}.
\newblock \bibinfo{note}{ArXiv:quant-ph/0209131}.

\bibitem{aharonov2003adiabaticquantumstategeneration}
\bibinfo{author}{Aharonov, D.} \& \bibinfo{author}{Ta-Shma, A.}
\newblock \bibinfo{title}{Adiabatic quantum state generation and statistical zero knowledge}.
\newblock Preprint at \href {https://arxiv.org/abs/quant-ph/0301023} {https://arxiv.org/abs/quant-ph/0301023} (\bibinfo{year}{2003}).

\bibitem{Low_hamiltonian_simulation_2019}
\bibinfo{author}{Low, G.~H.} \& \bibinfo{author}{Chuang, I.~L.}
\newblock \bibinfo{title}{Hamiltonian simulation by qubitization}.
\newblock \href {https://doi.org/10.22331/q-2019-07-12-163}{\emph{\bibinfo{journal}{Quantum}} \textbf{\bibinfo{volume}{3}}, \bibinfo{pages}{163}} (\bibinfo{year}{2019}).

\bibitem{Babbush_linear_T}
\bibinfo{author}{Babbush, R.} \emph{et~al.}
\newblock \bibinfo{title}{Encoding electronic spectra in quantum circuits with linear t complexity}.
\newblock \href {https://doi.org/10.1103/PhysRevX.8.041015}{\emph{\bibinfo{journal}{Phys. Rev. X}} \textbf{\bibinfo{volume}{8}}, \bibinfo{pages}{041015}} (\bibinfo{year}{2018}).

\bibitem{Ferris_spectral_2014}
\bibinfo{author}{Ferris, A.~J.}
\newblock \bibinfo{title}{Fourier transform for fermionic systems and the spectral tensor network}.
\newblock \href {https://doi.org/10.1103/physrevlett.113.010401}{\emph{\bibinfo{journal}{Phys. Rev. Lett.}} \textbf{\bibinfo{volume}{113}}, \bibinfo{pages}{010401}} (\bibinfo{year}{2014}).

\bibitem{orman_quantum_2024}
\bibinfo{author}{Orman, P.}, \bibinfo{author}{Gharibyan, H.} \& \bibinfo{author}{Preskill, J.}
\newblock \bibinfo{title}{Quantum chaos in the sparse syk model}.
\newblock Preprint at \href {https://arxiv.org/abs/2403.13884} {https://arxiv.org/abs/2403.13884} (\bibinfo{year}{2024}).

\bibitem{Schmoll_honeycomb_2017}
\bibinfo{author}{Schmoll, P.} \& \bibinfo{author}{Orús, R.}
\newblock \bibinfo{title}{Kitaev honeycomb tensor networks: Exact unitary circuits and applications}.
\newblock \href {https://doi.org/10.1103/physrevb.95.045112}{\emph{\bibinfo{journal}{Phys. Rev. B}} \textbf{\bibinfo{volume}{95}}, \bibinfo{pages}{045112}} (\bibinfo{year}{2017}).

\bibitem{fowler2019lowoverheadquantumcomputation}
\bibinfo{author}{Fowler, A.~G.} \& \bibinfo{author}{Gidney, C.}
\newblock \bibinfo{title}{Low overhead quantum computation using lattice surgery}.
\newblock Preprint at \href {https://arxiv.org/abs/1808.06709} {https://arxiv.org/abs/1808.06709} (\bibinfo{year}{2019}).

\bibitem{cain_correlated_decoding_2024}
\bibinfo{author}{Cain, M.} \emph{et~al.}
\newblock \bibinfo{title}{Correlated decoding of logical algorithms with transversal gates}.
\newblock \href {https://doi.org/10.1103/PhysRevLett.133.240602}{\emph{\bibinfo{journal}{Phys. Rev. Lett.}} \textbf{\bibinfo{volume}{133}}, \bibinfo{pages}{240602}} (\bibinfo{year}{2024}).

\bibitem{zhou2024algorithmicfaulttolerancefast}
\bibinfo{author}{Zhou, H.} \emph{et~al.}
\newblock \bibinfo{title}{Algorithmic fault tolerance for fast quantum computing}.
\newblock Preprint at \href {https://arxiv.org/abs/2406.17653} {https://arxiv.org/abs/2406.17653} (\bibinfo{year}{2024}).

\bibitem{Zhou_architecture_2025_ICSA}
\bibinfo{author}{Zhou, H.} \emph{et~al.}
\newblock \bibinfo{title}{Resource analysis of low-overhead transversal architectures for reconfigurable atom arrays}.
\newblock In \emph{\bibinfo{booktitle}{Proceedings of the 52nd Annual International Symposium on Computer Architecture}}, ISCA '25, \bibinfo{pages}{1432–1448} (\bibinfo{publisher}{Association for Computing Machinery}, \bibinfo{address}{New York, NY, USA}, \bibinfo{year}{2025}).
\newblock \urlprefix\url{https://doi.org/10.1145/3695053.3731039}.

\bibitem{Poulin_spectral_2018}
\bibinfo{author}{Poulin, D.}, \bibinfo{author}{Kitaev, A.}, \bibinfo{author}{Steiger, D.~S.}, \bibinfo{author}{Hastings, M.~B.} \& \bibinfo{author}{Troyer, M.}
\newblock \bibinfo{title}{Quantum algorithm for spectral measurement with a lower gate count}.
\newblock \href {https://doi.org/10.1103/physrevlett.121.010501}{\emph{\bibinfo{journal}{Phys. Rev. Lett.}} \textbf{\bibinfo{volume}{121}}, \bibinfo{pages}{010501}} (\bibinfo{year}{2018}).

\bibitem{corboz_fermionic_2009}
\bibinfo{author}{Corboz, P.} \& \bibinfo{author}{Vidal, G.}
\newblock \bibinfo{title}{Fermionic multiscale entanglement renormalization ansatz}.
\newblock \href {https://doi.org/10.1103/PhysRevB.80.165129}{\emph{\bibinfo{journal}{Phys. Rev. B}} \textbf{\bibinfo{volume}{80}}, \bibinfo{pages}{165129}} (\bibinfo{year}{2009}).

\bibitem{dai_fermionic_2025}
\bibinfo{author}{Dai, Z.}, \bibinfo{author}{Wu, Y.}, \bibinfo{author}{Wang, T.} \& \bibinfo{author}{Zaletel, M.~P.}
\newblock \bibinfo{title}{Fermionic {Isometric} {Tensor} {Network} {States} in {Two} {Dimensions}}.
\newblock \href {https://doi.org/10.1103/PhysRevLett.134.026502}{\emph{\bibinfo{journal}{Phys. Rev. Lett.}} \textbf{\bibinfo{volume}{134}}, \bibinfo{pages}{026502}} (\bibinfo{year}{2025}).

\bibitem{wan_matchgate_2023}
\bibinfo{author}{Wan, K.}, \bibinfo{author}{Huggins, W.~J.}, \bibinfo{author}{Lee, J.} \& \bibinfo{author}{Babbush, R.}
\newblock \bibinfo{title}{Matchgate {Shadows} for {Fermionic} {Quantum} {Simulation}}.
\newblock \href {https://doi.org/10.1007/s00220-023-04844-0}{\emph{\bibinfo{journal}{Commun. Math. Phys.}} \textbf{\bibinfo{volume}{404}}, \bibinfo{pages}{629--700}} (\bibinfo{year}{2023}).

\bibitem{Tantivasadakarn_2024}
\bibinfo{author}{Tantivasadakarn, N.}, \bibinfo{author}{Thorngren, R.}, \bibinfo{author}{Vishwanath, A.} \& \bibinfo{author}{Verresen, R.}
\newblock \bibinfo{title}{Long-range entanglement from measuring symmetry-protected topological phases}.
\newblock \href {https://doi.org/10.1103/physrevx.14.021040}{\emph{\bibinfo{journal}{Physical Review X}} \textbf{\bibinfo{volume}{14}}} (\bibinfo{year}{2024}).

\bibitem{Tantivasadakarn_shortest_2023}
\bibinfo{author}{Tantivasadakarn, N.}, \bibinfo{author}{Verresen, R.} \& \bibinfo{author}{Vishwanath, A.}
\newblock \bibinfo{title}{Shortest route to non-abelian topological order on a quantum processor}.
\newblock \href {https://doi.org/10.1103/physrevlett.131.060405}{\emph{\bibinfo{journal}{Physical Review Letters}} \textbf{\bibinfo{volume}{131}}} (\bibinfo{year}{2023}).

\bibitem{tantivasadakarn_hierarchy_2023}
\bibinfo{author}{Tantivasadakarn, N.}, \bibinfo{author}{Vishwanath, A.} \& \bibinfo{author}{Verresen, R.}
\newblock \bibinfo{title}{Hierarchy of {Topological} {Order} {From} {Finite}-{Depth} {Unitaries}, {Measurement}, and {Feedforward}}.
\newblock \href {https://doi.org/10.1103/PRXQuantum.4.020339}{\emph{\bibinfo{journal}{PRX Quantum}} \textbf{\bibinfo{volume}{4}}, \bibinfo{pages}{020339}} (\bibinfo{year}{2023}).

\bibitem{verresen2022efficientlypreparingschrodingerscat}
\bibinfo{author}{Verresen, R.}, \bibinfo{author}{Tantivasadakarn, N.} \& \bibinfo{author}{Vishwanath, A.}
\newblock \bibinfo{title}{Efficiently preparing schr\"odinger's cat, fractons and non-abelian topological order in quantum devices}.
\newblock Preprint at \href {https://arxiv.org/abs/2112.03061} {https://arxiv.org/abs/2112.03061} (\bibinfo{year}{2022}).

\bibitem{bravyi2022adaptiveconstantdepthcircuitsmanipulating}
\bibinfo{author}{Bravyi, S.}, \bibinfo{author}{Kim, I.}, \bibinfo{author}{Kliesch, A.} \& \bibinfo{author}{Koenig, R.}
\newblock \bibinfo{title}{Adaptive constant-depth circuits for manipulating non-abelian anyons} (\bibinfo{year}{2022}).
\newblock \urlprefix\url{https://arxiv.org/abs/2205.01933}.
\newblock Preprint at https://arxiv.org/abs/2205.01933.

\bibitem{Lu_meas_shortcut_2022}
\bibinfo{author}{Lu, T.-C.}, \bibinfo{author}{Lessa, L.~A.}, \bibinfo{author}{Kim, I.~H.} \& \bibinfo{author}{Hsieh, T.~H.}
\newblock \bibinfo{title}{Measurement as a shortcut to long-range entangled quantum matter}.
\newblock \href {https://doi.org/10.1103/PRXQuantum.3.040337}{\emph{\bibinfo{journal}{PRX Quantum}} \textbf{\bibinfo{volume}{3}}, \bibinfo{pages}{040337}} (\bibinfo{year}{2022}).

\bibitem{Ren_adaptive_2025}
\bibinfo{author}{Ren, Y.}, \bibinfo{author}{Tantivasadakarn, N.} \& \bibinfo{author}{Williamson, D.~J.}
\newblock \bibinfo{title}{Efficient preparation of solvable anyons with adaptive quantum circuits}.
\newblock \href {https://doi.org/10.1103/b9hf-gx4f}{\emph{\bibinfo{journal}{Physical Review X}} \textbf{\bibinfo{volume}{15}}} (\bibinfo{year}{2025}).

\bibitem{Christos_dwave_2023}
\bibinfo{author}{Christos, M.} \emph{et~al.}
\newblock \bibinfo{title}{A model of d-wave superconductivity, antiferromagnetism, and charge order on the square lattice}.
\newblock \href {https://doi.org/10.1073/pnas.2302701120}{\emph{\bibinfo{journal}{Proc. Natl. Acad. Sci. U.S.A.}} \textbf{\bibinfo{volume}{120}}, \bibinfo{pages}{e2302701120}} (\bibinfo{year}{2023}).

\bibitem{Arovas_hubbard_2022}
\bibinfo{author}{Arovas, D.~P.}, \bibinfo{author}{Berg, E.}, \bibinfo{author}{Kivelson, S.~A.} \& \bibinfo{author}{Raghu, S.}
\newblock \bibinfo{title}{The hubbard model}.
\newblock \href {https://doi.org/10.1146/annurev-conmatphys-031620-102024}{\emph{\bibinfo{journal}{Annu. Rev. Condens. Matter Phys.}} \textbf{\bibinfo{volume}{13}}, \bibinfo{pages}{239–274}} (\bibinfo{year}{2022}).

\bibitem{Qin_hubbard_2022}
\bibinfo{author}{Qin, M.}, \bibinfo{author}{Schäfer, T.}, \bibinfo{author}{Andergassen, S.}, \bibinfo{author}{Corboz, P.} \& \bibinfo{author}{Gull, E.}
\newblock \bibinfo{title}{The hubbard model: A computational perspective}.
\newblock \href {https://doi.org/10.1146/annurev-conmatphys-090921-033948}{\emph{\bibinfo{journal}{Annu. Rev. Condens. Matter Phys.}} \textbf{\bibinfo{volume}{13}}, \bibinfo{pages}{275–302}} (\bibinfo{year}{2022}).

\bibitem{Tan_DFPNA_2024}
\bibinfo{author}{Tan, D.~B.}, \bibinfo{author}{Bluvstein, D.}, \bibinfo{author}{Lukin, M.~D.} \& \bibinfo{author}{Cong, J.}
\newblock \bibinfo{title}{Compiling quantum circuits for dynamically field-programmable neutral atoms array processors}.
\newblock \href {https://doi.org/10.22331/q-2024-03-14-1281}{\emph{\bibinfo{journal}{Quantum}} \textbf{\bibinfo{volume}{8}}, \bibinfo{pages}{1281}} (\bibinfo{year}{2024}).

\bibitem{bluvstein2025architecturalmechanismsuniversalfaulttolerant}
\bibinfo{author}{Bluvstein, D.} \emph{et~al.}
\newblock \bibinfo{title}{Architectural mechanisms of a universal fault-tolerant quantum computer}.
\newblock Preprint at \href {https://arxiv.org/abs/2506.20661} {https://arxiv.org/abs/2506.20661} (\bibinfo{year}{2025}).

\bibitem{Childs_trotter_2021}
\bibinfo{author}{Childs, A.~M.}, \bibinfo{author}{Su, Y.}, \bibinfo{author}{Tran, M.~C.}, \bibinfo{author}{Wiebe, N.} \& \bibinfo{author}{Zhu, S.}
\newblock \bibinfo{title}{Theory of trotter error with commutator scaling}.
\newblock \href {https://doi.org/10.1103/physrevx.11.011020}{\emph{\bibinfo{journal}{Phys. Rev. X}} \textbf{\bibinfo{volume}{11}}, \bibinfo{pages}{011020}} (\bibinfo{year}{2021}).

\bibitem{Baumer_longrange_entangling_2025}
\bibinfo{author}{Bäumer, E.} \& \bibinfo{author}{Woerner, S.}
\newblock \bibinfo{title}{Measurement-based long-range entangling gates in constant depth}.
\newblock \href {https://doi.org/10.1103/physrevresearch.7.023120}{\emph{\bibinfo{journal}{Phys. Rev. Res.}} \textbf{\bibinfo{volume}{7}}, \bibinfo{pages}{023120}} (\bibinfo{year}{2025}).

\bibitem{Montanaro_iqp_poly_2017}
\bibinfo{author}{Montanaro, A.}
\newblock \bibinfo{title}{Quantum circuits and low-degree polynomials over ${{\mathbb{F}}_\mathsf{2}}$}.
\newblock \href {https://doi.org/10.1088/1751-8121/aa565f}{\emph{\bibinfo{journal}{J. Phys. A}} \textbf{\bibinfo{volume}{50}}, \bibinfo{pages}{084002}} (\bibinfo{year}{2017}).

\bibitem{Haldane_1988}
\bibinfo{author}{Haldane, F. D.~M.}
\newblock \bibinfo{title}{Model for a quantum hall effect without landau levels: Condensed-matter realization of the "parity anomaly"}.
\newblock \href {https://doi.org/10.1103/PhysRevLett.61.2015}{\emph{\bibinfo{journal}{Phys. Rev. Lett.}} \textbf{\bibinfo{volume}{61}}, \bibinfo{pages}{2015--2018}} (\bibinfo{year}{1988}).

\bibitem{Litinski_game_2019}
\bibinfo{author}{Litinski, D.}
\newblock \bibinfo{title}{A game of surface codes: Large-scale quantum computing with lattice surgery}.
\newblock \href {https://doi.org/10.22331/q-2019-03-05-128}{\emph{\bibinfo{journal}{Quantum}} \textbf{\bibinfo{volume}{3}}, \bibinfo{pages}{128}} (\bibinfo{year}{2019}).

\bibitem{kivlichan_improved_2020}
\bibinfo{author}{Kivlichan, I.~D.} \emph{et~al.}
\newblock \bibinfo{title}{Improved {Fault}-{Tolerant} {Quantum} {Simulation} of {Condensed}-{Phase} {Correlated} {Electrons} via {Trotterization}}.
\newblock \href {https://doi.org/10.22331/q-2020-07-16-296}{\emph{\bibinfo{journal}{Quantum}} \textbf{\bibinfo{volume}{4}}, \bibinfo{pages}{296}} (\bibinfo{year}{2020}).

\bibitem{Gidney_catalyzed_2019}
\bibinfo{author}{Gidney, C.} \& \bibinfo{author}{Fowler, A.~G.}
\newblock \bibinfo{title}{Efficient magic state factories with a catalyzed ccz to 2t transformation}.
\newblock \href {https://doi.org/10.22331/q-2019-04-30-135}{\emph{\bibinfo{journal}{Quantum}} \textbf{\bibinfo{volume}{3}}, \bibinfo{pages}{135}} (\bibinfo{year}{2019}).

\bibitem{Bravyi_universal_2005}
\bibinfo{author}{Bravyi, S.} \& \bibinfo{author}{Kitaev, A.}
\newblock \bibinfo{title}{Universal quantum computation with ideal clifford gates and noisy ancillas}.
\newblock \href {https://doi.org/10.1103/physreva.71.022316}{\emph{\bibinfo{journal}{Phys. Rev. A}} \textbf{\bibinfo{volume}{71}}, \bibinfo{pages}{022316}} (\bibinfo{year}{2005}).

\bibitem{knill2004postsel_threshold}
\bibinfo{author}{Knill, E.}
\newblock \bibinfo{title}{Fault-tolerant postselected quantum computation: Threshold analysis}.
\newblock Preprint at \href {https://arxiv.org/abs/quant-ph/0404104} {https://arxiv.org/abs/quant-ph/0404104} (\bibinfo{year}{2004}).

\bibitem{knill2004postsel_schemes}
\bibinfo{author}{Knill, E.}
\newblock \bibinfo{title}{Fault-tolerant postselected quantum computation: Schemes}.
\newblock Preprint at \href {https://arxiv.org/abs/quant-ph/0402171} {https://arxiv.org/abs/quant-ph/0402171} (\bibinfo{year}{2004}).

\bibitem{Litinski_msd_2019}
\bibinfo{author}{Litinski, D.}
\newblock \bibinfo{title}{Magic state distillation: Not as costly as you think}.
\newblock \href {https://doi.org/10.22331/q-2019-12-02-205}{\emph{\bibinfo{journal}{Quantum}} \textbf{\bibinfo{volume}{3}}, \bibinfo{pages}{205}} (\bibinfo{year}{2019}).

\bibitem{Paetznick_switching_2013}
\bibinfo{author}{Paetznick, A.} \& \bibinfo{author}{Reichardt, B.~W.}
\newblock \bibinfo{title}{Universal fault-tolerant quantum computation with only transversal gates and error correction}.
\newblock \href {https://doi.org/10.1103/physrevlett.111.090505}{\emph{\bibinfo{journal}{Phys. Rev. Lett.}} \textbf{\bibinfo{volume}{111}}, \bibinfo{pages}{090505}} (\bibinfo{year}{2013}).

\bibitem{Anderson_switching_2014}
\bibinfo{author}{Anderson, J.~T.}, \bibinfo{author}{Duclos-Cianci, G.} \& \bibinfo{author}{Poulin, D.}
\newblock \bibinfo{title}{Fault-tolerant conversion between the steane and reed-muller quantum codes}.
\newblock \href {https://doi.org/10.1103/physrevlett.113.080501}{\emph{\bibinfo{journal}{Phys. Rev. Lett.}} \textbf{\bibinfo{volume}{113}}, \bibinfo{pages}{080501}} (\bibinfo{year}{2014}).

\bibitem{bombin2015gaugefixing}
\bibinfo{author}{Bombin, H.}
\newblock \bibinfo{title}{Gauge color codes: Optimal transversal gates and gauge fixing in topological stabilizer codes}.
\newblock Preprint at \href {https://arxiv.org/abs/1311.0879} {https://arxiv.org/abs/1311.0879} (\bibinfo{year}{2015}).

\bibitem{bombin2016dimjump}
\bibinfo{author}{Bombin, H.}
\newblock \bibinfo{title}{Dimensional jump in quantum error correction}.
\newblock Preprint at \href {https://arxiv.org/abs/1412.5079} {https://arxiv.org/abs/1412.5079} (\bibinfo{year}{2016}).

\bibitem{Beverland_costof_2021}
\bibinfo{author}{Beverland, M.~E.}, \bibinfo{author}{Kubica, A.} \& \bibinfo{author}{Svore, K.~M.}
\newblock \bibinfo{title}{Cost of universality: A comparative study of the overhead of state distillation and code switching with color codes}.
\newblock \href {https://doi.org/10.1103/prxquantum.2.020341}{\emph{\bibinfo{journal}{PRX Quantum}} \textbf{\bibinfo{volume}{2}}, \bibinfo{pages}{020341}} (\bibinfo{year}{2021}).

\bibitem{Bomb_lowmagic_2024}
\bibinfo{author}{Bombín, H.}, \bibinfo{author}{Pant, M.}, \bibinfo{author}{Roberts, S.} \& \bibinfo{author}{Seetharam, K.~I.}
\newblock \bibinfo{title}{Fault-tolerant postselection for low-overhead magic state preparation}.
\newblock \href {https://doi.org/10.1103/prxquantum.5.010302}{\emph{\bibinfo{journal}{PRX Quantum}} \textbf{\bibinfo{volume}{5}}, \bibinfo{pages}{010302}} (\bibinfo{year}{2024}).

\bibitem{Chamberland_lowoverhead_2020}
\bibinfo{author}{Chamberland, C.} \& \bibinfo{author}{Noh, K.}
\newblock \bibinfo{title}{Very low overhead fault-tolerant magic state preparation using redundant ancilla encoding and flag qubits}.
\newblock \href {https://doi.org/10.1038/s41534-020-00319-5}{\emph{\bibinfo{journal}{npj Quantum Inf.}} \textbf{\bibinfo{volume}{6}}} (\bibinfo{year}{2020}).

\bibitem{gidney2024magicstatecultivationgrowing}
\bibinfo{author}{Gidney, C.}, \bibinfo{author}{Shutty, N.} \& \bibinfo{author}{Jones, C.}
\newblock \bibinfo{title}{Magic state cultivation: growing t states as cheap as cnot gates}.
\newblock Preprint at \href {https://arxiv.org/abs/2409.17595} {https://arxiv.org/abs/2409.17595} (\bibinfo{year}{2024}).

\bibitem{kliuchnikov2013fastefficientexactsynthesis}
\bibinfo{author}{Kliuchnikov, V.}, \bibinfo{author}{Maslov, D.} \& \bibinfo{author}{Mosca, M.}
\newblock \bibinfo{title}{Fast and efficient exact synthesis of single qubit unitaries generated by clifford and t gates}.
\newblock Preprint at \href {https://arxiv.org/abs/1206.5236} {https://arxiv.org/abs/1206.5236} (\bibinfo{year}{2013}).

\bibitem{ross2016optimalancillafreecliffordtapproximation}
\bibinfo{author}{Ross, N.~J.} \& \bibinfo{author}{Selinger, P.}
\newblock \bibinfo{title}{Optimal ancilla-free clifford+t approximation of z-rotations}.
\newblock Preprint at \href {https://arxiv.org/abs/1403.2975} {https://arxiv.org/abs/1403.2975} (\bibinfo{year}{2016}).

\bibitem{paetznick2014repeatuntilsuccessnondeterministicdecompositionsinglequbit}
\bibinfo{author}{Paetznick, A.} \& \bibinfo{author}{Svore, K.~M.}
\newblock \bibinfo{title}{Repeat-until-success: Non-deterministic decomposition of single-qubit unitaries}.
\newblock Preprint at \href {https://arxiv.org/abs/1311.1074} {https://arxiv.org/abs/1311.1074} (\bibinfo{year}{2014}).

\bibitem{Bocharov_RUS_2015}
\bibinfo{author}{Bocharov, A.}, \bibinfo{author}{Roetteler, M.} \& \bibinfo{author}{Svore, K.~M.}
\newblock \bibinfo{title}{Efficient synthesis of universal repeat-until-success quantum circuits}.
\newblock \href {https://doi.org/10.1103/physrevlett.114.080502}{\emph{\bibinfo{journal}{Phys. Rev. Lett.}} \textbf{\bibinfo{volume}{114}}} (\bibinfo{year}{2015}).

\end{thebibliography}

\clearpage
\onecolumngrid
\begin{center}

\newcommand{\beginsupplement}{%
        \setcounter{table}{0}
        \renewcommand{\thetable}{S\arabic{table}}%
        \setcounter{figure}{0}
        \renewcommand{\thefigure}{S\arabic{figure}}%
     }
\textbf{\large Supplemental Material}
\end{center}
\newcommand{\beginsupplement}{%
        \setcounter{table}{0}
        \renewcommand{\thetable}{S\arabic{table}}%
        \setcounter{figure}{0}
        \renewcommand{\thefigure}{S\arabic{figure}}%
     }

\appendix


\section{Fermionic Permutations}\label{appendix:fermionic_permutation}

The core technical result of this paper is a constructive procedure for efficiently compiling fermionic permutations $\mathcal{F}_p$ into qubit operations, while keeping the total number of Clifford gates and ancilla operations relatively small.

\subsection{Switching Jordan-Wigner encodings}

We start by defining in detail the operation that is required to switch between two Jordan-Wigner encodings.
The ordering of modes at time $t$ is described by $m_t(i)$.
For each $i$ and $t$, let $L_t(i) = \{ j | m_t(j) < m_t(i),j=1,...,N \}$ denote the set of all modes to the left of $i$ during step $t$.
The Jordan-Wigner encoding of the majorana fermionic operators is given by
\begin{align}
    \chi^{(t)}_{2i} = X_i \prod_{i' \in L_t(i)} Z_{i'}, \, \chi^{(t)}_{2i+1} = Y_i \prod_{i' \in L_t(i)} Z_{i'},
\end{align}
where the majorana operators are related to canonical fermionic operators by
\begin{align}
    c_i^{\vphantom{\dagger}} &= \frac{1}{2}\left(\chi_{2i} + i\chi_{2i+1} \right), \, c_i^{\dagger} = \frac{1}{2}\left(\chi_{2i} - i\chi_{2i+1} \right).
\end{align}

\newcommand{\sdiff}{\, \Delta \,} 

Define the target ordering during the next step as $m_{t+1}(i)$. The goal of the fermionic permutation circuit $\mathcal{F}_p$ is to transform the encoding, so that it is consistent with $m_{t+1}(i)$, and the corresponding left-function $L_{t+1}(i)$. In what follows, we simply consider two times $t=0$ and $t=1$, and their associated orderings $m_0$ and $m_1$. 

To track how the elements of the left-function change, we introduce the \textit{set difference} operation $A \sdiff B = (A \cup B) - (A \cap B)$, which identifies elements that are in $A$ or $B$ but not both. 

\begin{lemma}{\label{lem:cz_circuit}}
    A fermionic permutation $\mathcal{F}_p$ that transforms from a JW encoding with order $m_0(i)$ to a JW encoding with order $m_1(i)$, is equivalent to a CZ circuit containing two-qubit gates between all pairs of modes $(i, j)$ satisfying $i \in L_0(j) \Delta L_1(j)$. 
\end{lemma}

\begin{proof}
    First, we show that the inclusion definition is symmetric. So if we have a pair of modes $(i,j)$ satisfying $i \in L_0(j) \sdiff L_1(j)$, then it also satisfies $j \in L_0(i) \Delta L_1(i)$.
    Recall the definition that an element $x \in A \sdiff B$ if and only if $(x \in A) \oplus (x \in B)$, where $\oplus$ is the XOR operation. 
    Next, observe that $i \in L_0(j)$ if and only if $j \notin L_0(i)$, by the definition of left-inclusion. 
    Therefore, $(j \in L_0(i)) = \neg (i \in L_0(j))$ are related by negation.
    The XOR operation is invariant under negation, and therefore
    \begin{align}
        i \in L_0(j) \sdiff L_1(j) &= (i \in L_0(j)) \oplus (i \in L_1(j)) = \neg (i \in L_0(j)) \oplus \neg (i \in L_1(j)) \\
        &= (j \in L_0(i)) \oplus (j \in L_1(i)) = j \in L_0(i) \sdiff L_1(i)
    \end{align}
    This simplifies our analysis of the CZ circuit, because for any qubit $i$, we are guaranteed that it is only paired with qubits $j$ satisfying $j \in L_0(i) \sdiff L_1(i)$.

    Next, we use this observation to compute how the encoded Majorana fermions are transformed by the CZ circuit, denoted $C_Z$.
    \begin{align}
        C_Z \chi^{(0)}_{2i} C_Z = C_Z X_i C_Z \prod_{j \in L_0(i)} Z_{j} 
        = X_i \left( \prod_{j \in L_0(i) \sdiff L_1(i)} Z_{j}\right) \left( \prod_{j \in L_0(i)} Z_{j} \right) 
        = X_i \prod_{i' \in L_1(i)} Z_{i'} = \chi^{(1)}_{2i}
    \end{align}
    Hence, conjugation serves to multiply the Jordan-Wigner product by $L_0(i) \sdiff L_1(i)$ and transforms the non-local string from $L_0(i)$ to $L_1(i)$.
    This works because if $j$ is in $L_0(i) \sdiff L_1(i)$ and is in $L_0(i)$, then the Pauli identity $(Z_j)^2=I$ removes it from the product. Similarly, if $j$ is in $L_0(i) \sdiff L_1(i)$ and is not in $L_0(i)$, it is added to the product. An identical calculation shows $C_Z \chi^{(0)}_{2i+1} C_Z = \chi^{(1)}_{2i+1}$.

    Finally, we point out that the majorana operators $\chi_{i}$ form a complete basis for the space of operators. Therefore, all operations $\mathcal{F}_p$ which transforms $\chi_i^{(0)}$ to $\chi_i^{(1)}$ are equivalent up to trivial operations like global phases. Hence, $\mathcal{F}_p$ is equivalent to the $C_Z$ circuit described above.
\end{proof}

\subsection{Proof of Correctness of Interleave Circuit}

Next, we prove Lemma 1, by showing that the interleave circuit depicted in Figure.~\ref{fig:fig2}b implements a correct fermionic permutation $\mathcal{F}_p$. In particular, the approach we take here is to directly show equivalence with the CZ circuit defined in Lemma~\ref{lem:cz_circuit} using simple circuit identities.
Note that a more general framework is discussed below in the context of the general class of $O(1)$-overhead permutations, and a slightly different form of the interleave will be discussed again there.

\begin{proof} (of Lemma 1.)
Let $A$ be the contiguous group of input qubit with indices ranging from $i=1,...,|A|$, and $B$ the group from $j = |A|+1,...,N$, where $N = |A| + |B|$ is the total number of qubits.
Let $p$ be the target interleave permutation, such that $p(i) < p(i')$ for $i,i' \in A$, and $p(j) < p(j')$ for $j,j' \in B$.

The initial left-set is simple, 
$$L_0(i) = \{ i' | i' < i\},$$
and the final left-set is given by
$$L_1(i) = \{ i' | p(i') < p(i) \}.$$

Thererefore, we see that for any $i \in A$, the set-difference $L_0(i) \setdiff L_1(i) \in B$.
Similarly for any $j \in B$, we see that $L_0(j) \setdiff L_1(j) \in A$.
Thus, the CZ circuit we need to implement is \textit{bi-partite}, and only couples modes in $A$ to modes in $B$.
More specifically, for $i \in A$, the set-difference takes the form
\begin{align}
    L_0(i) \setdiff L_1(i) = \{j \in B | p(j) < p(i) \}.
\end{align}

We now show that the interleave circuit in Fig.~\ref{fig:fig2}b is equivalent to this CZ circuit. 
The key identity we require is the relationship between conjugation of a CZ gate with a CNOT gate.
Let $X_{i \rightarrow j}$ be the CNOT gate with control on site $i$ and target on site $j$.
Similarly let $Z_{ij}$ be a CZ gate on site $i$ and site $j$.
Then, it is straightforward to check the following identities hold, for when (i)  the target of the CNOT gate and (ii) the control of the CNOT gate intersects with the CZ gate at one location.
\begin{align}
    X_{1 \rightarrow 2} Z_{23} X_{1 \rightarrow 2} &= Z_{23}Z_{13}   \nonumber \\
    X_{2 \rightarrow 1} Z_{23} X_{2 \rightarrow 1} &= Z_{23}
\end{align}

Recall, the circuit starts by applying a CNOT cascade to group $B$. We number qubits in order in $B$ as $|A|+1, |A|+2,...,|A|+|B|$ of the form
\begin{align}
    U_X = \prod_{j=|A|+1,...,|A|+|B|-1}^{\rightarrow} X_{j, j+1},
\end{align}
where the product is generated from left-to-right.
Then, we apply a CZ circuit $C_Z$, which couples each mode $i \in A$ to the largest-index mode $j \in B$ such that $p(j) < p(i)$. 
Let $j_{max}(i)$ be the largest such $j$.
Then, the CZ circuit can be written as.
\begin{align}
    C_Z = \prod_{i=1,...,|A|} Z_{i, j_{max}(i)}
\end{align}

Since all $C_Z$ gates commute, it is sufficient to consider one representative pair $Z_{i,j_{max}(i)}$.
Then, we see that conjugation leads to
\begin{align}\label{eq:conjugated_circuit}
    U_X Z_{i,j} U_X^{\dagger} = Z_{i,j_{max}(i)} Z_{i,j_{max}(i)-1} ... Z_{i,|A|+1}
\end{align}
Further, by construction the set $\{ j \in B | j \leq j_{max}(i)\} = L_0(i) \setdiff L_1(i)$.
When extended to all pairs $Z_{i,j_{max}(i)}$, this realizes the target CZ circuit.
The CNOT cascade $U_X$ can be generated in $O(1)$ depth using $O(|B|) = O(N)$ gates and $O(|B|) = O(N)$ ancillas, using the techniques described in Ref.~\cite{Baumer_longrange_entangling_2025} and depicted in Fig.~\ref{fig:fig2}c. The classical processing involves executing a CNOT circuit with $O(N)$ operations arranged in depth-$O(N)$.

The $C_Z$ circuit by definition contains $|A| = O(N)$ gates. However, in some cases, there may be multiple sites $i$ which map to the same $j = j_{max}(i)$, potentially increasing the implementation depth.
However, this edge case can be easily accounted for.

Let $A_j = \{i \in A | j_{max}(i) = j \}$.
Further, by the interleave property, we see that $A_j$ is contiguous.
So, we label these modes $i_1,i_2,...,i_{|A_j|}$.
Then, we see that the CNOT circuit coupling each mode $i$ to each mode $j$ takes the form
\begin{align}
    C_Z[A_j] = Z_{i_1,j} Z_{i_2,j}...Z_{i_{N},j}
\end{align}
This takes the same form as a conjugated circuit Eq.~\ref{eq:conjugated_circuit}, but where the index being iterated lies in $A$ instead of $B$.
Hence, an identical argument shows that we can apply a CNOT cascade to $A_j$, in order to compress these $|A_j|$ CZ gates to a single CZ gate, conjugated by a size $O(|A_j|)$ CNOT cascade. Since this CNOT cascade is non-overlapping with the cascade applied to $B$, it can be performed in parallel.

Therefore, after making this simplification, we see that the CZ circuit can always be made constant depth.
This proves that in total, the circuit contains $O(N)$ gates, and can be arranged in $O(1)$ depth with $O(N)$ ancillas.
More strongly, we see that every mode in $A$ or $B$ is involved in at most two CNOT cascades, and one CZ gate.
Hence, the number of CNOT gates is bounded by $4N$, the number of CZ gates is bounded by $N$, and the number of ancillas is bounded by $N$.
\end{proof}

\section{Extended class of $O(1)$-overhead operations}

While the interleave is sufficient to prove our main theorem, in practice there are many additional structures of fermionic permutations that can be implemented with $O(1)$-overhead, that cannot be directly compiled into $O(1)$ interleave gates.
These include the 1D and 2D reflections depicted in Figure.~\ref{fig:fig3}, which are particularly important for the fast state-preparation techniques discussed in Figure.~\ref{fig:fig5}.

The construction of these operations is considerably more involved than the interleave operation, so to facilitate careful analysis, we start by introducing a simple matrix-based formalism.
Then, we show how generating fermionic permutations corresponds to a particular kind of matrix decomposition. 
Finally, we apply the techniques to uncover a few additional $O(1)$-structures, and present a heuristic algorithm which in some cases can aide in identifying such structures.

\subsection{Formalism for working with CNOT + CZ circuits}

To start, we introduce some formalism and definitions, which we will use to generate efficient algorithms and protocols for constructing CZ circuits relevant for implementing fast fermionic permutations. 
Our approach is to represent CZ circuits as quadratic polynomials~\cite{Montanaro_iqp_poly_2017}, whose coefficients are encoded into matrices.

\begin{definition}
    An $N$-qubit circuit $C_Z(A)$ composed entirely of Z and CZ gates can be uniquely specified by a symmetric matrix $A \in \mathds{F}_2^{N \times N}$. The circuit $C_Z(A)$ contains a $Z$ gate acting on qubit $i$ if and only if $A_{ii} = 1$, and contains a $CZ$ gate between a pair of qubits $(i,j)$ if and only if $A_{ij} = A_{ji} = 1/2$. 
    \begin{enumerate}
        \item We often decompose $A = A^0 + \frac{1}{2}(B + B^{T})$, where $A^0 \in \mathds{F}_2^{N}$ is a diagonal matrix and $B \in \mathds{F}_2^{N(N-1)/2}$ is a lower-triangular matrix.
        \item $C_Z(A)$ acts on a computational basis state $\vert x \rangle = \vert x_1 \rangle \vert x_2 \rangle ... \vert x_n \rangle$ by applying a phase $C_Z(A) \vert x \rangle = (-1)^{f_A(x)}\vert x \rangle$ where $f_A(x) = x^T A x = \sum_{ij} x_i A_{ij} x_j$.
        \item Composition of two circuits is described by matrix addition, $C_Z(A_1) + C_Z(A_2) = C_Z(A_1 + A_2)$. Diagonal entries are defined up to multiples of 2, so can be mapped to to $\{0,1\}$. Off-diagonal entries are defined up to integers, and can be mapped to $\{0,1/2\}$.
    \end{enumerate}
\end{definition}

\begin{definition}
    An $N$-qubit circuit $C_X(P)$ composed entirely of CNOT gates is specified by a matrix $P \in \mathds{F}_2^{N \times N}$. Two key properties of $C_X(P)$ are:
    \begin{enumerate}
        \item The matrix $P$ captures how $X$-operators transform. In particular, $C_X(P) X_j = \left[ \prod_i (X_i)^{P_{ij}} \right] C_X(P)$. 
        \item Similarly, $P$ captures how computational basis states transform. In particular, $C_X(P) \vert x \rangle = \vert x' \rangle$ where $(x')_i = \sum_j P_{ij} x_j$, or equivalently $x' = P x$.
        \item Composition of two $C_X$ circuits is equivalent to matrix multiplication, $C_X(P_1) C_X(P_2) = C_X(P_1 P_2)$.
    \end{enumerate}
\end{definition}

\begin{lemma}{\label{lem:conjugation_formula}}
    A CZ circuit $C_Z(A)$ conjugating by a CNOT circuit $C_X(P)$ is equivalent to a CZ circuit $C_Z(\overline{A})$ where $\overline{A} = P^T A P$ is defined below.
\end{lemma}

\begin{proof}
    Let $x$ be a bit-string. 
    By definition, the CZ circuit applies parity $(-1)^{f_A(x)} = \langle x \vert C_Z(A) \vert x \rangle$. 
    The effect of the CNOT circuit $C_X(P)$ is to transform $x \rightarrow Px$.
    Hence, for the combined circuit, we have
    \begin{align}
        (-1)^{f_{A,P}(x)} = \langle Px \vert C_Z(A) \vert Px \rangle
    \end{align}
    where
    \begin{align}
        f_{A,P}(x) &= \sum_{ij} \left( \sum_k P_{ik} x_k \right) A_{ij} \left( \sum_l P_{jl} x_l \right) 
        = \sum_{kl} x_k \left( P^T A P \right)_{kl} x_l = f_{\overline{A}}(x).
    \end{align}
    Thus, we see the effective CZ circuit is given by $C_Z(\overline{A})$ where
    \begin{align}
        \overline{A} = P^T A P.
    \end{align}
\end{proof}

\subsection{General purpose algorithm for compiling a fermionic permutation}

Our key technical result is a way of compressing the CZ circuits arising from fermionic permutation, into more efficient ones constructed by conjugating CZ circuits with CNOT circuits. In the main text, we presented one such compression for the interleave operation. Here, we instead work with a more general formalism, which enables us to construct a larger class of $O(1)$-overhead operations.
First, we present a general decomposition of a CZ circuit on $N$ qubits into $N$ CZ gates and a generic CNOT circuit.
The decomposition uses $N$ ancilla modes prepared in the state $\vert 0 \rangle$. 

\begin{theorem}{\label{thm:ancilla_cz}}
    Consider a CZ circuit on $N$ qubits $C_Z(A)$, where we assume the diagonal part $A^0=0$, and $A = \frac{1}{2}(B + B^T)$. We can implement $C_Z(A)$ using a circuit acting on $2N$ qubits, formed by conjugating $C_Z\left(\frac{\mathds{I} \otimes X}{2}\right)$ with $C_X(P)$, where
    \begin{align}
        \mathds{I} \otimes X &= \left(\begin{array}{cc}
                                0 & \mathds{I}\\
                                \mathds{I} & 0
                            \end{array}\right), \\
        P &= \left(\begin{array}{cc}
                                \mathds{I} & R_1\\
                                B & R_2
                            \end{array}\right),
    \end{align}
    and the additional $N$ ancilla qubits are prepared in $\vert 0 \rangle$. Here $R_1$ and $R_2$ can be an arbitrary matrices.
\end{theorem}

\begin{proof}
    First, we apply Lemma.~\ref{lem:conjugation_formula} to compute the form of the effective CZ circuit.
    \begin{align}
        P^T \left(\frac{\mathds{I} \otimes X}{2}\right) P &= \frac{1}{2}\left(\begin{array}{cc}
                    I & B^{T}\\
                    R_1^T & R_2^{T}
        \end{array}\right)\left(\begin{array}{cc}
                    0 & \mathds{I}\\
                    \mathds{I} & 0
        \end{array}\right)\left(\begin{array}{cc}
                    I & R_1\\
                    B & R_2
        \end{array}\right) \\
        &= \frac{1}{2}\left(\begin{array}{cc}
                    (B^{T}+B) & R_2\\
                    R_2^{T} & (R_1^T + R_1)
        \end{array}\right)
    \end{align}
    Next, we apply this to an initial state of the form $(x,0)$, where all ancilla qubits are initialized in zero. 
    \begin{align}
        f_{\mathds{I} \otimes X, P}((x,0)) = \frac{1}{2}
        \left(\begin{array}{cc}
                x & 0\end{array}\right)
        \left(\begin{array}{cc}
                B^{T}+B & R_2\\
                R_2^{T} & R_1^T + R_1
        \end{array}\right)
        \left(\begin{array}{c}
                x\\
                0
        \end{array}\right)
        =\frac{1}{2}\sum_{ij}x_{i}\left(B+B^{T}\right)_{ij}x_{j}=f_{A}(x)
    \end{align} 
    Hence, the initial state makes it so only the upper-left block of the matrix contributes. Therefore, independent of the specific form of $R_1$ and $R_2$, this circuit implements the target CZ circuit given by $A = \frac{1}{2}(B+B^T)$ on the first $N$ qubits.
\end{proof}

Next, we describe a general-purpose algorithm for decomposing $C_X(P)$ into two-qubit gates, for lower-triangular matrices $P$ like those used in Theorem.~\ref{thm:ancilla_cz}.
They key idea is to build up the matrix $P$ iteratively, row-by-row, by applying CNOT circuits.
Our building blocks will be two-qubit CNOT gates.
\begin{definition}
    Define $X_{i \rightarrow j}$ as the unitary associated with the two-qubit gate CNOT gate with control on site $i$ and target on $j$. The matrix $P$ specifying this circuit, such that $C_X(P) = X_{i \rightarrow j}$ is given by $P = Q^{i \rightarrow j}$ where
    \begin{align}
        Q^{i \rightarrow j} = (Q^{j\leftarrow i})_{ab}&=\begin{cases}
        1 & a=b\\
        1 & a=j,b=i\\
        0 & \mathrm{otherwise}
        \end{cases}.
    \end{align}
\end{definition}

As composition of CNOT circuits $C_X(P_0) C_X(P_1) = C_X(P_0 P_1)$ corresponds to matrix multiplication of $P_0$ and $P_1$, with arithmetic over $\mathbb{F}_2$, the effect of applying a CNOT gate from $i \rightarrow j$ to an existing circuit $C_X(P)$ is to add row $i$ to $j$.
Further, in Theorem.~\ref{thm:ancilla_cz}, we only care about the form of $P$ up to the second block of columns $R_1$ and $R_2$. Hence, the algorithm also takes as a condition, which columns of $P$ need to be matched.

\begin{algorithm}[ht]
\caption{Iterative CNOT Decomposition}
\label{alg:cx-match}
\KwIn{Target matrix $P\in\mathbb{F}_2^{N\times N}$, set of column indices \texttt{relevant\_columns}}
\KwOut{Two‐qubit CNOT circuit $C$ as a sequence of gates}
$C \leftarrow \{ \emptyset \}$                           \tcp{Empty circuit}
$P_C \leftarrow I_{N}$                       \tcp{Matrix representation of circuit}
\BlankLine

\For{$r \leftarrow 1$ \KwTo $N$}{  
    $b \leftarrow P[r,\texttt{relevant\_columns}] + P_C[r,\texttt{relevant\_columns}]$       \tcp*{Target pattern for row $r$}

    $x_r \leftarrow \arg\min_{\;x_r\in\{0,1\}^N\;:\;} \|x_r\|_1$
    subject~to $x\,P_C[:,\texttt{relevant\_columns}]=b$ \tcp*{By solving an integer program}  \label{line:ip}

  \ForEach{$i \in \mathrm{nonzero}(x)$}{  
    \If{$i \neq r$}{  
      Append $X_{i\to r}$ to $C$                       \tcp*{Update circuit: $C\!\leftarrow\!X_{i\to r}\,C$}  
      $P_C \leftarrow Q^{i \to r}\,P_C$               \tcp*{Update matrix: left‐multiply by gate}  
    }  
  }  
}
Return $C$
\end{algorithm}

It is straightforward to see that the algorithm is correct. The key step is identifying via integer programming the vector $x$, whose non-zero entries correspond to rows of $P_C$ which, when added to $P_C[r]$ ensure it matches the target matrix $P[r]$ on the relevant columns.
By iteratively matching one row at a time, this ensures the $P_C=P$ by the end of the algorithm.

This procedure allows us to move a lot of the overhead of synthesizing CZ circuits to synthesizing CNOT circuits. In general, synthesizing a generic $O(N)$-qubit CNOT circuit requires at least $O(N^2)$ two-qubit CNOT gates, so at first glance this procedure has not reduced the overhead.
However, we will find that the circuits that arise for specific fermionic permutations have remarkably efficient CNOT decompositions, in particular requiring only $O(N)$ total CNOT gates.

\subsection{$O(1)$ overhead structured permutations}

\begin{figure}
    \centering
    \includegraphics[width=0.85\linewidth]{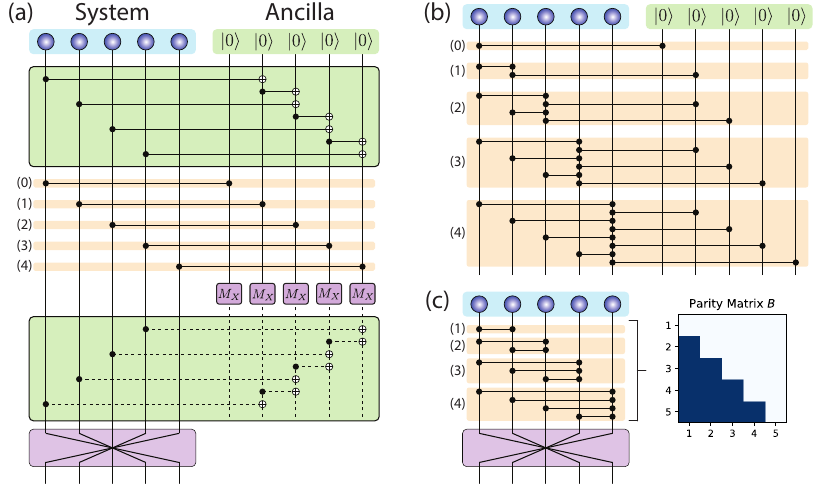}
    \caption{\textbf{1D reflection protocol}. (a) In the more general version of the protocol, we start by introducing an ancilla register of $N$ qubits, initialized in $\vert 0 \rangle$. Then, we apply CNOT circuits to compress the target FSWAP, such that the targets are always acting on ancilla qubits. Then, we apply a layer of transversal CZ's, followed by $X$-basis measurement of the ancilla. The constraint ensures the CNOT circuit can be undone by classical processing and Pauli-$Z$ feedforward. Finally, the qubits are permuted to their final position. (b) We track how the CZ gates are broadcasted under conjugation by the CNOT circuit. For the $i$-th system qubit, the CNOT circuit propagates each transversal CZ (labeled 0 to 4) onto all qubits which affect the output of the $i$-th ancilla. (c) The gates which act on an initial $\vert 0 \rangle$ states are trivial, and can be removed. What is left is an $N$-qubit all-to-all CZ circuit, which when combined with qubit permutation realizes the the fermionic permutation associated with reflection. In the language of Theorem~\ref{thm:ancilla_cz}, the lower triangular part $B$ of the CZ parity matrix $A = \frac{1}{2}(B + B^{T})$ of this operation is composed of all ones, since there are CZ's between all pairs. Hence, the entire procedure only uses $O(N)$ Clifford gates to simulate $O(N^2)$ CZ gates.}
    \label{fig:protocol_1d_reflection}
\end{figure}

\textbf{1D Reflection}. 
Consider an $N$-qubit fermionic permutation, with initial ordering $m_0(i) = i$ and final ordering $m_1(i) = N-i+1$. It is easy to see that the ordering of all pairs of modes changes, and hence the CZ circuit is specified by $A = \frac{1}{2}(\mathds{L} + \mathds{L}^T)$ where $\mathds{L}$ is a lower-triangular matrix with all ones, i.e. $\mathds{L}_{i,j} = 1$ for all $i > j$.

We can implement this CZ circuit using the decomposition from Theorem.~\ref{thm:ancilla_cz} and selecting $B = \mathds{L}$, so the problem reduces to implementing a CNOT circuit $C_X(P_{1D})$ on $2N$ qubits where
\begin{align}
    P_{1D} = \left(\begin{array}{cc}
                \mathds{I} & R_1\\
                \mathds{L} & R_2
        \end{array}\right),
\end{align}
where $R_1$ and $R_2$ are arbitrary binary matrix. Hence, we only match $P_{1D}$ on the first $N$ columns, denoted as $P_{1D}[:,0:N]$. We also note that the decomposition algorithm typically produces $R_1=0$, since the upper-right block of $P_{1D}$ is already equivalent to the trivial CNOT circuit $\mathds{I}$.

The key observation, is that when we run Algorithm~\ref{alg:cx-match}, the weight of the solution $x_r$ for each row is bounded, $||x_r||_1 \leq 2$.
To show this, we explicitly construct solutions $x_r$.

\begin{corollary}
    The CNOT circuit $C_X(P_{1D})$ acting on $2N$ qubits has an efficient decomposition into $\leq 2N$ CNOT gates. Further, one such decomposition is produced by Algorithm~\ref{alg:cx-match}.
\end{corollary}

\begin{proof}
We start with the trivial CNOT circuit, with $P_C = \mathds{I}$.
For $r=0,...,N$, we see that $P_C[r,0:N] = P[r,0:N]$, and so $x_r=0$ is a valid solution.
For $r = N+1$, $P[r,0:N] + P_C[r,0:N] = (1,0,...,0)$, which is equal to the first row of $P_C$. Hence, $x_{N+1} = (1,0,...,0)$, is a valid solution with $||x_{N+1}||_1 = 1$.
For the next row, $r=N+2$, we see that $P[r,0:N] + P_C[r,0:N] = (1,1,0,...,0) = b$. 
We see there are two equal weight solutions. We can either select rows $0$ and $1$, or rows $N+1$ and $1$. This is because row $N+1$ and row $0$ match on the first $N$ columns.
For both choices, the solution has weight $||x_{N+2}||_1 = 2$.
More generally, once we reach row $r = N+k$, we will have $P[r,0:N] + P_c[r,0:N] = (1,...,1,0,...0)$ where the string starts with $k$ ones and ends with $N-k$ zeros. This string can always be constructed by adding the previous row $N+k-1$ and row $k$. Hence, $||x_{N+k}||_1 \leq 2$ and $||x_{k}||_1 \leq 0$ for all $k$ from $1$ to $N$.
Therefore, the total number of CNOT gates used to implement $C_X(P)$ is $\leq 2N$, and the overall gate cost is $O(N)$.
\end{proof}

A visual depiction of this circuit is presented in Figure.~\ref{fig:protocol_1d_reflection}, where we explicitly show the CNOT + CZ decomposition of the fermionic reflection for $N=5$. The structure of the CNOT cascade is such that for the ancilla located at index $N+k$, there are two relevant CNOTs, one coupled to qubit $N+k-1$ and the other to qubit $k$. We further use circuit identities to show how conjugation of the transversal CZ circuit associated with $C_Z\left(\frac{I \otimes X}{2}\right)$ with $C_X(P_{\mathrm{1D}} )$ generates an all-to-all CZ circut on $N$ qubits, realizing the reflection. 
In particular the simple identity we use is the same as in the interleave example. Consider three qubits $i,j,k$. Then, \begin{align}
    X_{i \rightarrow j} Z_{jk} X_{i \rightarrow j} = Z_{ik} Z_{jk}.
\end{align}

\textbf{2D Reflection}
The next case we consider is the so-called ``2D reflection'' permutation, introduced in Figure.~\ref{fig:fig3}. 
This permutation switches from a ``row-first'' ordering of fermions to a ``column-first'' ordering.
In particular, if we orient fermionic modes in an $L_r \times L_c$ grid with row and column indexed by $(r,c)$, in the row-first ordering we have $m_0(r,c) = rL_c + c$, while in the column-first we have $m_1(r,c)=cL_r+r$, where the indexes run from $r=0,...,L_r - 1$ and $c = 0,...,L_c-1$ respectively. 
Hence, the structure of which modes cross during the operation is relatively intricate.
Nevertheless, we will show that by applying Algorithm~\ref{alg:cx-match} with a carefully selected ordering, we can generate a CNOT decomposition where the weight of the solution $||x_r||_1 \leq 4$ for each row. 

First, we understand the structure of the mode crossings, using intrinsically 2D coordinates. 
Consider two distinct modes $i=(r,c)$ and $i' = (r',c')$, and order them such that $m_0(r,c) < m_0(r',c')$.
Then, the two modes cross under 2D reflection, if and only if $m_1(r,c) > m_1(r',c')$.
Combining these two conditions, we can derive the relationship between $(r,r',c,c')$ that corresponds to the crossing condition.
\begin{align}
    rL_c + c &< r' L_c + c' \\
    cL_r + r &> c' L_r + r'
\end{align}
Rearranging these, we find the two conditions are
\begin{align}
    (r'-r)L_c + (c'-c) &> 0 \\
    (c - c') L_c + (r - r') &> 0.
\end{align}
For both conditions to be true, the variables must simultaneously satisfy $(r' > r)$ and $(c > c')$.

Just as in the case of 1D reflection, we see the condition $(r' > r)$ can be exactly encoded into a lower-triangular matrix, since
\begin{align}
    (\mathds{L})_{r',r} = \begin{cases}
        0 & r' \leq r \\
        1 & \mathrm{otherwise}
    \end{cases}.
\end{align}
Similarly, the condition $(c > c')$ can be encoded into a lower-triangular matrix, but the ordering of the rows/columns needs to be reversed. This is achieved by the transformation $o(c) = L_c - c - 1$.
\begin{align}
    (\mathds{L})_{o(c'),o(c)} = \begin{cases}
        0 & o(c') \leq o(c) \\
        1 & \mathrm{otherwise}
    \end{cases} = \begin{cases}
        0 & c \leq c' \\
        1 & \mathrm{otherwise}
    \end{cases}
\end{align}
Therefore, by choosing an ordering for the 2D coordinates as $o(i) = o(r,c) = r L_r + (L_c - c - 1)$, the combined crossing condition can be written as a tensor (kronecker) product of two lower-triangular matrices.
\begin{align}
    (B_{2D})_{o(i),o(i')} = \begin{cases}
        0 & r \leq r' \\
        0 & c' \leq c \\
        1 & \mathrm{otherwise}
    \end{cases} = (\mathds{L} \otimes \mathds{L})_{o(i), o(i')}
\end{align}

Hence, by first permuting the input modes using the map $o(i)$, we can reduce the problem to implementing the CNOT matrix,
\begin{align}
    P_{\mathrm{2D}} = \left(\begin{array}{cc}
                I & R_1\\
                B_{2D} & R_2
        \end{array}\right),
\end{align}
where $R_1$ and $R_2$ can be arbitrary. 
Next, we show this CNOT matrix can be implemented with constant overhead, using Algorithm~\ref{alg:cx-match}.

\begin{figure}
    \centering
    \includegraphics[width=1.0\linewidth]{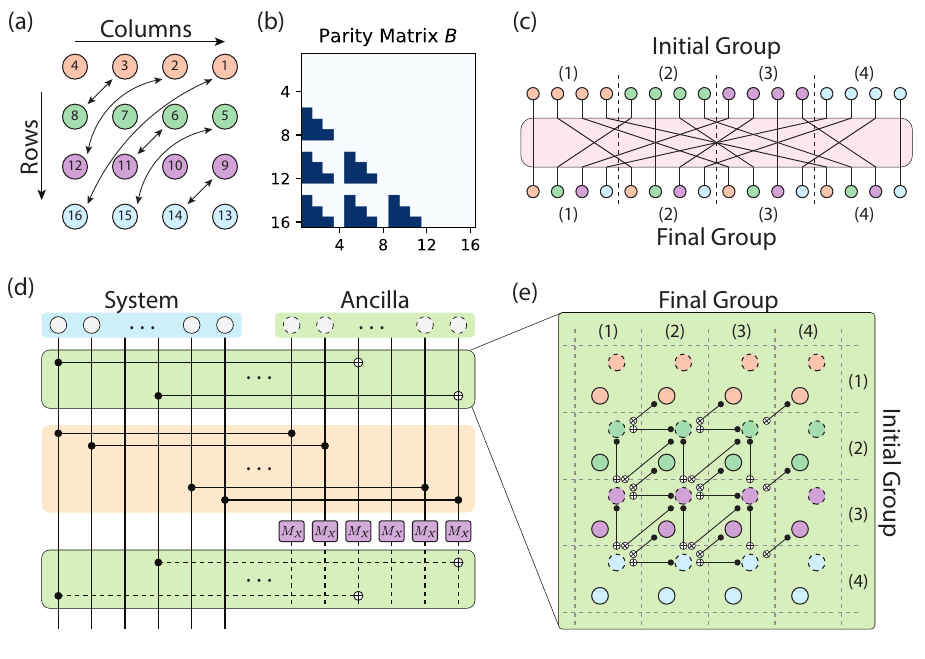}
    \caption{\textbf{2D Reflection Protocol}. (1) A 2D reflection is defined as follows. Consider fermionic modes laid out in a 2D  $L_r \times L_c$ grid, and initially encoded in a Jordan-Wigner encoding, with $m_0(r,c) = rL_c + c$ and $r=0,...,L_r-1, c=0,...,L_c-1$. The 2D reflection switches the role of the row and column, so $m_1(r,c) = cL_r + r$. In the figure depicted, $L_r = L_c = 4$. We further select an ordering $o(r,c) = rL_c + (L_c - c)$ for Algorithm~\ref{alg:cx-match}. Modes are labeled by $o(r,c)$. (b) With this ordering, the lower-triangular part $B$ of the CZ matrix $A$ exhibits an interesting structure. In particular, we see that $B = \mathds{L} \otimes \mathds{L}$ is the tensor product of two dense lower-triangular matrices. (c) Ultimately, we simulate a fermionic permutation circuit, as drawn here. Modes can be uniquely specified by their initial group (i.e. their row), and their final group (i.e. their column).
    (d) As before, we leverage the general structure where we introduce $N$ ancilla modes, run a CNOT circuit, applying transversal CZs, and then invert the CNOT circuit. Ancillas are denoted with dashed lines, while the original system qubits have solid lines. (e) The special structure of the parity matrix $B$ makes it possible to implement the analogous CNOT circuit using $\leq 4N$ CNOT gates. To uniquely specify the causal order, assume that all CNOT gates targeting a qubit are applied \textit{before} any CNOT gates controlled on that same qubit. We see the causal structure flows from the top-right to the bottom-left. Using the fact the ancilla's are initialized in $\vert 0 \rangle$, the circuit can also be simplified near the boundaries. In the end, the precise number of CNOT gates is given by $(2L_r-3)(2L_c-3)$. Note when $L_r=2$, this reduces to a structured interleave.}
    \label{fig:2d_reflection}
\end{figure}

\begin{corollary}
    The CNOT circuit $C_X(P_{\mathrm{2D}})$ acting on $2N$ qubits has an efficient decomposition into $\leq 4N$ CNOT gates, which can be produced by Algorithm~\ref{alg:cx-match}. 
\end{corollary}

\begin{proof}
    Here, we describe a simple decomposition for the target parity matrix $P = P_{2D}$ using the iterative Algorithm~\ref{alg:cx-match}.
    As before, we start with the trivial CNOT circuit $P_C = \mathds{I}$.
    For the first $N+1$ rows $a=0,...,N$ we see that $P_C[a,0:N] = P[a,0:N]$ already match. Hence, we simply select $x_a=0$ for these rows.

    Next, we inductively prove that we can find a solution $||x_{N+a}||_1 \leq 4$ for all $a=0,...,N-1$. 
    Further, we will always satisfy the constraint that $x_{N+a}[N+a:]=0$, so at each step we only use prior rows.
    To further clarify notation, we label the circuit $P_C^{(t)}$ as the value of $P_C$ when we start step $t$. Hence, $P_C^{(0)} = \mathds{I}$ and $P_C^{(2N)}[:,0:N] = P[:,0:N]$.

    A helpful identify is
    \begin{align}
        P[N+a,0:N] = B_{2D}[a],
    \end{align}
    and when we reach step $N+a$, the existing solution satisfies
    \begin{align}
        P_C^{(N+a)}[N+a,0:N] = 0.
    \end{align}
    So,when we reach step $N+a$, the target vector is
    \begin{align}
        P[N+a,0:N] + P_C^{(N+a)}[N+a,0:N] = P[N+a,0:N] = B_{2D}[a].
    \end{align}
    Further, $P_C^{(N+a)}[0:N,0:N]=\mathds{I}$ for all steps. Thus, at each step, we can always add an arbitrary unit vector to our solution.

    The inductive argument proceeds as follows.
    We can work directly with the 2D structure by writing the index as $a = xL_r + y$.
    We see that the $B_{2D}[a]$ can be written as the sum of up to four available rows
    \begin{align}{\label{eq:2d_decomp}}
        B_{2D}[(x,y)] &= 
        \begin{cases}
            B_{2D}[(x-1,y)] 
            + B_{2D}[(x,y-1)]
            + B_{2D}[(x-1,y-1)] + \mathds{I}[(x-1,y-1)] & \text{if } (x>0) \text{ and } (y>0) \\
            0 & \text{otherwise}
        \end{cases}
    \end{align}
    where $B_{2D}[(x,y)] \equiv B_{2D}[x L_r + y], \mathds{I}[(x,y)] \equiv \mathds{I}[x L_r + y]$.
    Therefore, assuming at step $a$ that $P_C^{(a)}[a'] = P[a']$ for all $a' < a$, then each term $B_{2D}[(x-1,y)], B_{2D}[(x,y-1)], B_{2D}[(x-1,y-1)], \mathds{I}[(x-1,y-1)]$ is available as a row of $P_C^{(a)}$, and Eq.~\ref{eq:2d_decomp} provides a solution with weight $||x_{a}||_1 \leq 4$. 

    The base case with $x=0$ or $y=0$ is trivial. Hence, the inductive construction provides a solution for all $x>0$ and $y>0$.
    Therefore, the total number of CNOT gates is $\leq 4N$.

    Lastly, we prove that this decomposition is correct.
    To do so, it is helpful to use the following identity for $B_{2D}$ and lower-triangular matrices,
    \begin{align}
        (B_{2D})_{a,a'} & = (\mathds{L})_{x,x'} (\mathds{L})_{y,y'}\\
        \mathds{I}_{x,x'} &= \delta_{x,x'} \\
        \mathds{L}_{x,x'} &= \mathds{L}_{x-1,x'} + \delta_{x-1,x'},
    \end{align}
    to simplify and regroup terms. 
    \begin{align}
        (B_{2D})_{(x,y),a'} &= (B_{2D})_{(x-1,y),a'} + (B_{2D})_{(x,y-1),a'} + (B_{2D})_{(x-1,y-1),a'} + (\mathds{I})_{(x-1,y-1),a'} \nonumber\\
        &= \mathds{L}_{x-1,x'} \mathds{L}_{y,y'} + \mathds{L}_{x,x'} \mathds{L}_{y-1,y'} + \mathds{L}_{x-1,x'} \mathds{L}_{y-1,y'} + \mathds{\delta}_{x-1,x'} \mathds{\delta}_{y-1,y'} \nonumber\\
        &= (\mathds{L}_{x-1,x'} \mathds{L}_{y-1,y'} + \mathds{L}_{x-1,x'} \delta_{y-1,y'}) + (\mathds{L}_{x-1,x'} \mathds{L}_{y-1,y'} + \delta_{x-1,x'} \mathds{L}_{y-1,y'}) + \mathds{L}_{x-1,x'} \mathds{L}_{y-1,y'} + \delta_{x-1,x'} \delta_{y-1,y'} \nonumber\\
        &= \mathds{L}_{x-1,x'} \delta_{y-1,y'} + \delta_{x-1,x'} \mathds{L}_{y-1,y'} + \mathds{L}_{x-1,x'} \mathds{L}_{y-1,y'} + \delta_{x-1,x'} \delta_{y-1,y'} \nonumber\\
        &= (\mathds{L}_{x-1,x'} + \delta_{x-1,x'})(\mathds{L}_{y-1,y'} + \delta_{y-1,y'}) \nonumber\\
        &= \mathds{L}_{x,x'} \mathds{L}_{y,y'}
    \end{align}
    Note that between the third and fourth line, we also use the fact that addition is modulo 2.

    The detailed circuit associated with this decomposition is depicted in Figure.~\ref{fig:2d_reflection}.
    There, we further use the fact that when $x=0$ or $y=0$, $(B_{2D})[(x,y)]=0$ to simplify the circuit, and careful counting shows the number of CNOT gates is $(2L_r - 3)(2 L_c - 3)$.
\end{proof}

\textbf{Structured Interleave}. It turns out, that the interleave can be understood as a special case of the 2D reflection. 
In particular, consider a structured interleave acting on $2N$ modes, where $A$ constitutes the first half of the modes, and $B$ constitutes the second half of the modes.
Let $m_0(i) = i$ be the original ordering.
Arranged as a 2D grid, let $L_r = 2$ and $L_c=N$.
Then, we can equivalently write $m_0(i) = m_0(r,c) = rN+c$.
Thus, $A$ corresponds to the row where $r=0$ and $B$ the row where $r=1$.
The structured interleave permutation is given by $m_1(i) = m_1(r,c) = 2c + r$, so each mode in $A$ is mapped to an even index, and each mode in $B$ to an odd index.
This is exactly the interleave depicted in Fig.~\ref{fig:fig3}a.
It is easy to see from this construction, that this is exactly equivalent to a 2D reflection.
Hence, the structure interleave is a special case of the 2D reflection.

Below, we will discuss a large class of deformations which can be efficiently performed. We will see that the generic interleave presented in Figure.~\ref{fig:fig2} is a deformation of the structured interleave. Hence, this constitutes an alternate route to performing interleave-type fermionic permutations.

\subsection{Deformations of structured permutations}

The examples above are specific operations that can be performed with $O(1)$ overhead. 
In practice, to compile generic permutations, and more complex structures, it is helpful to consider permutations which are simple deformations of these structures. 
To enable this, we define two kinds of deformations: deletions and duplications.
A deletion is a simple modification of a permutation, which simply removes one of the modes. A duplication is another simple modification, that involves \textit{copying} or doubling a fermionic mode. 
Formally, both are defined as follows.

\begin{definition} {\bf (Deletions)}
    Consider a fermionic permutation $\mathcal{F}_p$ of $N$ modes which maps $i \mapsto p(i)$. A permutation $\mathcal{F}_{p'}$ of $N-1$ modes is related to $\mathcal{F}_p$ by deletion of mode $i_0$, if $p'(i) = p(i)$ for $i < i_0$ and $p'(i) = p(i+1)$ for $i > i_0$. Let $C_Z(A)$ be the CZ circuit that implements $\mathcal{F}_p$. Let $C_Z(A')$ be the CZ circuit that implements $\mathcal{F}_{p'}$.
    Let $p'$ be related to $p$ by deletion of mode $i_0$. Then, $A'$ is related to $A$ be deletion of row $i_0$ and column $i_0$.
\end{definition}

\begin{definition} {\bf (Duplications)}
    Consider a fermionic permutation $\mathcal{F}_p$ of $N$ modes which maps $i \mapsto p(i)$. A permutation $\mathcal{F}_{p'}$ of $N+1$ modes is related to $\mathcal{F}_p$ by \textit{duplication} of mode $i_0$ if $p'(i) = p(i)$ for $i < i_0$, $p'(i_0) = p(i_0)$, and $p'(i) = p(i-1)+1$ for $i > i_0$. Let $C_Z(A)$ be the CZ circuit that implements $\mathcal{F}_p$ and $C_Z(A')$ be the CZ circuit that implements $\mathcal{F}_{p'}$.
    Then, $A'$ is related to $A$ by duplicating row $i_0$ and column $i_0$.
\end{definition}

\begin{figure}
    \centering
    \includegraphics[width=1.0\linewidth]{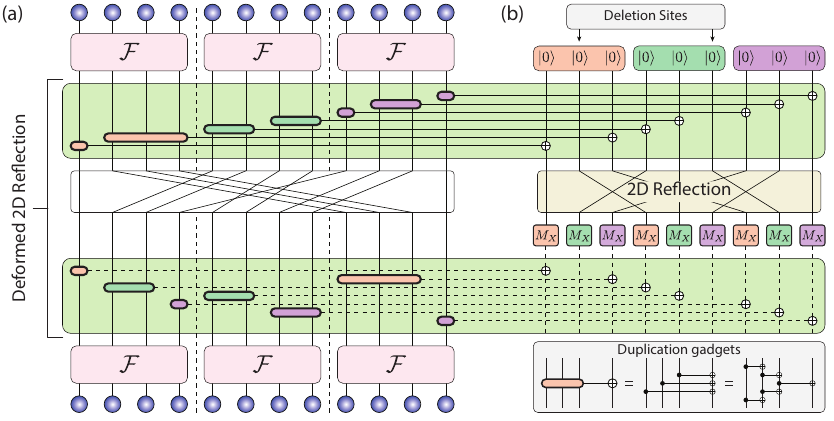}
    \caption{\textbf{Protocol for deformed permutations}. Here, we illustrate the deformation gadgets (deletion and duplication) by showing how to implement a deformation of a 2D reflection. The key idea is to introduce an ancilla register on which we will apply the standard (i.e. undeformed) 2D reflection. Certain modes in the standard 2D reflection are not present in the deformed ones. These are the deletion sites. Handling deletions is simple, because these modes are simply uncoupled to the system qubits.
    In contrast, certain modes in the standard reflection are present multiple times in the deformed one. These are the duplicated modes. Each duplicated mode is coupled via a single CNOT, represented here by introducing a duplication gadget. The modes which are neither deleted or duplicated are simply coupled with a single CNOT. After performing the 2D reflection, the CNOT circuit can be inverted and the ancilla discarded, using measurement and feedforward.}
    \label{fig:deformations}
\end{figure}

Note that the definition of a duplication implies that $p'(i_0+1)=p(i_0)+1$, so in essence the new mode has been inserted to the right of $i_0$ before permutation, and remains adjacent to $i_0$ after the permutation.
It is also easy to see that these operations commute, so it doesn't matter which order the deformations and duplications are performed, the final circuit $C_Z(A)$ will be identical.
With these definitions, we present a general-purpose construction for implementing deformations of simple CZ circuits with constant overhead (see Figure.~\ref{fig:deformations}).

\begin{theorem}
Consider a fermionic permutation of $N$ modes $\mathcal{F}_p$, which is related to another permutation of $N'$ modes $\mathcal{F}_{p'}$ by $n_E$ deletions and $n_P$ duplications. Suppose that we have access to a CNOT + CZ circuit that implements $\mathcal{F}_p$. 
Then, there exists a CNOT + CZ circuit implementing $\mathcal{F}_{p'}$ with an additional $O(N')$ qubits, $\leq N'$ CZ gates, and $O(N')$ CNOT gates. Further, the additional circuit elements can be arranged in $O(1)$-depth.
\end{theorem}

\begin{proof}
    Let $A$ be the $N \times N$  matrix such that $C_Z(A)$ generates the fermionic permutation $\mathcal{F}_p$.
    Our goal is to construct a $C_X$ circuit $C_X(P)$ which, when conjugating $C_Z(A)$, generates the target fermionic permutation $\mathcal{F}_{p'}$.

    As $P$ is an $(N + N') \times (N + N')$ matrix, we can decompose it in block-diagonal form as
    \begin{align}
        P&=\left(\begin{array}{cc}
            I & 0\\
            B & R
        \end{array}\right)
    \end{align}
    where $B$ is an $N \times N'$ sub-matrix.
    As we apply $C_Z(A)$ to the lower-block, the overall parity function takes the form
    \begin{align}
        f_{B,A'}(x)&=\left(\begin{array}{cc}
            x^{T} & 0\end{array}\right)\left(\begin{array}{cc}
            I & B^{T}\\
            0 & R^{T}
            \end{array}\right)\left(\begin{array}{cc}
            0 & 0\\
            0 & A
            \end{array}\right)\left(\begin{array}{cc}
            I & 0\\
            B & R
            \end{array}\right)\left(\begin{array}{c}
            x\\
            0
            \end{array}\right)\\&=x^{T}B^{T}A'Bx.
    \end{align}
    Therefore, our goal is to construct $B$ such that $A = B^T A' B$ realizes the $N \times N$ parity matrix of $C_Z(A)$ that implements the target fermionic permutation $\mathcal{F}_p$.

    By construction, we chose $\mathcal{F}_{p'}$ to be a deformation of $\mathcal{F}_p$, related by some number of deletions and duplications.
    Therefore, we can map each mode $i=1,...,N'$ of $\mathcal{F}_{p'}$ to a mode $j=1,...,N$ of $\mathcal{F}_p$.
    Let $\sigma(i)$ be the map from $N'$ modes to $N$ modes.
    Then, we can set matrix elements of $B$ such that $B_{\sigma(i),i}=1$ for all $i$, and all other matrix elements are zero.

    By inspection, we see that $B$ constructed in this manner generates the appropriate duplication and deletions of $A$ under conjugation.
    Start by considering right multiplication, $A \rightarrow A B$.
    The columns of the output matrix are given by
    \begin{align}
        (AB)_{\cdot, j} = (A)_{\cdot, \sigma(j)}.
    \end{align}
    Therefore, any column of $A$ that is not in the image of $\sigma$ is effectively deleted, and each column $\sigma(i)$ is duplicated to all columns $i'$ satisfying $\sigma(i') = \sigma(i)$.

    Left multiplication performs the analagous operation.
    Consider,
    \begin{align}
        (B^T A B)_{i, j} = (A)_{\sigma(i), \sigma(j)}.
    \end{align}
    Therefore, the same duplication and deletion rule is applied to the rows and columns under conjugation.
    This proves that $B^T A B = A'$ by construction.

    What is left to show, is that $P[:,0:N']$ can be efficiently implemented using $N'$ CNOT gates.
    This follows directly from Algorithm~\ref{alg:cx-match}.
    As usual, the first $N'$ rows of $P$ match the trivial $C_X$ circuit, and hence these steps produce trivial solution vectors.
    Then, consider the next $N$ rows, corresponding go the $B$ sub-matrix.
    For every row $B[r]$ of $B$, the non-zero entries are located at columns $i$ satisfying $\sigma(i) = r$.
    Hence, the weight of the solution $||x_r||_1 \leq | \{ i | \sigma(i)=r\} |$ is at most the number of times $r$ was duplicated.
    Since each column appears once in $B$, this further guarantees that $\sum_r ||x_r||_1 \leq \sum_r | \{ i | \sigma(i)=r\} | = N'$, and so the total number of CNOT gates is $\leq N'$.

    Finally, we show explicitly that the circuit $C_X(P)$ can be implemented in constant depth, using $O(N')$ ancilla modes and $O(N')$ CNOT gates. 
    This is non-trivial, because the row-weight of $||x_r||_1$ for each individual row may in principle scale as $O(N')$.
    However, using the construction depicted in Figure.~\ref{fig:deformations} under "duplication gadgets", we see how a sequence of CNOT gates with the same target but different control can be parallelized, by transforming it into two CNOT cascade.
    As show in Ref.~\cite{Baumer_longrange_entangling_2025}, each CNOT cascade of size $||x_r||_1$ can be implemented with constant-depth and $O(||x_r||_1)$ ancilla qubits.
    Thus, every row can be compiled into constant depth, with a total number of ancilla qubits scaling as $O(\sum_r ||x_r||_1) = O(N')$.
\end{proof}

In practice, we note that the construction described here may not be an optimal utilization of resources, and further reductions can certainly be acheived depending on the target application and underlying hardware. Nevertheless, it illustrates, that asymptotically, there is a large degree of flexibility in constructing fermionic circuits with asymptotic $O(1)$-encoding overhead.
Next, we illustrate two applications of the deformation construction.

\hspace{0.1in}

{\bf (General Interleave from Structured Interleave).}
The general interleave we presented in Figure.~\ref{fig:fig2} is a fermionic permutation $\mathcal{F}_p$, where the input modes can be partitioned into two contiguous groups $A$ and $B$, such that all crossings occur between groups (i.e. the relative order within each group is preserved).
Here we show an alternate compilation of a general interleave with $O(1)$ overhead, by reducing it to the structured interleave using the deletion and duplication operations discussed above.
The construction is simple. Consider the output function $p(i)$. 
Any time two consecutive elements from the same group map to adjacent outputs (i.e. $p(i)+1 = p(i+1)$), they can be collapsed to a single mode. This corresponds to an inverse duplication.
After doing this for all modes, we are guaranteed that the resulting output modes are alternating between groups $A$ and $B$.
Then, we can add an additional modes to ensure $|A|=|B|$ if needed. This corresponds to an inverse deletion.

\hspace{0.1in}

{\bf (Higher-Dimensional Transpositions).} A powerful application of the deformation technique, is that it enables us to intuitively construct efficient transpositions of higher-dimensional lattices.
Consider for example, a set of $N = L^D$ modes arranged in an $L \times L \times ... \times L$ D-dimensional lattice, such that the initial JW encoding is given by
$$m_0(i) = m_0(x^{D-1},..., x^1, x^0) = x^{D-1} L^{D-1} + ... + x^1 L + x^0.$$

Then, assuming access to a 2D reflection as the building block, we show we can efficiently construct fermionic permutations that swap ``adjacent'' dimensions, with $O(N)$ additional CNOT gates, arranged in depth $O(1)$.
Specifically we can swap any pair of indexes $x^{d}, x^{d-1}$ where $d=1,...,D-1$. So for example, if we swap the middle two indexes, the output ordering would be
$$m_1(i) = m_1(x^{D-1},...,x^{d}, x^{d-1},...,x^0) = m_0(x^{D-1},...,x^{d}, x^{d-1},...,x^0)$$.

First, observe that modes with differing \textit{outer} indices (i.e. $x^{D-1},...,x^{d+1}$) do not cross during the transformation.
Hence, the transformation is equivalent to $L^{D-d-1}$ independent fermionic permutations of $L^{d+1}$ modes satisfying
\begin{align}
    m_1(x^{d}, x^{d-1},..., x^0) = m_0(x^{d-1}, x^{d},..., x^0).
\end{align}

Next, observe that the $L$ modes which only differ in the \textit{inner} indices (i.e. $x^{d-1},...,x^0$) remain adjacent before and after the permutation with the same relative ordering.
Hence, we can further reduce each permutation of $L^{d+1}$ modes to $L^2$ modes.
In particular, we can duplicate each $(x^{d}, x^{d-1})$ mode exactly $L^{d-2}$ times, and define $(x^{d-1},...,x^0)$ as the index of the duplicated mode.
This completes the reduction from a $D$-dimensional transposition to a 2D permutation.
Both steps only require $O(N)$ operations and can be arranged with $O(1)$-overhead, ensuring the entire procedure remains efficient.
The case where indices have differing sizes $L_D, ..., L_1$ is also a simple generalization of this argument. In particular, the reduction proceeds identically. First observe that outer-indices do not cross during the permutation. Second, observe that inner-indices are generated by a simple duplication.
The set of pairwise swaps is sufficient to generate all permutations of $D$ indices, and hence any $D$-dimensional index permutation can be generated sequentially using this technique.

\section{Proof of Construction for Majorana Permutations}

Next, we show that if we can efficiently realize the fermionic permutation $\mathcal{F}_p$, then we can use this to efficiently generate the analogous majorana permutation $\mathcal{M}_p$ with minimal overhead, using the construction shown in Figure.~\ref{fig:majorana_permutation}. 

\begin{theorem}
    Consider a Majorana permutation $\mathcal{M}_p$ of $2N$ majoranas, and the analogous fermionic permutation $\mathcal{F}_p$ of $2N$ fermions. Given a circuit for implementing $\mathcal{F}_p$ with $g_F$ Clifford gates and $N_F$ ancilla qubits, there exists a circuit implementing $\mathcal{M}_p$ with $N_F + N$ ancillas and $g_F + 2N$ Clifford gates. 
\end{theorem}

\begin{proof}
    We prove this by providing a constructive algorithm for implementing $\mathcal{M}_p$, using $\mathcal{F}_p$ as a subroutine.
    A visual depiction of the algorithm is shown in Figure.~\ref{fig:majorana_permutation}. 
    Recall we start with $N$ fermion modes, described by $2N$ majorana operators, associated with even indexes $\chi_{2i}$ and odd indexes $\chi_{2i+1}$ for $i=1,...,N$.

    The first step of the algorithm involves introducing one ancilla per fermion mode, and prepare the ancilla in $\vert 0 \rangle$.
    Since the ancillas are in zero, they can be trivially inserted while preserving the Jordan-Wigner encoding. This can be seen by considering the FSWAP circuit required to interleave the original $N$ modes with a register of $\vert 0 \rangle^{N}$ ancilla states. The zero states ensure all $CZ$'s act trivially, so a simple permutation is sufficient to insert them.
    After this step, the index of the original majorana operators are modified to $\chi_{2i} \rightarrow \chi_{4i}$ and $\chi_{2i+1} \rightarrow \chi_{4i+1}$, where now the majorana indexes runs from $0,...,4N-1$.

    In the second step, we evolve the system under pairwise two-qubit gates, and locally swap the majorana modes. In particular, the evolution operator is
    \begin{align}
        U_{lms} &= \prod_{i=0}^{N-1} \exp\left( -i \frac{\pi}{2} X_{2i} X_{2i+1} \right) 
        = \prod_{i=0}^{N-1} \exp\left( -i \frac{\pi}{2} \chi_{4i+1} \chi_{4i+2} \right)
    \end{align}
    This operator is designed to separate the original majorana modes, so they are supported on different qubits. 
    In particular, conjugation by $U_{lms}$ preserves $\chi_{4i} \rightarrow \chi_{4i}$ and transforms $\chi_{4i+1} \rightarrow \chi_{4i+2}$. 
    It turns out, this operator is a Clifford circuit. Hence, the second step requires applying $N$ Clifford gates.

    In the third step, now that the $2N$ majorana operators are supported on $2N$ distinct qubits, a fermionic permutation $\mathcal{F}_p$ can be applied. The cost of this step is set by the cost of $\mathcal{F}_p$.
    This takes $\chi_{4i} = \chi_{2(2i)} \rightarrow \chi_{2p(2i)}$ and $\chi_{4i+2} = \chi_{2(2i+1)} \rightarrow \chi_{2p(2i+1)}$, where recall $p$ is a permutation of the $2N$ indexes.

    In the fourth step, we apply $U_{lms}^{\dagger}$ to recombine the original $2N$ majorana operators into $N$ qubits.
    Note that the fermionic permutation can change whether a majorana operator is supported on a mode with odd or even index, i.e. we do not know a prior if $p(2i)$ and $p(2i+1)$ are odd or even.
    However, in general if $p(i)$ is even, it remains fixed, but if $p(i)$ is odd, it moves one majorana site to the left. 
    We can use the remainder ${p(2i) \bmod 2}$ to determine if it's odd or even.
    Then, the transformation is written as $\chi_{2p(i)} \rightarrow \chi_{2p(i) - (p(i)\bmod 2)}$.

    The fifth and final step of the procedure is to disentangle the ancillas from the system.
    This is now a non-trivial task, because during the permutation, the ancilla transforms from a trivial state $\vert 0 \rangle^{N}$ to a highly-entangled state. 
    Hence, to separate out the original fermions from the ancillas, we need to carefully track the fermionic statistics. One option would be to use an inverse interleave operation, to move all ancillas to the right of all system qubits. However, since we do not care to preserve the ancilla state, a cheaper option is to measure each ancilla in the $Z$-basis, and apply the inverse interleave with classical computation and feedforward Pauli operators. 
    Indeed, the final step is to simulate the CZ circuit associated with the inverse interleave, by applying a conditional-$Z$ for each ancilla measured in $\vert 1 \rangle$, on all output modes to the right of the ancilla. In the depicted circuit, we depict it concisely by drawing a CNOT cascade acting purely on the measured ancillas, followed by pairwise CZs, as in Figure.~\ref{fig:fig2}.
    After removing the ancillas, the index of the original modes change as 
    $\chi_{2p(i)} \rightarrow \chi_{2p(i) - (p(i)\bmod 2)} \rightarrow \chi_{p(i)}$.

    To summarize, during the five steps, the majorana operators transform as,
    \begin{align}
        &\chi_{2i} \rightarrow \chi_{4i} \rightarrow \chi_{4i} \rightarrow \chi_{2p(2i)} \rightarrow \chi_{2p(2i) - (p(2i)\bmod 2)} \rightarrow \chi_{p(2i)} \\
        &\chi_{2i+1} \rightarrow \chi_{4i+1} \rightarrow \chi_{4i+2} \rightarrow \chi_{2p(2i+1)} \rightarrow \chi_{2p(2i+1) - (p(2i+1)\bmod 2)} \rightarrow \chi_{p(2i+1)},
    \end{align}
    showing the operation realizes the target majorana permutation.
\end{proof}

\section{Efficient preparation of translationally-invariant free-fermion (TIFF) states}

TIFF states on a $D$-dimensional periodic lattice with $L^D$ sites can be prepared exactly using $O(\log L)$-depth circuits, by first preparing them in momentum space, and then transforming them to real space using a FFFT. Before presenting the general algorithm, we discuss some relevant examples in detail. For that, we follow the constructions presented in Refs.~\cite{verstraete_quantum_2009, jiang_quantum_2018}, noting that the speedups obtained here compared to previous approaches are based on the $O(\log L)$-depth implementation of the 2D FFFT presented in the main text, together with additional $O(1)$-depth fermionic permutations required to pair modes in momentum space.

\begin{figure}
    \centering
    \includegraphics[width=0.9\linewidth]{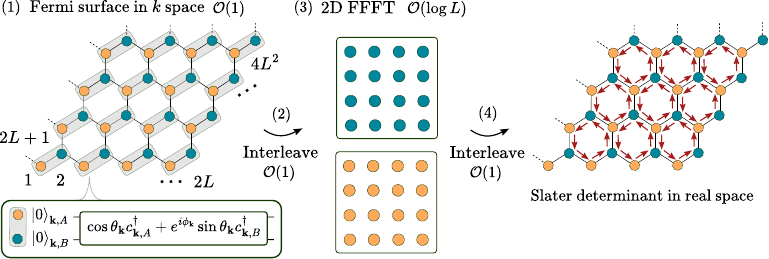}
    \caption{\textbf{State preparation circuits for translationally-invariant Slater determinants}. Translational-invariant Slater determinants, corresponding to ground states of particle-number conserving non-interacting Hamiltonians, can be prepared at any filling using a $O(\log L)$-depth circuit. In the figure, we exemplify the protocol for the case of a two-mode unit cell, denoted by A (yellow circles) and B (blue circles). (1) The state is first prepared in momentum space, by applying a $O(1)$-depth circuit (gray rectangles) to the different unit cell, which can be done in parallel. The circuit depth will depend on the size of the unit cell, but it is independent on $L$ for a TI state. To create a Fermi surface with certain filling, this operation is applied only to the corresponding filled modes, leaving the rest empty. In the figure, a half-filled state can be obtained by applying these circuits everywhere, each of them creating one fermionic mode. (2) The initial fermionic order is modified with an interleave operation, separating the modes within the unit cell. After a 2D FFFT is applied to each set of modes, the original order is restored with another interleave operation, resulting in the target Slater determinant in real space.}
    \label{fig:SI_chern_insulator}
\end{figure}

\subsection{Particle-number conserving TIFF states}

Consider first the case of a particle-number conserving TIFF state, i.e., a Slater determinant, with a unit cell composed of a finite number of modes that repeat periodically in space. Such a state corresponds, e.g., to the ground state of a non-interacting Hamiltonian with the same periodic structure, which can be partially diagonalized in momentum space by applying a FFFT in each spatial direction. The Hamiltonian can be then fully diagonalize by acting independently on each finite-size unit cell. As an example, take the Haldane model on a hexagonal lattice~\cite{Haldane_1988},
\begin{equation}
H = -t \sum_{\langle i,j \rangle} c_i^\dagger c_j + i t_2 \sum_{\langle\langle i,j \rangle\rangle} \nu_{ij} c_i^\dagger c_j + M \sum_i \eta_i c_i^\dagger c_i \,,
\end{equation}
where $t$ is the nearest-neighbor (NN) tunneling amplitude, $t_2$ is the next-nearest-neighbor (NNN) tunneling with phase factors $\nu_{ij} = \pm 1$, and $M$ is the sublattice potential, with $\eta = +1$ ($-1$) for modes associated to the A (B) sublattice. By applying a Fourier transform to the modes associated to each sublattice independently, we can write the Hamiltonian in the momentum space corresponding to a square lattice, with an internal spin-$1/2$ degree of freedom associated to the sublattice of the original model. Specifically,
\begin{equation}
\begin{aligned}
&H = \sum_\mathbf{k} \Psi_\mathbf{k}^\dagger H(\mathbf{k}) \Psi_\mathbf{k}, & H(\mathbf{k}) = \begin{pmatrix}
d_z(\mathbf{k}) & d_x(\mathbf{k}) - i d_y(\mathbf{k}) \\
d_x(\mathbf{k}) + i d_y(\mathbf{k}) & -d_z(\mathbf{k})
\end{pmatrix},
\end{aligned}
\end{equation}
where $\Psi_\mathbf{k}$ is the corresponding Nambu spinor,
\begin{equation}
\Psi_\mathbf{k} = \begin{pmatrix} c_{\mathbf{k}, A} \\ c_{\mathbf{k}, B} \end{pmatrix},
\end{equation}
and
\begin{align}
d_x(\mathbf{k}) &= -t \sum_{i=1}^{3} \cos(\mathbf{k} \cdot \mathbf{a}_i), \\
d_y(\mathbf{k}) &= -t \sum_{i=1}^{3} \sin(\mathbf{k} \cdot \mathbf{a}_i), \\
d_z(\mathbf{k}) &= M - 2 t_2 \sum_{i=1}^{3} \sin(\mathbf{k} \cdot \mathbf{b}_i),
\end{align}
with $\mathbf{a}_i$ and $\mathbf{b}_i$ the NN and NNN lattice vectors of the hexagonal lattice, respectively.

Note that the Hamiltonian can be now brought into a full diagonal form by diagonalizing each $H({\bf k})$, which can be done independently with a product of $\mathbf{k}$-dependent unitaries $\mathcal{U}_{\mathbf{k}}$, each of them acting on two neighboring modes. The ground state of the full Hamiltonian can be then written in momentum space as
\begin{equation}
\ket{\psi_0} = \bigotimes_{\mathbf{k}}\left(\mathcal{U}_{\mathbf{k}} \ket{0}_{\mathbf{k}, A}\ket{0}_{\mathbf{k}, B}\right)\equiv\prod_{\mathbf{k}} \left( \cos \theta_\mathbf{k} c_{\mathbf{k}, A}^\dagger + e^{i \phi_\mathbf{k}} \sin \theta_\mathbf{k} c_{\mathbf{k}, B}^\dagger \right) \ket{\Omega},
\end{equation}
where $\theta_\mathbf{k}$ and $\phi_\mathbf{k}$ are determined by the Hamiltonian parameters, and $\ket{\Omega}$ is the fermionic vacuum. The ground state in real space is then obtained after applying the Fourier transform to $\ket{\psi_0}$.

The protocol to prepare the ground state of the Haldane model in a quantum computer follows this procedure, as described in detail in Fig.~\ref{fig:SI_chern_insulator}. We start with a hexagonal array of qubits in the $\ket{0}$ state, and a JW encoding for the fermionic modes with the usual order. The ground state  in momentum space $\ket{\psi_0}$ is first prepared by acting in parallel with the two-qubit unitaries $\mathcal{U}_{\mathbf{k}}$ on each unit cell. Note that the latter vary locally with $\mathbf{k}$ (corresponding to the location of the modes in the qubit array), and that they can be decomposed into a finite-depth circuit. Depending on the target filling, these operations are applied to the modes that are filled, and nothing is done to those that should remain empty. To obtain the half-filled state for the Haldane model, which can correspond to a chiral topological insulator, we need to fill every unit cell with one fermion, so this operation should be applied everywhere. To obtain the real-space ground state, we implement a 2D FFFT independently to each sublattice, as described above. As shown in the figure, to do this we first need to reorder the modes into two groups, which can be done in constant depth with an interleave operation, and then merge them again after the FFFT. This protocol can be easily generalized to any other unit cell. The circuit required to implement $\mathcal{U}_{\mathbf{k}}$ can be generated by decomposing it into Givens rotations~\cite{jiang_quantum_2018}, with depth growing linearly with the size of the unit cell. This is however independent of the system size for a TI system. The interleave operations should then split and merge the modes associated to each sublattice, which again can be done in constant depth for any unit cell.

\begin{figure}
    \centering
    \includegraphics[width=0.9\linewidth]{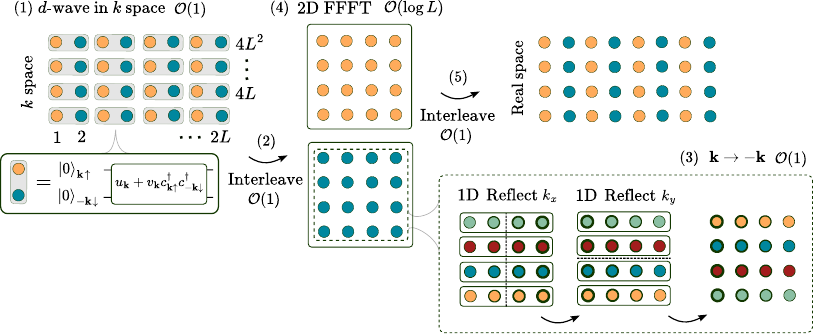}
    \caption{\textbf{State preparation circuit for translationally-invariant d-wave superconductors}. (1) The state is first prepared in momentum space by applying a two-mode circuit corresponding to a Bogoliubov transformation for every pair of modes $\mathbf{k}\uparrow$ (yellow) and $\mathbf{-k}\downarrow$ (blue). (2) These two sets of modes are separated with an interleave operation, and the second is reordered according to (3) $\mathbf{k}\rightarrow\mathbf{-k}$, corresponding to a sequence of $O(1)$-depth reflections. Note that $k_x\rightarrow-k_x$ can be implemented with $L$ independent 1D reflections, and $k_y\rightarrow-k_y$ can be implemented using one 1D reflection duplicated $L$ times. (4) After a 2D FFFT is applied to each set of modes, (5) these are merged again using an interleave operation, resulting in a $d$-wave superconductor in real space.}
    \label{fig:SI_d_wave}
\end{figure}

\subsection{Superconducting TIFF states}

With a slight modification, our protocol can be also used to prepare TI superconducting states in $O(\log L)$ depth. These states involve pairing of modes in momentum space, related by reflection $\textbf{k},\textbf{-k}$.
To see this, let as first consider the mean-field Hamiltonian of a $d$-wave superconductor,
\begin{equation}
H = -t \sum_{\langle i,j \rangle, \sigma} c_{i,\sigma}^\dagger c_{j,\sigma} - \sum_{\langle i,j \rangle} \Delta_{ij}\left(c_{i,\uparrow}^\dagger c_{j,\downarrow}^\dagger - c_{i,\downarrow}^\dagger c_{j,\uparrow}^\dagger  + \text{h.c.} \right),
\end{equation}
where $\Delta_{ij} = \Delta/2$ ($-\Delta/2$) is the pairing coupling in the horizontal (vertical) direction. Transforming again to momentum space we obtain the following Hamiltonian,
\begin{equation}
\begin{aligned}
&H = \sum_\mathbf{k} H({\bf k}),  &H({\bf k}) = \Psi_\mathbf{k}^\dagger \begin{pmatrix}
\xi_{\bf k} & -\Delta_{\bf k} \\
-\Delta_{\bf k} & -\xi_{\bf k}
\end{pmatrix} \Psi_\mathbf{k}\,,
\end{aligned}
\end{equation}
with
\begin{equation}
\begin{aligned}
&\xi_{\bf k} = -2t (\cos k_x + \cos k_y)\,, &
&\Delta_{\bf k} = \Delta (\cos k_x - \cos k_y)\,,
\end{aligned}
\end{equation}
and Nambu spinors defined now as
\begin{equation}
\Psi_\mathbf{k} = \begin{pmatrix} c_{\mathbf{k},\uparrow} \\ c_{-\mathbf{k},\downarrow}^\dagger \end{pmatrix}.
\end{equation}

The difference with the particle-number conserving case considered above is that, in order to diagonalize $H({\bf k})$, which can be done with a Bogoliubov transformation, the corresponding unitary $\mathcal{U}_{\bf k}$ couples now two modes of opposite momenta. The ground state in momentum space can be then written as
\begin{equation}
\ket{\psi_0} = \bigotimes_{\mathbf{k}}\left(\mathcal{U}_{\mathbf{k}} \ket{0}_{\mathbf{k}, \uparrow}\ket{0}_{-\mathbf{k}, \downarrow}\right)\equiv\prod_{\mathbf{k}} \left( u_\mathbf{k}+ v_\mathbf{k} c_{\mathbf{k}, \uparrow}^\dagger c_{-\mathbf{k}, \downarrow}^\dagger \right) \ket{\Omega},
\end{equation}
where $u_{\bf k}$ and $v_{\bf k}$ are the usual Bogoliubov coefficients.

The protocol to prepare the ground state of the $d$-wave SC Hamiltonian is shown in Fig.~\ref{fig:SI_d_wave}. In this case, we start with a JW encoding where pairs of modes of opposite momenta correspond initialy to adjacent qubits, such that the Bogoliubov transformation can be applied in parallel by acting with the two-qubit unitaries $\mathcal{U}_{\bf k}$. The difference with the previous protocol is that, before applying the 2D FFFT to bring the state to real space, an extra permutation is required to reorder the $\downarrow$ modes according to ${\bf k}\rightarrow -{\bf k}$. As shown in the figure, this can be done in $O(1)$ depth using a sequence of 1D and 2D reflections.

\begin{figure}
    \centering
    \includegraphics[width=0.9\linewidth]{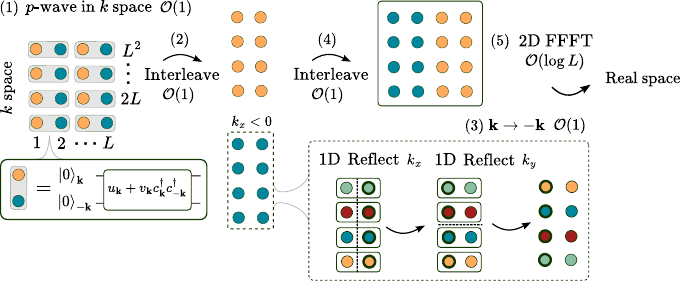}
    \caption{\textbf{State preparation circuit for translationally-invariant p-wave superconductors}. (1) The state is first prepared in momentum space, where two-modes circuits (gray rectangles) are applied in parallel to create local superpositions of $0$ and $2$ fermions shared between $\mathbf{k}$ (yellow circles) and $\mathbf{-k}$ modes (Bogoliubov  transformation). To capture every pair of modes, these are first ordered such that every $k_x>0$ mode is adjacent to a $k_x<0$ mode. (2) These two sets of modes are then separated using an interleave operation, and (3) the order for the $k_x<0$ is modified following a sequence $O(1)$-depth reflections. Finally, the modes in momentum space are arranged in the right order with an interleave operation, and the state is brought to real space with a 2D FFFT.}
    \label{fig:SI_p_wave}
\end{figure}

As a final example, consider the mean-field $p$-wave SC Hamiltonian for spinless fermions,
\begin{equation}
H = -t \sum_{\langle i,j \rangle} c_i^\dagger c_j + \sum_{\langle i,j \rangle} \left( \Delta_{ij} c_i^\dagger c_j^\dagger + \text{h.c.} \right)
\end{equation}
where $\Delta_{ij} = \Delta e^{i \theta_{ij}}$ represents the $p$-wave pairing with phase $\theta_{ij} = 0$ ($\pi / 2$) in the horizontal (vertical direction). In the right parameter regime, the ground state of the model corresponds to a chiral topological superconductor. In momentum space, the Hamiltonian can be written as
\begin{equation}
\begin{aligned}
&H = \sum_{\mathbf{k} | k_x>0} H({\bf k})\,, & &H({\bf k}) = \Psi_\mathbf{k}^\dagger \begin{pmatrix}
\xi_k & \Delta_k \\
\Delta_k^* & -\xi_k
\end{pmatrix} \Psi_\mathbf{k}\,,
\end{aligned}
\end{equation}
with
\begin{equation}
\begin{aligned}
&\xi_{\bf k} = -2t (\cos k_x + \cos k_y)\,, &
&\Delta_k = \Delta (\sin k_x + i \sin k_y)\,,
\end{aligned}
\end{equation}
and Nambu spinor
\begin{equation}
\Psi_\mathbf{k} = \begin{pmatrix} c_\mathbf{k} \\ c_{-\mathbf{k}}^\dagger \end{pmatrix}.
\end{equation}

The only difference with the $d$-wave SC case described above, is that now we should couple ${\bf k}$ and $-{\bf k}$ modes from the same species, since we are dealing with spinless fermions (note the restriction to $k_x>0$ in the sum above to avoid double counting). The protocol to prepare the $p$-wave superconducting state, described in detail in Fig.~\ref{fig:SI_p_wave}, is therefore very similar, with the subtlety that only $k_x<0$ modes undergo the ${\bf k}\rightarrow -{\bf k}$ permutation before the 2D FFFT.

\subsection{Generic TIFF state}

Finally, we consider the most general quadratic fermionic Hamiltonian in real space, given by
\begin{equation}
H = \sum_{\mathbf{R},\mathbf{R}'} \sum_{\alpha,\beta} \left[ t_{\alpha\beta}(\mathbf{R}-\mathbf{R}') c_{\mathbf{R},\alpha}^\dagger c_{\mathbf{R}',\beta} + \frac{1}{2} \left( \Delta_{\alpha\beta}(\mathbf{R}-\mathbf{R}') c_{\mathbf{R},\alpha}^\dagger c_{\mathbf{R}',\beta}^\dagger + \text{H.c.} \right) \right].
\end{equation}
Here, $c_{\mathbf{R},\alpha}^\dagger$ and $c_{\mathbf{R},\alpha}$ are the fermionic creation and annihilation operators, respectively, at unit cell $\mathbf{R}$ and internal index $\alpha\in \{1,\dots N_{\text orb}\}$, with $N_{\text orb}$ the number of orbitals within the unit cell, while $t_{\alpha\beta}(\mathbf{R}-\mathbf{R}')$ and $\Delta_{\alpha\beta}(\mathbf{R}-\mathbf{R}')$ represent the translational invariant tunneling and pairing amplitudes. In momentum space, the Hamiltonian can be written as
\begin{equation}
H = \sum_{\mathbf{k} \in \text{BZ}} \sum_{\alpha, \beta} \left[ c_{\mathbf{k},\alpha}^\dagger H_{\alpha\beta}(\mathbf{k}) c_{\mathbf{k},\beta} + \frac{1}{2} \left( c_{\mathbf{k},\alpha}^\dagger \Delta_{\alpha\beta}(\mathbf{k}) c_{-\mathbf{k},\beta}^\dagger + \text{H.c.} \right) \right],
\end{equation}
where the Bloch Hamiltonian matrices are defined as
\begin{align}
H_{\alpha\beta}(\mathbf{k}) &= \sum_{\mathbf{R}} t_{\alpha\beta}(\mathbf{R}) e^{-i \mathbf{k} \cdot \mathbf{R}}, \\
\Delta_{\alpha\beta}(\mathbf{k}) &= \sum_{\mathbf{R}} \Delta_{\alpha\beta}(\mathbf{R}) e^{-i \mathbf{k} \cdot \mathbf{R}}.
\end{align}
In terms of generalized Nambu spinors,
\begin{equation}
\Psi_{\mathbf{k}} =
\begin{bmatrix} 
c_{\mathbf{k},1} \\ 
c_{\mathbf{k},2} \\ 
\vdots \\ 
c_{\mathbf{k},N_{\text{orb}}} \\ 
c_{-\mathbf{k},1}^\dagger \\ 
c_{-\mathbf{k},2}^\dagger \\ 
\vdots \\ 
c_{-\mathbf{k},N_{\text{orb}}}^\dagger
\end{bmatrix},
\end{equation}
the Hamiltonian takes the Bogoliubov–de Gennes (BdG) form,
\begin{equation}
H = \frac{1}{2} \sum_{\mathbf{k} \in \text{BZ}} \Psi_{\mathbf{k}}^\dagger
\begin{bmatrix}
H(\mathbf{k}) & \Delta(\mathbf{k}) \\
\Delta^\dagger(\mathbf{k}) & -H^T(-\mathbf{k})
\end{bmatrix}
\Psi_{\mathbf{k}}.
\end{equation}

The protocol to prepare the ground state of this generalized translationally-invariant quadratic Hamiltonian follows the steps of the examples shown above. First, we consider $L^D N_{\text orb}$ qubits, one for each fermionic mode, where $L$ is the number of unit cells in each of the $D$ spatial dimensions. We also start with a JW encoding chosen such that, for every ${\mathbf k}$ with first component $k_0>0$, the set of $2N_{\text orb}$ modes corresponding to ${\mathbf k}$ and $-{\mathbf k}$ are contiguous. In that situation, we can prepare the ground state in momentum space by applying ${\bf k}$-dependent $2N_{\text orb}$-qubit unitaries, acting on parallel across the array. These unitaries can be decomposed in terms of Givens rotations, and implemented with a qubit circuit of depth $O(N_{\text orb})$, independent on the system size $L^D$. 

Finally, to bring the state to real space, we need to apply FFFTs independently to the $N_{\text orb}$ sets of $L^D$ $\mathbf{k}$ modes. To do this, the JW ordering has to be first modified such that the modes within each set are now contiguous, which can be done using one interleave operation. Moreover, the order of the $\mathbf{k}$ modes within each set has to be modified such that those with $k_0<0$ undergo a $\mathbf{k}\rightarrow -\mathbf{k}$ permutation, which can be done in $O(D)$ depth using a sequence of $O(1)$ reflections. After implementing in parallel the $N_{\text orb}$ $O(\log L)$-depth FFFTs, the modes are reordered with one final interleave operation such that modes within each unit cell are contiguous. The asymptotic scaling of the protocols corresponds therefore to a $O(\log L)$-depth circuit.

\subsection{$\log(L)$ depth preparation of Kitaev $B$ phase}

The ability to prepare arbitrary free-fermion states in $\log(L)$-depth also has applications for preparation of non-abelian topological orders.
Consider as a paradigmatic example the Kitaev honeycomb $B$ phase.
An exactly solvable state in the phase can be generated by combining a $C=1$ free fermion state, with a topologically-ordered toric code state.

The Kitaev honeycomb model is a spin-model defined on a honeycomb lattice, with qubits on the vertices~\cite{kitaev_anyons_2006}. The links on the lattice are divided into three groups, $x,y,z$, by their orientation.
The Hamiltonian takes the form
\begin{align}
    H_{KH} = -J_x \sum_{x-\mathrm{links}} X_i X_j - J_y \sum_{y-\mathrm{links}} Y_i Y_j - J_z \sum_{z-\mathrm{links}} Z_i Z_j.
\end{align}

As discussed in Ref.~\cite{kells_description_2009, Schmoll_honeycomb_2017} and others, the qubit operators can be re-written into non-interacting and interacting fermionic degrees of freedom. 
The non-interacting part of the Hamiltonian takes the form of the parent hamiltonian of a family of $p$-wave superconductors.
In contrast, the interacting part are plaquette operators, which are conversed integrals of motion that commute with each other and the non-interacting parts. These plaquette terms are essentially the parent Hamitonian of a Kitaev toric code.

As Ref.~\cite{Schmoll_honeycomb_2017} shows, this decomposition enables an efficient description of the Kitaev $B$ phase, using fermionic tensor networks.
When translated to circuits, their tensor network is equivalent to independently preparing a $p$-wave superconducting state, and the Kitaev toric code, and then intertwining the two using a fermionic permutation.
This fermionic permutation turns out to be an \textit{interleave}.
Therefore, each step can be efficiently implemented using our toolbox. The $p$-wave state can be prepared in $\log(N)$-depth. The toric code state can be prepared in $O(1)$-depth~\cite{raussendorf_measurement-based_2003,bluvstein_logical_2024, tantivasadakarn_hierarchy_2023, verresen2022efficientlypreparingschrodingerscat}.
The final intertwining operation can be prepared in $O(1)$-depth.
Hence the overall procedure only requires $O(\log L)$ depth, and these techniques provide a remarkably efficient route to generating long-range, non-abelian (chiral) topological order.

\section{Fault-Tolerant Compilation}
Ultimately, to accurately execute quantum circuits at scale, and simulate interesting fermionic systems, it will be crucial to incorporate quantum error-correction into the algorithm design.
The techniques introduced here for fast fermionic simulation are naturally compatible with state-of-the-art schemes for error-corrected and fault-tolerant quantum computing.

\subsection{Encoding fermionic permutations into error-correcting codes}

When working with encoded information, not all single and two-qubit operations are equal in terms of resources~\cite{Nielsen_Chuang_2010,bluvstein2025architecturalmechanismsuniversalfaulttolerant}.
It turns out that fermionic permutations $\mathcal{F}_p$ are \textit{Clifford circuits}, which are among the least-expensive operations in standard fault-tolerant architectures~\cite{Litinski_game_2019,fowler2019lowoverheadquantumcomputation}.
Furthermore, our approach leverages ancilla qubits and mid-circuit measurement to ensure the circuit depth remains small.
Recent work has shown that in platforms supporting non-local connectivity, all operations involving (1) Pauli-basis preparation of ancilla qubits (2) Clifford gates and (3) Pauli-basis measurements, can be performed fault-tolerantly, using only $O(1)$ rounds of stabilizer measurement between each operation~\cite{bluvstein_logical_2024,cain_correlated_decoding_2024,zhou2024algorithmicfaulttolerancefast}. 
This is because non-local connectivity enables \textit{transversal gates} to directly interact disjoint logical blocks.
The circuits produced by our method to generate $\mathcal{F}_p$ are immediately compatible with these techniques.

\begin{figure}
    \centering
    \includegraphics[width=0.95\linewidth]{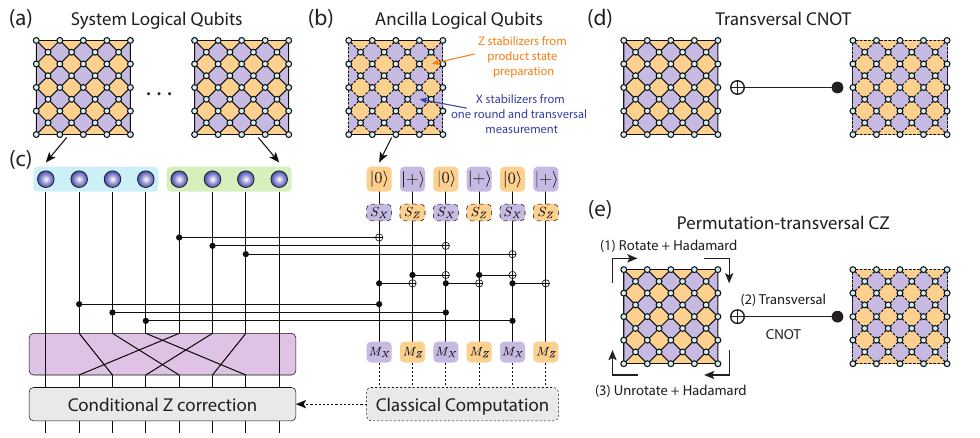}
    \caption{\textbf{Encoding fermionic permutations into error-correcting codes}. (a) System qubits, which are long-lived and persist through the computation, should be stored in an error-correcting code, such as the standard surface code. (b) In contrast, the temporary ancilla qubits do not need to be encoded fault-tolerantly into a QEC code in order to preserve fault-tolerance. Instead, it is sufficient to prepare them in a product state of the same size as the encoded qubits they will interact with, and apply one-round of stabilizer measurement~\cite{bluvstein_logical_2024}. (c) Then, the fermionic permutation circuit can be implemented by applying a sequence of CNOT and CZ gates, followed by a transversal measurement of the ancilla register, where all qubits are measured in the denoted basis. Ultimately the stabilizer information from state preparation and measurement propagates and affects the encoded logical qubits through the transversal gates. The circuit as a whole remains fualt-tolerant~\cite{cain_correlated_decoding_2024,zhou2024algorithmicfaulttolerancefast}.
    (d) Logical CNOT gates can be easily implemented using transversal CNOT gates, which act pairwise between qubits of the two codes. 
    (e) When working with surface codes, CZ gates are not strictly transversal, but can still be implemented fault-tolerantly using qubit permutation, physical Hadamard, and transversal CNOT.
    }
    \label{fig:qec_compatibility}
\end{figure}

Taking a close look at the structure of such fault-tolerant circuits, we note that the overall cost of the fermionic encoding is quite small, and comparable to the cost of a couple of rounds of stabilizer measurements.
We depict the interleave encoded into surface codes in Figure.~\ref{fig:qec_compatibility}, using an ancilla block, and with the CNOT cascade compiled into an $O(1)$ depth circuit.
We see that overall, the fermionic encoding part corresponds to a depth-4 Clifford circuit.
The ancilla qubits undergo one round of stabilizer measurement upon initialization, a depth-4 circuit as well, so in total this circuit requires a depth-8 circuit, following by transversal measurement.
The transversal CZ gates are more complex than CNOT gates, but can be applied using fold-transversal operations in the surface code.

\subsection{Compiling the fermionic fast fourier transform (FFFT) into a fault-tolerant gateset}

\begin{figure}
    \centering
    \includegraphics[width=0.95\linewidth]{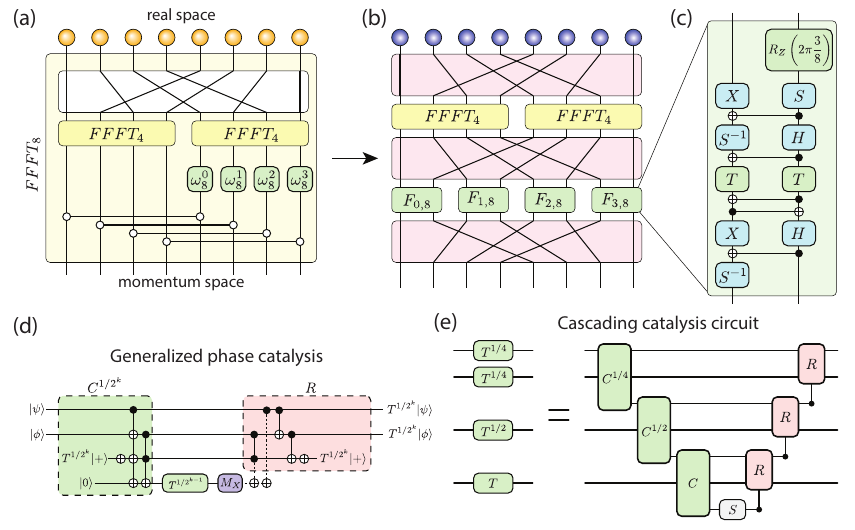}
    \caption{\textbf{Compilation of the FFFT}. (a) The FFFT can be constructed recursively~\cite{Ferris_spectral_2014}. Consider a 1D array with $N=8$ fermions (orange circles). The recursion starts by separating the even and odd index sites via a permutation. Then two FFFT on $N/2$ modes each is performed. Finally, single-mode phase gates are applied to the right side of the array, with rotation angles $R_z(\omega_N^{a})$ where $\omega_N^a = 2\pi\frac{a}{N}$. Then, long-range tunneling gates coupling the left and right half of the array are performed, which equally mix the two fermionic modes (see text).
    (b) The various parts of the circuit can be efficiently mapped to our qubit architecture, by using interleaves to perform the fermionic permutation, and compile the long-range gates. The resulting circuit contains three interleave layers at each level of recursion. The single-mode phase gate and long-range tunneling are combined into a single operation $F_{a,N}$~\cite{kivlichan_improved_2020,schuckert_fermion}.
    (c) In a fault-tolerant architecture, we need to further compile the two-qubit gates $F_{a,N}$ into an error-corrected gateset. In particular, $F_{a,N}$ breaks into the single-axis rotation $R_z(2\pi \frac{a}{N})$, and a short circuit implementing essentially a $\sqrt{\text{SWAP}}$ to simulate the tunneling gate. The short circuit contains Clifford operations and two $T$ gates, which are a standard component of the fault-tolerant gateset.
    (d) To efficiently synthesize single-axis rotations, we consider the generalized phase catalysis approach from Ref.~\cite{Gidney_catalyzed_2019}. In particular, the depicted circuit uses one Toffoli (or CCZ with $H$ gates), one $R_Z(2\theta)$ rotation, and a catalyst state $R_Z(\theta) \vert + \rangle$, to generate two rotations $R_Z(\theta)$. Notice that the catalyst state is not consumed in the process. 
    (e) When $N = 2^n$ is a power-of-two, the relevant single-axis rotations $R_z(2\pi \frac{a}{N})$ form a particularly nice set of angles for this approach~\cite{kivlichan_improved_2020}. Note that $T, T^{1/2}$, and $T^{1/4}$ correspond to $R_z(2\pi \frac{1}{8}),R_z(2\pi \frac{1}{16}),R_z(2\pi \frac{1}{32})$ respectively. 
    These gates are special, because after a few steps of the transformation $\theta \rightarrow 2\theta$ we arrive at $R(\theta = 2\pi \frac{1}{4}) = S$, which is a simple Clifford gate. As such, assuming access to the relevant catalyst states, we can chain together generalized phase catalysis circuits  to construct cascading structures, which require only CCZ and Clifford gates, to generate in parallel many single-axis rotations. Notice that any collection of single-axis rotations of the form $R_z(2\pi \frac{1}{2^n})$, whose rotation angles sum to an integer multiple of $2\pi \frac{1}{4}$, can be efficiently synthesized with $\leq$ one CCZ per operation on average. This is especially useful to generate the single-axis rotations associated with each layer of the FFFT, as discussed in the main text.
    }
    \label{fig:ft_compiling}
\end{figure}

Our approach to implementing fermionic permutations enables an asymptotic reduction in the Clifford cost of implementing the FFFT, by enabling the interleave fermionic permutation of $N$ modes to be performed with $\leq N$ ancilla, $\leq 2N$ Clifford gates, and depth $4$. 
This is in sharp contrast to an architecture based on local FSWAP networks, which requires depth $N/2$ and $O(N^2)$ total Clifford gates.
However, our discussion so far has not considered the costs associated with the non-Clifford rotations required to implement the two-fermion entangling gates. 
In particular, in the FFFT circuit, there are $n = \log_2(N)$ layers of alternating interleave operations, and two-qubit tunneling gates.  

Unlike the fermionic permutations, the tunneling gates typically require non-Clifford operations, which are generally more expensive in a fault-tolerant architecture.
There are two aspects which make non-Clifford gates expensive.
The first is that arbitrary rotations need to be further compiled into an gateset compatible with error correction, typically composed of Clifford gates along with either a T or CCZ gate.
The second is that fault-tolerantly generating T or CCZ gates is fundamentally more expensive than the associated Clifford operations, requiring e.g. distillation~\cite{Bravyi_universal_2005,knill2004postsel_threshold,knill2004postsel_schemes,Litinski_msd_2019}, code-switching~\cite{Paetznick_switching_2013,Anderson_switching_2014,bombin2015gaugefixing,bombin2016dimjump,Beverland_costof_2021}, or cultivation~\cite{Bomb_lowmagic_2024,Chamberland_lowoverhead_2020,gidney2024magicstatecultivationgrowing}.
As such, the combination of compilation overhead and fault-tolerant preparation, makes it such that for many quantum simulation algorithms, non-Clifford gates often dominate the cost of computation.

In this section, we build off the discussion in Ref.~\cite{kivlichan_improved_2020}, and show that for the FFFT, the non-Clifford \textit{compilation} cost is relatively small. In particular, we show that one CCZ gate and a handful of Clifford gates are sufficient to generate each two-fermion gate. 
Hence, the non-Clifford cost is minimal in the FFFT.
We achieve this by applying the phase rotation catalysis circuits from Ref.~\cite{Gidney_catalyzed_2019,kivlichan_improved_2020}. 
In particular, we show how catalysis circuits can be applied in a cascaded structure, to very efficiently generate many small angle rotations in parallel. Our analysis goes beyond Ref.~\cite{kivlichan_improved_2020}, as we consider rotations smaller than $\sqrt{T}$.

We further show that asymptotically, the depth of the catalysis circuit for each FFFT layer grows slowly as $O(\log N)$.
This is a minimal increase, and is considerably smaller then the depth overhead $O(N)$ associated with using FSWAP gates to generate the interleave.
In terms of number of gates, the number of non-Clifford gates required for the catalysis circuit scales as $O(N)$, in contrast to the $O(N^2)$ gate count for the FSWAP gates in the interleave.
Ultimately, the cost of generating T and CCZ gates to a fixed logical error rate will be a fixed multiple more expensive than Clifford gates, and understanding in detail this tradeoff is an area of significant active research~\cite{Litinski_msd_2019,gidney2024magicstatecultivationgrowing}.
Nevertheless, without our techniques, it is clear the Clifford cost associated with FSWAP networks would rapidly become the the dominant source of overhead, whereas with our techniques, the Clifford and non-Clifford cost will remain on equal footing. 

We start by presenting the recursive form of the FFFT construction (see Fig.~\ref{fig:ft_compiling}a).
An FFFT on $N$ modes, denoted $FFFT_N$, can be constructed via the following decomposition (see Ref.~\cite{Ferris_spectral_2014} for more details).
\begin{align}
    \sum_{x=0}^{N-1} e^{\frac{2\pi i k x}{N}} c_x^{\dagger} = \sum_{x'=0}^{N/2 - 1} e^{\frac{2\pi i k x'}{N/2}} c_{2x'}^{\dagger} + e^{\frac{2\pi i k}{N}} \sum_{x'=0}^{N/2 - 1} e^{\frac{2\pi i k x'}{N/2}} c_{2x'+1}^{\dagger}
\end{align}

From the point-of-view of a fermionic circuit, this can be implemented using one interleave permutation, two $FFFT_{N/2}$ operations, single-mode phase gates, and a layer of non-local tunneling operations. 
When compiling to a qubit circuit (see Fig.~\ref{fig:ft_compiling}b), the non-local tunneling can be generated using two interleave circuits, and two qubit gates.
The two qubit gates are spatially inhomogenous, and take the form 
\begin{align}
    F_{a,N} = F_2 \left(I \otimes R_Z(2\pi \frac{a}{N})\right),
\end{align}
i.e. a single-site rotation followed by a two-qubit gate $F_2$. The $F_2$ operation equally mixes the two fermionic modes
\begin{align}
    F_2 = \left(\begin{array}{cccc}
        1\\
        & \frac{1}{\sqrt{2}} & \frac{1}{\sqrt{2}}\\
        & \frac{1}{\sqrt{2}} & \frac{-1}{\sqrt{2}}\\
        &  &  & -1
    \end{array}\right),
\end{align}
and can be implemented with a short sequence of Clifford gates and two $T$ gates (see Fig.~\ref{fig:ft_compiling}).

To generate single-axis rotations, we consider the generalized phase catalysis approach of Ref.~\cite{Gidney_catalyzed_2019}. In this approach, a Toffoli gate (CCZ + Hadamard), rotation gate $R_Z(2\theta)$, and catalyst state $R(\theta) \vert + \rangle$, are used to generate two rotations $R_Z(\theta)$ (see Fig.~\ref{fig:ft_compiling}d). 
For example, assuming access to a $T$ catalyst, the two $T$ gates in $F_2$ can be generated with just one CCZ gate and Clifford operations. This is because the double-angle rotation $T^2 = S$ is Clifford.  
Indeed, when $N = 2^n$ is an integer multiple of two, all the single-axis rotations $R_Z(2\pi \frac{1}{N})$ eventually form $S$ gates when combined. 
We can leverage this to efficiently generate all of the single-site rotations associated with one layer of the FFFT together, by chaining together generalized phase catalysis circuits in a cascading structure (see Fig.~\ref{fig:ft_compiling}e).

We make the following simple observation. 
At layer $n$, the FFFT acts on $2^n$ modes, and the phases we need to generate take the form $R_z(2\pi \frac{a}{2^n})$, where $a = 0,...,2^{n-1}-1$.
Pairs of rotations $(a,2^{n-1}-a)$ can be catalyzed simultaneously by making the following substitution,
\begin{align}
    R_z(2\pi \frac{a}{b}) = Z R_z(2\pi (\frac{a}{b}-1)) &= Z R_z(2\pi \frac{a-b}{b}) \\
    X R_z(2\pi \frac{a-b}{b}) X &= R_z(2\pi \frac{b-a}{b})
\end{align}
so when $b = 2^{n-1}$, we see that $R_z(2\pi \frac{a}{2^{n-1}})$ and $R_z(2\pi \frac{2^{n-1} - a}{2^{n-1}})$ are equivalent up to global Pauli's.

Hence, to generate all non-trivial rotations, we can synthesize two of each gate $R_z(2\pi \frac{a}{2^{n-1}})$ for $a=0,...,2^{n-2}-1$, using $\leq 2^{n-2}$ CCZ gates, $\leq 2^{n-2}$ catalyst states, and the double-angle rotations $R_z(2\pi \frac{a}{2^{n-2}})$.
These double-angle rotations are exactly the same angles that are required at level $n-1$ of the FFFT.
Hence, the argument proceeds recursively, and gates at level $n-1$ can be analogously paired up using $2^{n-3}$ CCZ gates and catalyst states, to reduce the problem down to synthesizing the angles at level $n-2$.
Starting with $N=2^n$ and repeating this procedure recursively until we reach the base level $k=1$, the overall procedure uses $\frac{N}{4} + \frac{N}{8} + \frac{N}{16} + ... \leq N/2$ CCZ gates and catalyst states. 

In principle, the $N$ catalyst states can be prepared ahead of time, and used repeatedly throughout the computation. Preparing a generic state $R_Z(\theta) \vert + \rangle$ can be performed with rotation synthesis, whose asymptotic cost depends on the target precision $~\log 1/\epsilon$~\cite{kliuchnikov2013fastefficientexactsynthesis,ross2016optimalancillafreecliffordtapproximation,paetznick2014repeatuntilsuccessnondeterministicdecompositionsinglequbit}. In practice, tens of non-Clifford operations are required to generate a single rotation~\cite{Bocharov_RUS_2015}. Hence, by performing this expensive computation once, and using the cascaded catalysis structure described above, future rotations can be generated using only one CCZ gate. Therefore, the expensive $\log 1/\epsilon$ cost is \textit{added} to overall the circuit volume of the simulation ($NT$), instead of \textit{multiplying} the circuit volume. The dominant cost will scale with circuit volume, which is determined by the number of catalyst steps required. Here, we have shown the number of CCZ gates arising from the catalysis in the FFFT is $\leq NT$.
However, in practice, there will also be cost associated with storing $N$ catalyst states during the computation. Ultimately, the tradeoff between storage and synthesis cost will depend on architectural considerations.

Next, we explicitly compute and tabulate the number of CCZ gates per mode required to generate the two-fermion gates $F_{a,N}$ at level $N=2^n$ of the FFFT recursion, up to $N = 64$. 
We define the CCZ cost $N_{CCZ} = N_{CCZ}^{(F_2)} + N_{CCZ}^{(R_Z)}$ as having two contributions, associated with the fixed portion $F_2$ and inhomogeneous portion, $R_Z(2\pi \frac{a}{N})$.
For the fixed portion $F_2$ which mixes the two modes, there are always $N/2$ CCZ gates required to generate the pair of T states, and $N_{CCZ}^{(F_2)} = N/2$.
For the inhomogeneous portion, $R_Z(2\pi \frac{a}{N})$, we use the cascading catalysis circuit described above, with tree-like structure, but stop the cascade when we reach $S$ gates.
As such, notice for $N=2,4$, $N_{CCZ}^{(R_Z)}=0$ because all gates are Clifford.
Beyond this base case, the cost $N^{(F_2)}_{CCZ}$ is described by a simple recursion formulas
\begin{align}
    N_{CCZ}^{(R_Z)}(n) = (2^{n-2}-1) + N_{CCZ}^{(R_Z)}(n-1), \,
    N_{CCZ}^{(R_Z)}(2) = 0.
\end{align}
In particular, $2^{n-2}-1$ is the number of non-trivial, non-Clifford rotations that are introduced at level $n$ of the tree structure. Using this approach, we see the average CCZ cost per qubit of $N^{(R_Z)}_{CCZ}$ starts small, and nears $0.5$ asymptotically (see Table~\ref{tab:ccz_estimates}).

\begin{table}
    \centering
    \begin{tabular}{c|cccccccc}
    $N$ &         2&  4&    8&   16&   32&    64&  128&  256\\
    \hline
    $N^{(R_Z)}_{CCZ}$&    0&  0&    1&    4&   11&    26&   57&  120\\
    $N^{(R_Z)}_{CCZ}/N$&  0&  0&0.125& 0.25&0.344& 0.406&0.445&0.469\\
    \end{tabular}
    \caption{Total number and average number of CCZ gates required for the outermost layer of $FFFT_{N}$.}
    \label{tab:ccz_estimates}
\end{table}

We contrast this with the cost of an FSWAP network, for implementing the three interleave operations, as also discussed in Ref.~\cite{kivlichan_improved_2020}.
We denote this by $N^{(FSWAP)}$
As the first non-trivial base case, consider $N=4$. We need one FSWAP gate to switch modes $1 \leftrightarrow 2$, and hence would have $N^{(FSWAP)}(n=2)/3=1$.
To understand the recursion, next consider $N=8$, as depicted in (Figure.~\ref{fig:ft_compiling}).
We see that
\begin{align}
    N^{(FSWAP)}(n=3)/3 = 3 + 2 + 1 = 6
\end{align}
because the mode that travels the furthest intersects three modes, the next two modes, and the next one mode.
From this, we can generalize to arbitrary $n$,
\begin{align}
    N^{(FSWAP)}(n \geq 2)/3 = (2^{n-1}-1) + (2^{n-1}-2) + ... + 1 = 2^{2n-3} - 2^{n-2}
\end{align}

For reference, we also compute the Clifford cost associated with our dynamical JW encoding approach for these system sizes. We denote this cost $N^{(DJW)}$, and following the construction in Fig.~\ref{fig:qec_compatibility}, get the asymptotic expression
\begin{align}
    N^{(DJW)}(n \geq 3)/3 = 2*(2^{n-1}-1) + 2*(2^{n-1}-2) = 2^{n+1} - 6
\end{align}
which clearly shows the asymptotic overhead per qubit is per interleave is $\leq 2$.
After tabulating the numbers, we see that the DJW begins to outperform a swap network based approach around $N = 16$, and by $N=256$, generates already a 15x reduction in gate cost. 

Finally, we note that by combining dynamical JW to perform the interleaves, and cascading catalysis to perform the two fermion gates, the total gate cost of each layer of the FFFT can be kept constant. Further, after including the Clifford gates in $F_2$, the ratio of Clifford gates to CCZ gates is roughly 20-to-1, suggesting that in practice the relative costs of these two circuit subroutines may be comparable.

\begin{table}
    \centering
    \begin{tabular}{c|cccccccc}
    $N$ &                2&   4&    8&   16&    32&    64&  128&  256\\
    \hline
    $N^{(FSWAP)}/3$ &    0&   1&    6&   28&   120&    496&   2016&  8128\\
    $N^{(FSWAP)}/N/3$&   0&0.25& 0.75& 1.75&  3.75&   7.75&  15.75&  31.75\\
    $N^{DJW}/3$&         0&   2&   10&   26&    58&    122&    250&    506\\
    $N^{DJW}/N/3$&       0& 0.5& 1.25& 1.63&  1.81&   1.91&  1.95 &   1.98
    \end{tabular}
    \caption{\textbf{1D Interleave, Clifford}. Total and average gate count using an FSWAP network, or our dynamic Jordan-Wigner (DJW) encoding.}
    \label{tab:ffft_fswap_interleave}
\end{table}

\subsection{Considerations arising from 2D FFFT}

The advantage of the dynamical JW approach over FSWAP networks becomes even more dramatic when considering more complex applications of the FFFT, for example to perform higher-dimensional FFFT. 
In Figure.~\ref{fig:fig4}, we discuss the 2D FFFT, and expand on the construction here.

Consider an $L_r \times L_c$ lattice of modes indexed by $(r,c)$. Further, assume that each side length $L_r, L_c \in O(\sqrt{N})$ scales as the square-root of the total number of qubits.
In the row-wise indexing, $i = r L_c + c$, we can treat each row as a separate 1D chain and apply the 1D FFFT.
While this can be done with constant-overhead using the JW approach, using FSWAP networks the overhead is $O(L) \sim O(\sqrt{N})$. While this is expensive, as we saw in the previous section for $L \sim 20$ the additional overhead is not too large.

In contrast, to perform the 2D FFFT, we then need to apply 1D FFFT to each column.
Using FSWAP networks to generate long-range connectivity between rows leads to $O(N)$ overhead, scaling with the system size. In the dynamical JW approach, we could instead apply a 2D reflection to switch from row- to column- indexing, which can be performed with $O(1)$ overhead. An intermediate approach is to leverage a static fermion-to-qubit encoding with native 2D connectivity, like the Verstrate-Cirac-Kitaev encoding~\cite{verstraete_mapping_2005,kitaev_anyons_2006}, or the compact encoding~\cite{derby_compact_2021}. In this encoding, local operations can be efficiently performed in both row-wise and column-wise directions. Hence, the asymptotic scaling of the 2D FFFT is limited only be the side-length $O(L)$. 
However, these encodings also come with some cost, as two-body fermionic operators map to higher-body qubit operators. For example, in the compact encoding, each two-fermion tunneling operator is mapped to a three-body qubit operator, increasing the Clifford overhead by a small multiplicative factor.
Therefore, for sufficiently small systems, the static encoding may be the simplest  method for generating a 2D FFFT. Nevertheless, asymptotically, the cost of our method is $O(L / \log(L))$-times lower then other known techniques, and determining the crossover point will require a more careful analysis.

\end{document}